\input harvmac

%
\message{S-Tables Macro v1.0, ACS, TAMU (RANHELP@VENUS.TAMU.EDU)}
%
%
\newhelp\stablestylehelp{You must choose a style between 0 and 3.}%
\newhelp\stablelinehelp{You should not use special hrules when
stretching
a table.}%
\newhelp\stablesmultiplehelp{You have tried to place an S-Table
inside another S-Table.  I would recommend not going on.}%
%
%
\newdimen\stablesthinline
\stablesthinline=0.4pt
\newdimen\stablesthickline
\stablesthickline=1pt
%
%
\newif\ifstablesborderthin
\stablesborderthinfalse
\newif\ifstablesinternalthin
\stablesinternalthintrue
\newif\ifstablesomit
\newif\ifstablemode
\newif\ifstablesright
\stablesrightfalse
%
%
\newdimen\stablesbaselineskip
\newdimen\stableslineskip
\newdimen\stableslineskiplimit
%
%
\newcount\stablesmode
\newcount\stableslines
\newcount\stablestemp
\stablestemp=3
\newcount\stablescount
\stablescount=0
\newcount\stableslinet
\stableslinet=0
%
%
%
\newcount\stablestyle
\stablestyle=0
%
%
\def\stablesleft{\quad\hfil}%
\def\stablesright{\hfil\quad}%
%
%
\catcode`\|=\active%
%
%
\newcount\stablestrutsize
\newbox\stablestrutbox
\setbox\stablestrutbox=\hbox{\vrule height10pt depth5pt width0pt}
\def\stablestrut{\relax\ifmmode%
                         \copy\stablestrutbox%
                       \else%
                         \unhcopy\stablestrutbox%
                       \fi}%
%
%
\newdimen\stablesborderwidth
\newdimen\stablesinternalwidth
\newdimen\stablesdummy
\newcount\stablesdummyc
\newif\ifstablesin
\stablesinfalse
%
%
%
%
%
\def\stablesadj{%
  \ifcase\stablestyle%
    \hbox to \hsize\bgroup\hss\vbox\bgroup%
  \or%
    \hbox to \hsize\bgroup\vbox\bgroup%
  \or%
    \hbox to \hsize\bgroup\hss\vbox\bgroup%
  \or%
    \hbox\bgroup\vbox\bgroup%
  \else%
    \errhelp=\stablestylehelp%
    \errmessage{Invalid style selected, using default}%
    \hbox to \hsize\bgroup\hss\vbox\bgroup%
  \fi}%
\def\stablesend{\egroup%
  \ifcase\stablestyle%
    \hss\egroup%
  \or%
    \hss\egroup%
  \or%
    \egroup%
  \or%
    \egroup%
  \else%
    \hss\egroup%
  \fi}%
\def\stablestart{%
  \ifstablesin%
    \errhelp=\stablesmultiplehelp%
    \errmessage{An S-Table cannot be placed within an S-Table!}%
  \fi
  \global\stablesintrue%
  \global\advance\stablescount by 1%
  \message{<S-Tables Generating Table \number\stablescount}%
  \begingroup%
  \stablestrutsize=\ht\stablestrutbox%
  \advance\stablestrutsize by \dp\stablestrutbox%
  \ifstablesborderthin%
    \stablesborderwidth=\stablesthinline%
  \else%
    \stablesborderwidth=\stablesthickline%
  \fi%
  \ifstablesinternalthin%
    \stablesinternalwidth=\stablesthinline%
  \else%
    \stablesinternalwidth=\stablesthickline%
  \fi%
  \tabskip=0pt%
  \stablesbaselineskip=\baselineskip%
  \stableslineskip=\lineskip%
  \stableslineskiplimit=\lineskiplimit%
  \offinterlineskip%
  \def\borderrule{\vrule width \stablesborderwidth}%
  \def\internalrule{\vrule width \stablesinternalwidth}%
  \def\thinline{\noalign{\hrule height \stablesthinline}}%
  \def\thickline{\noalign{\hrule height \stablesthickline}}%
  \def\trule{\omit\leaders\hrule height \stablesthinline\hfill}%
  \def\ttrule{\omit\leaders\hrule height \stablesthickline\hfill}%
  \def\tttrule##1{\omit\leaders\hrule height ##1\hfill}%
  \def\stablesel{&\omit\global\stablesmode=0%
    \global\advance\stableslines by 1\borderrule\hfil\cr}%
  \def\el{\stablesel&}%
  \def\elt{\stablesel\thinline&}%
  \def\eltt{\stablesel\thickline&}%
  \def\elttt##1{\stablesel\noalign{\hrule height ##1}&}%
  \def\elspec{&\omit\hfil\borderrule\cr\omit\borderrule&%
              \ifstablemode%
              \else%
                \errhelp=\stablelinehelp%
                \errmessage{Special ruling will not display properly}%
              \fi}%
  \def\stmultispan##1{\mscount=##1 \loop\ifnum\mscount>3
\stspan\repeat}%
  \def\stspan{\span\omit \advance\mscount by -1}%
  \def\multicolumn##1{\omit\multiply\stablestemp by ##1%
     \stmultispan{\stablestemp}%
     \advance\stablesmode by ##1%
     \advance\stablesmode by -1%
     \stablestemp=3}%
  \def\multirow##1{\stablesdummyc=##1\parindent=0pt\setbox0\hbox\bgroup%
    \aftergroup\emultirow\let\temp=}
  \def\emultirow{\setbox1\vbox to\stablesdummyc\stablestrutsize%
    {\hsize\wd0\vfil\box0\vfil}%
    \ht1=\ht\stablestrutbox%
    \dp1=\dp\stablestrutbox%
    \box1}%
%
  \def\stpar##1{\vtop\bgroup\hsize ##1%
     \baselineskip=\stablesbaselineskip%
     \lineskip=\stableslineskip%

\lineskiplimit=\stableslineskiplimit\bgroup\aftergroup\estpar\let\temp=}%
  \def\estpar{\vskip 6pt\egroup}%
  \def\stparrow##1##2{\stablesdummy=##2%
     \setbox0=\vtop to ##1\stablestrutsize\bgroup%
     \hsize\stablesdummy%
     \baselineskip=\stablesbaselineskip%
     \lineskip=\stableslineskip%
     \lineskiplimit=\stableslineskiplimit%
     \bgroup\vfil\aftergroup\estparrow%
     \let\temp=}%
  \def\estparrow{\vfil\egroup%
     \ht0=\ht\stablestrutbox%
     \dp0=\dp\stablestrutbox%
     \wd0=\stablesdummy%
     \box0}%
  \def|{\global\advance\stablesmode by 1&&&}%
  \def\|{\global\advance\stablesmode by 1&\omit\vrule width 0pt%
         \hfil&&}%
\def\vt{\global\advance\stablesmode
by 1&\omit\vrule width \stablesthinline%
          \hfil&&}%
  \def\vtt{\global\advance\stablesmode by 1&\omit\vrule width
\stablesthickline%
          \hfil&&}%
  \def\vttt##1{\global\advance\stablesmode by 1&\omit\vrule width ##1%
          \hfil&&}%
  \def\vtr{\global\advance\stablesmode by 1&\omit\hfil\vrule width%
           \stablesthinline&&}%
  \def\vttr{\global\advance\stablesmode by 1&\omit\hfil\vrule width%
            \stablesthickline&&}%
\def\vtttr##1{\global\advance\stablesmode
 by 1&\omit\hfil\vrule width ##1&&}%
  \stableslines=0%
  \stablesomitfalse}
\def\stablesdef{\bgroup\stablestrut\borderrule##\tabskip=0pt plus 1fil%
  &\stablesleft##\stablesright%
  &##\ifstablesright\hfill\fi\internalrule\ifstablesright\else\hfill\fi%
  \tabskip 0pt&&##\hfil\tabskip=0pt plus 1fil%
  &\stablesleft##\stablesright%
  &##\ifstablesright\hfill\fi\internalrule\ifstablesright\else\hfill\fi%
  \tabskip=0pt\cr%
  \ifstablesborderthin%
    \thinline%
  \else%
    \thickline%
  \fi&%
}%
\def\endtable{\advance\stableslines by 1\advance\stablesmode by 1%
   \message{- Rows: \number\stableslines, Columns:
\number\stablesmode>}%
   \stablesel%
   \ifstablesborderthin%
     \thinline%
   \else%
     \thickline%
   \fi%
   \egroup\stablesend%
\endgroup%
\global\stablesinfalse}
%

\overfullrule=0pt \abovedisplayskip=12pt plus 3pt minus 3pt
\belowdisplayskip=12pt plus 3pt minus 3pt

\noblackbox
\input epsf
\newcount\figno
\figno=0
\def\fig#1#2#3{
\par\begingroup\parindent=0pt\leftskip=1cm\rightskip=1cm\parindent=0pt
\baselineskip=11pt \global\advance\figno by 1 \midinsert
\epsfxsize=#3 \centerline{\epsfbox{#2}} \vskip 12pt
\centerline{{\bf Figure \the\figno:} #1}\par
\endinsert\endgroup\par}
\def\figlabel#1{\xdef#1{\the\figno}}

\def\IR{\relax{\rm I\kern-.18em R}}


\font\cmss=cmss10 \font\cmsss=cmss10 at 7pt
\def\rlx{\relax\leavevmode}
\def\inbar{\vrule height1.5ex width.4pt depth0pt}
\def\IC{\relax\,\hbox{$\inbar\kern-.3em{\rm C}$}}
\def\IN{\relax{\rm I\kern-.18em N}}
\def\IP{\relax{\rm I\kern-.18em P}}
\def\ZZ{\rlx\leavevmode\ifmmode\mathchoice{\hbox{\cmss Z\kern-.4em Z}}
 {\hbox{\cmss Z\kern-.4em Z}}{\lower.9pt\hbox{\cmsss Z\kern-.36em Z}}
 {\lower1.2pt\hbox{\cmsss Z\kern-.36em Z}}\else{\cmss Z\kern-.4em
 Z}\fi}
\def\IZ{\relax\ifmmode\mathchoice
{\hbox{\cmss Z\kern-.4em Z}}{\hbox{\cmss Z\kern-.4em Z}}
{\lower.9pt\hbox{\cmsss Z\kern-.4em Z}} {\lower1.2pt\hbox{\cmsss
Z\kern-.4em Z}}\else{\cmss Z\kern-.4em Z}\fi}

\def\narrowplus{\kern -.04truein + \kern -.03truein}
\def\narrowminus{- \kern -.04truein}
\def\narrowminussub{\kern -.02truein - \kern -.01truein}

\def\e{{\epsilon}}

\def\frac#1#2{{#1\over #2}}

\def\IZ{\relax\ifmmode\mathchoice
{\hbox{\cmss Z\kern-.4em Z}}{\hbox{\cmss Z\kern-.4em Z}}
{\lower.9pt\hbox{\cmsss Z\kern-.4em Z}} {\lower1.2pt\hbox{\cmsss
Z\kern-.4em Z}}\else{\cmss Z\kern-.4em Z}\fi}
\def\IB{\relax{\rm I\kern-.18em B}}
\def\IC{{\relax\hbox{$\inbar\kern-.3em{\rm C}$}}}
\def\ID{\relax{\rm I\kern-.18em D}}
\def\IE{\relax{\rm I\kern-.18em E}}
\def\IF{\relax{\rm I\kern-.18em F}}
\def\IG{\relax\hbox{$\inbar\kern-.3em{\rm G}$}}
\def\IGa{\relax\hbox{${\rm I}\kern-.18em\Gamma$}}
\def\IH{\relax{\rm I\kern-.18em H}}
\def\II{\relax{\rm I\kern-.18em I}}
\def\IK{\relax{\rm I\kern-.18em K}}
\def\IP{\relax{\rm I\kern-.18em P}}

\font\cmss=cmss10 \font\cmsss=cmss10 at 7pt
\def\IR{\relax{\rm I\kern-.18em R}}

\def\1{{\bf 1}}
\def\3{{\bf 3}}
\def\7{{\bf 7}}
\def\2{{\bf 2}}
\def\8{{\bf 8}}

\def\hat{\widehat}
\def\quabla{{\sqcap}\!\!\!\!{\sqcup}}

\def\o{\over}
%

%
%
\def\eqnn#1{\xdef #1{(\secsym\the\meqno)}\writedef{#1\leftbracket#1}%
\global\advance\meqno by1\wrlabeL#1}
\def\eqna#1{\xdef #1##1{\hbox{$(\secsym\the\meqno##1)$}}
\writedef{#1\numbersign1\leftbracket#1{\numbersign1}}%
\global\advance\meqno by1\wrlabeL{#1$\{\}$}}
\def\eqn#1#2{\xdef #1{(\secsym\the\meqno)}\writedef{#1\leftbracket#1}%
\global\advance\meqno by1$$#2\eqno#1\eqlabeL#1$$}



\lref\vafai{C.~Vafa, ``Superstrings and topological strings at
large N,'' J.\ Math.\ Phys.\  {\bf 42}, 2798 (2001),
hep-th/0008142.}

\lref\civ{F.~Cachazo, K.~A.~Intriligator and C.~Vafa, ``A large N
duality via a geometric transition,'' Nucl.\ Phys.\ B {\bf 603}, 3
(2001), hep-th/0103067.}

\lref\grayone{A. Gray, L. Hervella, ``The sixteen classes of
almost Hermitian manifolds and their linear invariants,'' Ann.
Mat. Pura Appl.(4) {\bf 123} (1980) 35.}

\lref\salamon{ S. Chiossi, S. Salamon, ``The intrinsic torsion of
$SU(3)$ and $G_2$ structures,''
 Proc. conf. Differential Geometry Valencia 2001.}

\lref\gauntlett{J.~P.~Gauntlett, D.~Martelli and D.~Waldram,
``Superstrings with intrinsic torsion,'' Phys.\ Rev.\ D {\bf 69},
086002 (2004), hep-th/0302158.}

\lref\gukov{S.~Gukov, ``Solitons, superpotentials and
calibrations,'' Nucl.\ Phys.\ B {\bf 574}, 169 (2000),
hep-th/9911011.}

\lref\tp{G.~Papadopoulos and A.~A.~Tseytlin, ``Complex geometry of
conifolds and 5-brane wrapped on 2-sphere,'' Class.\ Quant.\
Grav.\  {\bf 18}, 1333 (2001).hep-th/0012034.}

\lref\lust{G.~L.~Cardoso, G.~Curio, G.~Dall'Agata, D.~Lust,
P.~Manousselis and G.~Zoupanos, ``Non-Kaehler string backgrounds
and their five torsion classes,'' Nucl.\ Phys.\ B {\bf 652}, 5
(2003), hep-th/0211118.}

\lref\louis{S.~Gurrieri, J.~Louis, A.~Micu and D.~Waldram,
``Mirror symmetry in generalized Calabi-Yau compactifications,''
Nucl.\ Phys.\ B {\bf 654}, 61 (2003), hep-th/0211102.}

\lref\rstrom{A.~Strominger, ``Superstrings with torsion,'' Nucl.\
Phys.\ B {\bf 274}, 253 (1986).}

\lref\mal{J.~M.~Maldacena, ``The large N limit of superconformal
field theories and supergravity,'' Adv.\ Theor.\ Math.\ Phys.\
{\bf 2}, 231 (1998) [Int.\ J.\ Theor.\ Phys.\  {\bf 38}, 1113
(1999), hep-th/9711200.}

\lref\mn{J.~M.~Maldacena and C.~Nunez, ``Towards the large N limit
of pure N = 1 super Yang Mills,'' Phys.\ Rev.\ Lett.\  {\bf 86},
588 (2001), hep-th/0008001.}

\lref\bdkt{M.~Becker, K.~Dasgupta, A.~Knauf and R.~Tatar,
``Geometric transitions, flops and non-Kaehler manifolds. I,''
hep-th/0403288.}

\lref\civ{F.~Cachazo, K.~A.~Intriligator and C.~Vafa, ``A large N
duality via a geometric transition,'' Nucl.\ Phys.\ B {\bf 603}, 3
(2001), hep-th/0103067.}

\lref\syz{A.~Strominger, S.~T.~Yau and E.~Zaslow, ``Mirror
symmetry is T-duality,'' Nucl.\ Phys.\ B {\bf 479}, 243 (1996),
hep-th/9606040.}

\lref\tduality{E.~Bergshoeff, C.~M.~Hull and T.~Ortin, ``Duality
in the type II superstring effective action,'' Nucl.\ Phys.\ B
{\bf 451}, 547 (1995), hep-th/9504081; P.~Meessen and T.~Ortin,
``An Sl(2,Z) multiplet of nine-dimensional type II supergravity
theories,'' Nucl.\ Phys.\ B {\bf 541}, 195 (1999),
hep-th/9806120.}

\lref\eot{J.~D.~Edelstein, K.~Oh and R.~Tatar, ``Orientifold,
 geometric transition and large N duality for SO/Sp gauge  theories,''
JHEP {\bf 0105}, 009 (2001), hep-th/0104037.}

\lref\dotu{K.~Dasgupta, K.~Oh and R.~Tatar, {``Geometric
transition, large N dualities and MQCD dynamics,''} Nucl.\ Phys.\
B {\bf 610}, 331 (2001), hep-th/0105066; {``Open/closed string
dualities and Seiberg duality from geometric transitions in
M-theory,''} JHEP {\bf 0208}, 026 (2002), hep-th/0106040.}

\lref\dotd{K.~Dasgupta, K.~h.~Oh, J.~Park and R.~Tatar,
``Geometric transition versus cascading solution,'' JHEP {\bf
0201}, 031 (2002), hep-th/0110050.}

\lref\ohta{K.~Ohta and T.~Yokono, ``Deformation of conifold and
intersecting branes,'' JHEP {\bf 0002}, 023 (2000),
hep-th/9912266.}

\lref\dott{K.~h.~Oh and R.~Tatar, ``Duality and confinement in N =
1 supersymmetric theories from geometric  transitions,'' Adv.\
Theor.\ Math.\ Phys.\  {\bf 6}, 141 (2003), hep-th/0112040.}

\lref\edelstein{J.~D.~Edelstein and C.~Nunez, ``D6 branes and
M-theory geometrical transitions from gauged supergravity,'' JHEP
{\bf 0104}, 028 (2001), hep-th/0103167.}

\lref\candelas{P.~Candelas and X.~C.~de la Ossa, ``Comments on
conifolds,'' Nucl.\ Phys.\ B {\bf 342}, 246 (1990).}

\lref\chsw{P.~Candelas, G.~T.~Horowitz, A.~Strominger and
E.~Witten, ``Vacuum Configurations For Superstrings,'' Nucl.\
Phys.\ B {\bf 258}, 46 (1985).}

\lref\minasianone{R.~Minasian and D.~Tsimpis, ``Hopf reductions,
fluxes and branes,'' Nucl.\ Phys.\ B {\bf 613}, 127 (2001),
hep-th/0106266.}

\lref\gkp{S.~B.~Giddings, S.~Kachru and J.~Polchinski,
``Hierarchies from fluxes in string compactifications,'' Phys.\
Rev.\ D {\bf 66}, 106006 (2002), hep-th/0105097.}

\lref\uranga{A.~M.~Uranga, ``Brane configurations for branes at
conifolds,'' JHEP {\bf 9901}, 022 (1999), hep-th/9811004.}

\lref\tatar{R.~de Mello Koch, K.~Oh and R.~Tatar, ``Moduli space
for conifolds as intersection of orthogonal D6 branes,'' Nucl.\
Phys.\ B {\bf 555}, 457 (1999), hep-th/9812097.}
 
\lref\stef{
B.~J.~Stefanski,
``Gravitational couplings of D-branes and O-planes,''
Nucl.\ Phys.\ B {\bf 548}, 275 (1999), hep-th/9812088.}

\lref\moral{
J.~F.~Morales, C.~A.~Scrucca and M.~Serone,
Nucl.\ Phys.\ B {\bf 552}, 291 (1999), hep-th/9812071.}

\lref\serone{M.~Serone and M.~Trapletti,
``String vacua with flux from freely-acting obifolds,''
JHEP {\bf 0401}, 012 (2004), hep-th/0310245.}

\lref\djm{K.~Dasgupta, D.~P.~Jatkar and S.~Mukhi, ``Gravitational
couplings and Z(2) orientifolds,'' Nucl.\ Phys.\ B {\bf 523}, 465
(1998), hep-th/9707224.}

\lref\dmconi{K.~Dasgupta and S.~Mukhi, ``Brane constructions,
conifolds and M-theory,'' Nucl.\ Phys.\ B {\bf 551}, 204 (1999),
hep-th/9811139.}

\lref\senFone{A.~Sen, ``F-theory and Orientifolds,'' Nucl.\ Phys.\
B {\bf 475}, 562 (1996), hep-th/9605150; `Orientifold limit of
F-theory vacua,'' Phys.\ Rev.\ D {\bf 55}, 7345 (1997),
hep-th/9702165; ``Orientifold limit of F-theory vacua,'' Nucl.\
Phys.\ Proc.\ Suppl.\  {\bf 68}, 92 (1998) [Nucl.\ Phys.\ Proc.\
Suppl.\  {\bf 67}, 81 (1998)], hep-th/9709159.}

\lref\banks{T.~Banks, M.~R.~Douglas and N.~Seiberg, ``Probing
F-theory with branes,'' Phys.\ Lett.\ B {\bf 387}, 278 (1996),
hep-th/9605199.}

\lref\dmconstant{K.~Dasgupta and S.~Mukhi, ``F-theory at constant
coupling,'' Phys.\ Lett.\ B {\bf 385}, 125 (1996),
hep-th/9606044.}

\lref\imamura{Y.~Imamura, ``Born-Infeld action and Chern-Simons
term from Kaluza-Klein monopole in M-theory,'' Phys.\ Lett.\ B
{\bf 414}, 242 (1997), hep-th/9706144; A.~Sen, ``Dynamics of
multiple Kaluza-Klein monopoles in M and string theory,'' Adv.\
Theor.\ Math.\ Phys.\  {\bf 1}, 115 (1998), hep-th/9707042; ``A
note on enhanced gauge symmetries in M- and string theory,'' JHEP
{\bf 9709}, 001 (1997), hep-th/9707123.}

\lref\robbins{K.~Dasgupta, G.~Rajesh, D.~Robbins and S.~Sethi,
``Time-dependent warping, fluxes, and NCYM,'' JHEP {\bf 0303}, 041
(2003), hep-th/0302049; K.~Dasgupta and M.~Shmakova, ``On branes
and oriented B-fields,'' Nucl.\ Phys.\ B {\bf 675}, 205 (2003),
hep-th/0306030.}

\lref\svw{S.~Sethi, C.~Vafa and E.~Witten, ``Constraints on
low-dimensional string compactifications,'' Nucl.\ Phys.\ B {\bf
480}, 213 (1996), hep-th/9606122; K.~Dasgupta and S.~Mukhi, ``A
note on low-dimensional string compactifications,'' Phys.\ Lett.\
B {\bf 398}, 285 (1997), hep-th/9612188.}

\lref\kachruone{S.~Kachru, M.~B.~Schulz, P.~K.~Tripathy and
S.~P.~Trivedi, ``New supersymmetric string compactifications,''
JHEP {\bf 0303}, 061 (2003), hep-th/0211182; S.~Kachru,
M.~B.~Schulz and S.~Trivedi, ``Moduli stabilization from fluxes in
a simple IIB orientifold,'' JHEP {\bf 0310}, 007 (2003),
hep-th/0201028.}

\lref\hitchin{N. ~Hitchin, ``Stable forms and special metrics'',
Contemp. Math., {\bf 288}, Amer. Math. Soc. (2000).}

\lref\giveon{S.~S.~Gubser,
 ``Supersymmetry and F-theory realization of the deformed conifold with
three-form flux,'' hep-th/0010010; A.~Giveon, A.~Kehagias and
H.~Partouche, ``Geometric transitions, brane dynamics and gauge
theories,'' JHEP {\bf 0112}, 021 (2001), hep-th/0110115.}

\lref\rrj{R.~Roiban, R.~Tatar and J.~Walcher, ``Massless flavor in
geometry and matrix models,'' Nucl.\ Phys.\ B {\bf 665}, 211
(2003), hep-th/0301217.}

\lref\bonan{E. Bonan, ``Sur le varietes remanniennes a groupe
d'holonomie $G_2$ ou Spin(7),'' C. R. Acad. Sci. paris {\bf 262}
(1966) 127.}



\lref\amv{M.~Atiyah, J.~M.~Maldacena and C.~Vafa, ``An M-theory
flop as a large N duality,'' J.\ Math.\ Phys.\  {\bf 42}, 3209
(2001), hep-th/0011256.}

\lref\syz{A.~Strominger, S.~T.~Yau and E.~Zaslow, ``Mirror
symmetry is T-duality,'' Nucl.\ Phys.\ B {\bf 479}, 243 (1996),
hep-th/9606040.}

\lref\hullt{C.~M.~Hull, ``Sigma Model Beta Functions And String
Compactifications,'' Nucl.\ Phys.\ B {\bf 267}, 266 (1986).}

\lref\HULL{C.~M.~Hull, ``Superstring Compactifications With
Torsion And Space-Time Supersymmetry,'' Print-86-0251 (CAMBRIDGE),
Published in Turin Superunif.1985:347; C.~M.~Hull and E.~Witten,
``Supersymmetric Sigma Models And The Heterotic String,'' Phys.\
Lett.\ B {\bf 160}, 398 (1985).}

\lref\tduality{E.~Bergshoeff, C.~M.~Hull and T.~Ortin, ``Duality
in the type II superstring effective action,'' Nucl.\ Phys.\ B
{\bf 451}, 547 (1995), hep-th/9504081; P.~Meessen and T.~Ortin,
``An Sl(2,Z) multiplet of nine-dimensional type II supergravity
theories,'' Nucl.\ Phys.\ B {\bf 541}, 195 (1999),
hep-th/9806120.}

\lref\adoptone{J.~D.~Edelstein, K.~Oh and R.~Tatar, ``Orientifold,
geometric transition and large N duality for SO/Sp gauge
theories,'' JHEP {\bf 0105}, 009 (2001), hep-th/0104037.}

\lref\ohta{K.~Ohta and T.~Yokono, ``Deformation of conifold and
intersecting branes,'' JHEP {\bf 0002}, 023 (2000),
hep-th/9912266.}

\lref\adoptfo{K.~h.~Oh and R.~Tatar, ``Duality and confinement in
N = 1 supersymmetric theories from geometric transitions,'' Adv.\
Theor.\ Math.\ Phys.\  {\bf 6}, 141 (2003), hep-th/0112040.}

\lref\minasianone{R.~Minasian and D.~Tsimpis, ``Hopf reductions,
fluxes and branes,'' Nucl.\ Phys.\ B {\bf 613}, 127 (2001),
hep-th/0106266.}

\lref\pandoz{L.~A.~Pando Zayas and A.~A.~Tseytlin, ``3-branes on
resolved conifold,'' JHEP {\bf 0011}, 028 (2000), hep-th/0010088.}

\lref\bb{K.~Becker and M.~Becker, ``M-Theory on eight-manifolds,''
Nucl.\ Phys.\ B {\bf 477}, 155 (1996), hep-th/9605053.}

\lref\bbdgs{K.~Becker, M.~Becker, P.~S.~Green, K.~Dasgupta and
E.~Sharpe, ``Compactifications of heterotic strings on
non-K\"ahler complex manifolds. II,'' Nucl.\ Phys.\ B {\bf 678},
19 (2004), hep-th/0310058.}

\lref\bbdg{K.~Becker, M.~Becker, K.~Dasgupta and P.~S.~Green,
``Compactifications of heterotic theory on non-K\"ahler complex
manifolds. I,'' JHEP {\bf 0304}, 007 (2003), hep-th/0301161.}

\lref\hari{B.~de Wit, D.~J.~Smit and N.~D.~Hari Dass, ``Residual
Supersymmetry Of Compactified D = 10 Supergravity,'' Nucl.\ Phys.\
B {\bf 283}, 165 (1987).}

\lref\GP{E.~Goldstein and S.~Prokushkin, ``Geometric model for
complex non-K\"ahler manifolds with SU(3) structure,''
hep-th/0212307.}

\lref\bbdp{K.~Becker, M.~Becker, K.~Dasgupta and S.~Prokushkin,
``Properties of heterotic vacua from superpotentials,'' Nucl.\
Phys.\ B {\bf 666}, 144 (2003), hep-th/0304001.}

\lref\townsend{P.~K.~Townsend, ``D-branes from M-branes,'' Phys.\
Lett.\ B {\bf 373}, 68 (1996), hep-th/9512062.}

\lref\drs{K.~Dasgupta, G.~Rajesh and S.~Sethi, ``M theory,
orientifolds and G-flux,'' JHEP {\bf 9908}, 023 (1999),
hep-th/9908088.}

\lref\gv{R.~Gopakumar and C.~Vafa, ``On the gauge theory/geometry
correspondence,'' Adv.\ Theor.\ Math.\ Phys.\  {\bf 3}, 1415
(1999),hep-th/9811131.}

\lref\dvu{R.~Dijkgraaf and C.~Vafa, ``Matrix models, topological
strings, and supersymmetric gauge theories,'' Nucl.\ Phys.\ B {\bf
644}, 3 (2002), hep-th/0206255.}

\lref\ps{J.~Polchinski and M.~J.~Strassler, ``The String Dual of a
Confining Four-Dimensional Gauge Theory ,'' hep-th/0003136.}

\lref\rara{I.~Bena, R.~Roiban and R.~Tatar, ``Baryons, boundaries
and matrix models,'' Nucl.\ Phys.\ B {\bf 679}, 168 (2004),
hep-th/0211271; I.~Bena, H.~Murayama, R.~Roiban and R.~Tatar,
``Matrix model description of baryonic deformations,'' JHEP {\bf
0305}, 049 (2003), hep-th/0303115.}

\lref\dvd{R.~Dijkgraaf and C.~Vafa, ``A perturbative window into
non-perturbative physics,'' hep-th/0208048.}

\lref\civu{F.~Cachazo, S.~Katz and C.~Vafa, ``Geometric
transitions and N = 1 quiver theories,'' hep-th/0108120.}

\lref\civd{F.~Cachazo, B.~Fiol, K.~A.~Intriligator, S.~Katz and
C.~Vafa, ``A geometric unification of dualities,'' Nucl.\ Phys.\ B
{\bf 628}, 3 (2002), hep-th/0110028.}

\lref\radu{R.~Roiban, R.~Tatar and J.~Walcher, ``Massless flavor
in geometry and matrix models,'' Nucl.\ Phys.\ B {\bf 665}, 211
(2003), hep-th/0301217.}

\lref\radd{K.~Landsteiner, C.~I.~Lazaroiu and R.~Tatar,
``(Anti)symmetric matter and superpotentials from IIB
orientifolds,'' JHEP {\bf 0311}, 044 (2003),
hep-th/0306236;~``Chiral field theories, Konishi anomalies and
matrix models,'' JHEP {\bf 0402}, 044 (2004),
hep-th/0307182;~``Chiral field theories from conifolds,'' JHEP
{\bf 0311}, 057 (2003), hep-th/0310052;~ ``Puzzles for matrix
models of chiral field theories,'' Fortsch.\ Phys.\  {\bf 52}, 590
(2004), hep-th/0311103.}

\lref\ber{M.~Bershadsky, S.~Cecotti, H.~Ooguri and C.~Vafa,
 ``Kodaira-Spencer theory of gravity and exact results for quantum string
 amplitudes,''
Commun.\ Math.\ Phys.\  {\bf 165}, 311 (1994), hep-th/9309140.}

\lref\bismut{J. M. Bismut, ``A local index theorem for
non-K\"ahler manifolds,'' Math. Ann. {\bf 284} (1989) 681.}

\lref\monar{F. Cabrera, M. Monar and A. Swann, ``Classification of
$G_2$ structures,'' J. London Math. Soc. {\bf 53} (1996) 98; F.
Cabrera, ``On Riemannian manifolds with $G_2$ structure,''
Bolletino UMI A {\bf 10} (1996) 98.}

\lref\kath{Th. Friedrich and I. Kath, ``7-dimensional compact
Riemannian manifolds with killing spinors,'' Comm. Math. Phys.
{\bf 133} (1990) 543; Th. Friedrich, I. Kath, A. Moroianu and U.
Semmelmann, ``On nearly parallel $G_2$ structures,'' J. geom.
Phys. {\bf 23} (1997) 259; S. Salamon, ``Riemannian geometry and
holonomy groups,'' Pitman Res. Notes Math. Ser., {\bf 201} (1989);
V. Apostolov and S. Salamon, ``K\"ahler reduction of metrics with
holonomy $G_2$,'' math-DG/0303197.}

\lref\ooguri{H.~Ooguri and C.~Vafa, ``The C-deformation of gluino
and non-planar diagrams,'' Adv.\ Theor.\ Math.\ Phys.\  {\bf 7},
53 (2003), hep-th/0302109; ``Gravity induced C-deformation,''
Adv.\ Theor.\ Math.\ Phys.\  {\bf 7}, 405 (2004), hep-th/0303063.}

\lref\tv{T.~R.~Taylor and C.~Vafa, ``RR flux on Calabi-Yau and
partial supersymmetry breaking,'' Phys.\ Lett.\ B {\bf 474}, 130
(2000), hep-th/9912152.}

\lref\kat{K.~Becker and K.~Dasgupta, ``Heterotic strings with
torsion,'' JHEP {\bf 0211}, 006 (2002), hep-th/0209077.}

\lref\kachru{S.~Kachru, M.~B.~Schulz and S.~Trivedi, ``Moduli
stabilization from fluxes in a simple IIB orientifold,'' JHEP {\bf
0310}, 007 (2003), hep-th/0201028; S.~Kachru, M.~B.~Schulz,
P.~K.~Tripathy and S.~P.~Trivedi, ``New supersymmetric string
compactifications,'' JHEP {\bf 0303}, 061 (2003), hep-th/0211182;
M.~B.~Schulz, ``Superstring orientifolds with torsion: O5
orientifolds of torus fibrations and their massless spectra,''
hep-th/0406001}

\lref\sw{N.~Seiberg and E.~Witten, ``Electric - magnetic duality,
monopole condensation, and confinement in N=2 supersymmetric
Yang-Mills theory,'' Nucl.\ Phys.\ B {\bf 426}, 19 (1994)
[Erratum-ibid.\ B {\bf 430}, 485 (1994)], hep-th/9407087;
``Monopoles, duality and chiral symmetry breaking in N=2
supersymmetric QCD,'' Nucl.\ Phys.\ B {\bf 431}, 484 (1994),
hep-th/9408099}

\lref\afm{O.~Aharony, A.~Fayyazuddin and J.~M.~Maldacena, ``The
large N limit of N = 2,1 field theories from three-branes in
F-theory,'' JHEP {\bf 9807}, 013 (1998), hep-th/9806159.}

\lref\grana{M.~Grana and J.~Polchinski, ``Gauge / gravity duals
with holomorphic dilaton,'' Phys.\ Rev.\ D {\bf 65}, 126005
(2002), hep-th/0106014.}

\lref\lustu{G.~L.~Cardoso, G.~Curio, G.~Dall'Agata and D.~Lust,
``BPS action and superpotential for heterotic string
compactifications  with fluxes,'' JHEP {\bf 0310}, 004
(2003),hep-th/0306088.}

\lref\lustd{G.~L.~Cardoso, G.~Curio, G.~Dall'Agata and D.~Lust,
``Heterotic string theory on non-K\"ahler manifolds with H-flux
and gaugino condensate,'' hep-th/0310021.}

\lref\bd{M.~Becker and K.~Dasgupta, ``K\"ahler versus non-K\"ahler
compactifications,'' hep-th/0312221.}

\lref\micutwo{S.~Gurrieri, A.~Lukas, A.~Micu, ``Heterotic on
Half-flat'' hep-th/0408121.}

\lref\micu{S.~Gurrieri and A.~Micu, ``Type IIB theory on half-flat
manifolds,'' Class.\ Quant.\ Grav.\  {\bf 20}, 2181 (2003),
hep-th/0212278.}

\lref\dal{G.~Dall'Agata and N.~Prezas, ``N = 1 geometries for
M-theory and type IIA strings with fluxes,'' hep-th/0311146;
G.~Dall'Agata, ``On Supersymmetric Solutions of Type IIB
Supergravity with General Fluxes'', hep-th/0403220.}

\lref\douo{M.~R.~Douglas,``The statistics of string / M theory
vacua,'' JHEP {\bf 0305}, 046 (2003), hep-th/0303194.}

\lref\fg{A.~F.~Frey and A.~Grana,``Type IIB solutions with
interpolating supersymetries ,'' hep-th/0307142.}

\lref\bbs{K.~Becker, M.~Becker and R.~Sriharsha,``PP-waves,
M-theory and fluxes,'' hep-th/0308014.}

 \lref\min{P.~Kaste, R.~Minasian, M.~Petrini and A.~Tomasiello,
``Nontrivial RR two-form field strength and SU(3)-structure,''
Fortsch.\ Phys.\  {\bf 51}, 764 (2003), hep-th/0301063 ;
S.~Fidanza, R.~Minasian,A.~Tomasiello,''Mirror Symmetric SU(3)
Structure Manifolds with NS fluxes'', hep-th/0311122; M.~Grana,
R.~Minasian, M.~Petrini and A.~Tomasiello, ``Supersymmetric
backgrounds from generalized Calabi-Yau manifolds,''
hep-th/0406137.}

\lref\kleb{I.~R.~Klebanov and M.~J.~Strassler, ``Supergravity and
a confining gauge theory: Duality cascades and $\chi$ -
SB-resolution of naked singularities,'' JHEP {\bf 0008}, 052
(2000), hep-th/0007191.}

\lref\kklt{S.~Kachru, R.~Kallosh, A.~Linde and S.~P.~Trivedi, ``De
Sitter vacua in string theory,'' Phys.\ Rev.\ D {\bf 68}, 046005
(2003), hep-th/0301240.}

\lref\dolo{M.~R.~Douglas, D.~A.~Lowe and J.~H.~Schwarz, ``Probing
F-theory with multiple branes,'' Phys.\ Lett.\ B {\bf 394}, 297
(1997), hep-th/9612062}

\lref\wisd{ E.~Witten, ``Small Instantons in String Theory,''
Nucl.\ Phys.\ B {\bf 460}, 541 (1996), hep-th/9511030.}

\lref\rohm{M.~Dine, R.~Rohm, N.~Seiberg and E.~Witten, ``Gluino
Condensation In Superstring Models,'' Phys.\ Lett.\ B {\bf 156},
55 (1985).}

\lref\gukkac{S.~Gukov, S.~Kachru, X.~Liu and L.~McAllister,
``Heterotic moduli stabilization with fractional Chern-Simons
invariants,'' Phys.\ Rev.\ D {\bf 69}, 086008 (2004),
hep-th/0310159.}

\lref\mn{J.~M.~Maldacena and C.~Nunez, ``Towards the large N limit
of pure N = 1 super Yang Mills,'' Phys.\ Rev.\ Lett.\  {\bf 86},
588 (2001), hep-th/0008001.}

\lref\witt{E.~Witten, ``Topological Sigma Models,'' Commun.\
Math.\ Phys.\  {\bf 118}, 411 (1988).}

\lref\antoniad{I.~Antoniadis, E.~Gava, K.~S.~Narain and
T.~R.~Taylor, ``Topological Amplitudes in Heterotic Superstring
Theory,'' Nucl.\ Phys.\ B {\bf 476}, 133 (1996), hep-th/9604077.}

\lref\kapustin{A.~Kapustin and Y.~Li,
 ``Topological sigma-models with H-flux and twisted generalized complex
 manifolds,''hep-th/0407249}

\lref\blu{R.~Blumenhagen, R.~Schimmrigk and A.~Wisskirchen,
``(0,2) mirror symmetry,'' Nucl.\ Phys.\ B {\bf 486}, 598 (1997),
hep-th/9609167.}

\lref\as{A.~Adams, A.~Basu and S.~Sethi, ``(0,2) duality,'' Adv.\
Theor.\ Math.\ Phys.\  {\bf 7}, 865 (2004), hep-th/0309226.}

\lref\sk{S.~Katz and E.~Sharpe, ``Notes on certain (0,2)
correlation functions,'' hep-th/0406226.}

\lref\gvw{S.~Gukov, C.~Vafa and E.~Witten, ``CFT's from Calabi-Yau
four-folds,'' Nucl.\ Phys.\ B {\bf 584}, 69 (2000) [Erratum-ibid.\
B {\bf 608}, 477 (2001), hep-th/9906070.}

\lref\ura{J.~F.~G.~Cascales and A.~M.~Uranga, ``Branes on
generalized calibrated submanifolds,'' hep-th/0407132.}

\lref\pandovaman{B.~A.~Burrington, J.~T.~Liu, L.~A.~Pando Zayas
and D.~Vaman, ``Holographic duals of flavored N = 1 super
Yang-Mills: Beyond the probe approximation,'' hep-th/0406207.}

\lref\karchlust{A.~Karch, D.~Lust and D.~J.~Smith,
 ``Equivalence of geometric engineering and Hanany-Witten via fractional
 branes,''
Nucl.\ Phys.\ B {\bf 533}, 348 (1998), hep-th/9803232.}

\lref\bercve{K.~Behrndt and M.~Cvetic,
``General N = 1 supersymmetric flux vacua of (massive) type IIA
string theory,''~hep-th/0403049;~``General N = 1 supersymmetric
fluxes in massive type IIA string theory,'' hep-th/0407263.}

\lref\Schulz{ M.~B.~Schulz, ``Superstring orientifolds with
torsion: O5 orientifolds of torus fibrations and their massless
spectra,'' hep-th/0406001.}

\lref\bck{M.~Becker, G.~Curio and A.~Krause, ``De Sitter vacua
from heterotic M Theory,'' Nucl.\ Phys.\ B {\bf 693}, 223 (2004),
hep-th/0403027.}

\Title{\vbox{\hbox{hep-th/0408192}
\hbox{SLAC-PUB-10670, SU-ITP-04/32}\hbox{UCB-PTH-04/22, UMD-PP-05/015}
\hbox{LBNL-54768I}}} {\vbox{ \vskip-2.5in \hbox{\centerline{In
the Realm of the Geometric Transitions }}}}

\vskip-.2in \centerline{\bf Stephon Alexander${}^{1,2}$, Katrin
Becker${}^3$, Melanie Becker${}^4$} \vskip.02in \centerline{\bf
Keshav Dasgupta${}^1$, ~Anke Knauf${}^{4,5}$, ~Radu Tatar${}^6$}
\vskip.16in \centerline{\it ${}^1$~Department of Physics, Stanford
University, Stanford CA 94305}
\vskip.02in \centerline{\it ${}^2$~SLAC, Stanford
University, Stanford CA 94309}
\vskip.02in \centerline{\tt
stephon,keshav@itp.stanford.edu} \vskip.02in \centerline{\it
${}^3$~Department of Physics, University of Utah, Salt Lake City,
UT 84112} \centerline{\tt katrin@physics.utah.edu} \vskip.02in
\centerline{\it ${}^4$~Department of Physics, University of
Maryland, College Park, MD 20742} \vskip.02in \centerline{\tt
melanieb@physics.umd.edu, anke@umd.edu}\vskip.02in \centerline{\it
${}^5$~II. Institut f\"ur Theoretische Physik, Universit\"at
Hamburg} \centerline{\it ~~~~~ Luruper Chaussee 149, 22761
Hamburg, Germany} \vskip.02in \centerline{\it ${}^6$~Theoretical
Physics Group, LBL Berkeley, CA 94720} \vskip.02in \centerline{\tt
rtatar@socrates.Berkeley.EDU}

\centerline{\bf Abstract}

\noindent We complete the duality cycle 
by constructing the geometric transition duals in the type IIB,
type I and heterotic theories. We show that in the type IIB theory
the background on the closed string side is a K\"ahler deformed
conifold, as expected, even though the mirror type IIA backgrounds
are non-K\"ahler (both before and after the transition). On the
other hand, the Type I and heterotic backgrounds are non-K\"ahler.
Therefore, on the heterotic side these backgrounds give rise to
new torsional manifolds that have not been studied before. We show
the consistency of these backgrounds by verifying the torsional
equation.

\Date{}

\centerline{\bf Contents}\nobreak\medskip{\baselineskip=12pt
\parskip=0pt\catcode`\@=11

\noindent {1.} {Introduction} \leaderfill{1} \par \noindent {2.}
{Supergravity Analysis in Type IIA} \leaderfill{7} \par \noindent
\quad{2.1.} {Background before geometric transition}
\leaderfill{8} \par \noindent \quad{2.2.} {Background after
geometric transition} \leaderfill{9} \par \noindent {3.}
{Supergravity Analysis in Type IIB} \leaderfill{10} \par \noindent
\quad{3.1.} {Background before geometric transition}
\leaderfill{10} \par \noindent \quad{3.2.} {Background after
geometric transition} \leaderfill{11} \par \noindent {4.}
{Supergravity Analysis in Type I} \leaderfill{23} \par \noindent
\quad{4.1.} {Background before geometric transition}
\leaderfill{26} \par \noindent \quad{4.2.} {Background after
geometric transition} \leaderfill{41} \par \noindent {5.}
{Supergravity Analysis in Heterotic Theory} \leaderfill{45} \par
\noindent \quad{5.1.} {Background before geometric transition}
\leaderfill{46} \par \noindent \quad{5.2.} {Background after
geometric transition} \leaderfill{48} \par \noindent {6.} {Field
Theory and Geometric Transition in Type I and Heterotic String}
\leaderfill{50}\par \noindent \quad{6.1.}
{Geometric transition} \leaderfill{52} \par \catcode`\@=12
\bigbreak\bigskip}

\newsec{Introduction}
One of the most active areas of research during the last year has
been directed towards deepening the links between string theory
and realistic supersymmetric gauge theories. Four years ago three
different approaches linked non conformal ${\cal N} =1$
supersymmetric gauge theories with D--branes wrapped on cycles of
Calabi--Yau manifolds and flux compactifications of
ten--dimensional string theories \kleb\vafai\mn.~The approach of
\kleb\ dealt with the conifold supergravity solution describing a
cascade of Seiberg dualities from the UV to the IR. The IR theory
is a deformed conifold, where the conifold singularity was
replaced by an $S^3$ cycle.

The approach of \vafai\ was based on a type IIA duality \gv\
between open topological strings on Lagrangian submanifolds and
closed topological strings on a resolved conifold. By identifying
IR gauge invariant quantities (gluino condensate) with volumes in
the resolved conifold, a powerful duality between wrapped D6
branes and flux compactifications was developed. The type IIA
duality was then extended to many examples of mirror type IIB
duals \civ\eot\civu\civd\dott\dvd\dvu. In the type IIB theory the
duality relates the IR dynamics on D5 branes wrapped on resolution
$P^1$ cycles and fluxes on $S^3$ in a deformed conifold.

In \mn\ the supergravity solution of an $NS5$ brane wrapped on a
two cycle of a non-trivial background was analyzed. This approach
mainly discussed the open string side of the type IIB duality,
where supersymmetry was broken to ${\cal N} = 1$ by performing a
twist. These type IIB developments were stimulated by the fact
that there is an already well--known and accepted form for the
flux superpotential \gvw \eqn\potgvw{W_{IIB} = \int (H_{RR} + \tau
H_{NS}) \wedge \Omega,} where $H_{RR}$ is the RR flux, $H_{NS}$ is
the NS flux, $\tau$ is the axion-dilaton and $\Omega$ is the
holomorphic three--form. For most cases that were considered, the
3--cycles of the manifold were $P^1$ fibres over elliptic curves
and this allowed a direct connection to Seiberg--Witten curves
\civ\ and matrix models \dvd\dvu.

The duality on the type IIA side is harder to study because
Lagrangian submanifolds that can be wrapped by D6 branes are not
easy to obtain. One important ingredient in the discussion of
\vafai\ was the fact that the duality required a closed string
side with $d \Omega \ne 0$, i.e. a non--complex, non--K\"ahler
manifold. As the mathematical apparatus to describe these
manifolds has not yet been developed their construction turns out
to be difficult as well. The first step in this direction was
taken in \louis\ where the results of \salamon\ were used to
describe half--flat manifolds as examples of backgrounds with $d
\Omega \ne 0$.

One very explicit example of type IIA duality was worked out in
\bdkt\ (other type II models with torsion appeared in
\min\micu\dal\bercve). Starting with a conformal Calabi--Yau
solution on the type IIB side corresponding to D5 branes wrapped
on the resolution $P^1$ cycle of the resolved conifold \pandoz\
the mirror type IIA picture was obtained performing three
T--dualities using the Strominger, Yau and Zaslow (SYZ) technique
\syz\foot{Brane configurations obtained by one T--duality from
configurations with  D5 branes wrapped on $P^1$ cycles have been
discussed in \dotu\dotd\rara\radu\radd. }. Since the SYZ technique
works properly when the base of the manifold is much bigger than
the $T^3$ fiber, the limit of large complex structure was
taken\foot{There is a subtlety though. The technique of \syz\ is
valid in the limit where the base is large, whereas geometric
transitions occur in exactly the opposite limit. Therefore using
three T--dualities, we should not expect to get the right metric.
In fact, this is exactly what was shown in \bdkt\ as the metric
obtained by naive T--dualities was missing some crucial
components. The correct metric was only extracted after some
non--trivial coordinate transformations were performed \bdkt.}. By
identifying three angular directions, the type IIA dual with D6
branes wrapped on 3--cycles inside a non--K\"ahler deformation of
the deformed conifold\foot{Examples with D6 branes wrapped on
generalized calibrated submanifolds inside non--K\"ahler spaces
have been recently discussed in \ura.} was obtained.

Furthermore, the non--K\"ahler geometry with wrapped D6 branes was
connected to another non--K\"ahler geometry with fluxes \bdkt.
This was done by lifting the type IIA configuration with D6 branes
to a manifold with torsion and $G_2$--structure by performing a
flop in this seven--dimensional manifold. This allowed us to
replace D6 branes by RR two--form fluxes and the torsion classes
of the deformed geometry with the torsion classes of the resolved
geometry. The final result was an ${\cal N} = 1$ compactification
with RR fluxes on a non--K\"ahler manifold with $dJ \ne 0$ and $d
\Omega \ne 0$ with a superpotential \louis \eqn\pota{W_{IIA} =
\int (J+i B) \wedge d \Omega.}

The goal of the present paper is to connect the closed type IIB
theory to the closed type IIA theory, something that was not done
in \bdkt. We shall show that the result is a complex K\"ahler
manifold with a closed three--form $\Omega$ and a superpotential
given by \potgvw.~By taking the same limit of large complex
structure, we perform three T--dualities to go to a type IIB
geometry where the limits taken in \bdkt\ are naturally
reconstructed. In other words, the limits of large complex
structure between the type IIB geometries with D5 branes and with
fluxes are naturally identified. This is a very important check of
our duality cycle and confirms our approach.

Having completed the cycle of geometric transitions in the type
IIA and type IIB theories we will extend this cycle to the type I
and heterotic strings. To do so we use Sen's idea \senFone\ of
going to an orientifold limit of the type IIB theory and from
there to the type I theory by performing two T--dualities. More
precisely, we first need to build a metric which is invariant
under the orientifold action by eliminating terms of the type IIB
metric that are not invariant under the orientifold operation. Our
metric is valid only at the orientifold point, in the perturbative
regime. When non--perturbative effects become relevant, the
orientifold plane splits into ($p,q$) seven branes and when these,
together with the initial D7 branes, are moved to infinity, the
usual type IIB metric should be recovered\foot{It remains an open
problem how to describe the intermediate region, when there is a
non--trivial axion--dilaton and the metric is to be determined
from the Seiberg--Witten curve.}. This assumption will be crucial
later when we determine the metric at the orientifold point.

When the type I string is compactified on a non--K\"ahler
manifold\foot{In principle it could be also non--complex, as in
type IIA, but the Donaldson--Uhlenbeck--Yau equation for the gauge
bundle requires the manifold to be complex.} the superpotential
\potgvw\ is mapped into \bbdp \eqn\potu{W_{I} = \int (H_{RR} + i d
J) \wedge \Omega,} and the extra contribution from the D7 branes
in type IIB is mapped to the holomorphic Chern--Simons term for
the SO(32) gauge field.

The natural final step is to connect this theory to the heterotic
string compactified on a non--K\"ahler manifold. This theory has
been studied in great detail over the last few years
\drs\kat\bbdg\bbdgs \lustu \lustd.  The manifold obtained on the
heterotic side is a non--K\"ahler manifold and the fluxes will
appear as torsion \rstrom. In this theory the moduli stabilization
is achieved via the new superpotential proposed in \bbdp\ (see
also \lustu) \eqn\poth{W_{het}=\int(H+i d J)\wedge \Omega,} where
$H$ is the heterotic three--form satisfying $dH = {\rm tr} ~R
\wedge R - {1 \o 30} {\rm tr}~F \wedge F$.

An important aspect of our discussion is that we provide geometric
transition duals for both the type I and heterotic theories.
Whether this implies transitions for $(0,2)$ theories is not yet
clear\foot{There are results related to definitions of mirror
symmetry for such theories \blu\as\sk,~but a geometric transition
has not yet been proposed.}. In order to check the existence of
such a transition one should identify the ``sources'' for the
holomorphic superpotentials as topological strings. Some results
for the $(0,2)$ topological strings have appeared in \antoniad,
but more progress is needed in order to clarify this issue.

Another rather important aspect of our U--dual type I/heterotic
string configurations might be their relevance for string
cosmology, along the lines of \kklt\ for the type IIB theory.
Indeed, within all of the corners of the M--theory moduli space,
the heterotic string is the most studied one and developed towards
making contact with the standard model of particle physics. In
light of these developments an important goal of string theory is
to find consistent de Sitter vacua and a realization of inflation
in the context of the heterotic theory by identifying the inflaton
field as a natural extension of the standard model.

In order to stabilize all the moduli fields of the heterotic
string, compactification on non--K\"ahler manifolds have to be
considered. When the non--K\"ahler manifold is compact, the radius
can be shown to be stabilized at tree level by balancing the
fluxes with the non--K\"ahlerity of the internal space \bbdp. In
the simplest case only the dilaton remains to be stabilized by
including non--perturbative effects. On the other hand, for a
Calabi--Yau compactification, both the radius and the dilaton have
to be fixed by non--perturbative effects. This has been recently
discussed in \gukkac.

Interesting solutions of strongly coupled heterotic string theory
in de Sitter space have recently been obtained in \bck. Here it
was shown that the balancing of two non--perturbative effects,
coming from open membrane instantons stretching between the
boundaries of heterotic M--theory and gluino condensation on the
hidden boundary, leads to meta--stable de Sitter vacua. It was
possible to predict the masses of charged scalar matter fields,
the scale of supersymmetry breaking and the gravitino mass in a
phenomenologically interesting TeV scale.

Cosmological relevant models might be constructed by considering
compactifications of the heterotic on non--K\"ahler manifolds. The
above observations are rather encouraging because the manifolds
that we will construct here are all non--K\"ahler. It remains to
be seen if our concrete Klebanov--Strassler type IIB solutions are
relevant for cosmology. This will be left for future work.

Our analysis in this paper will follow the road outlined in the
figure below: \vskip.1in

\centerline{\epsfbox{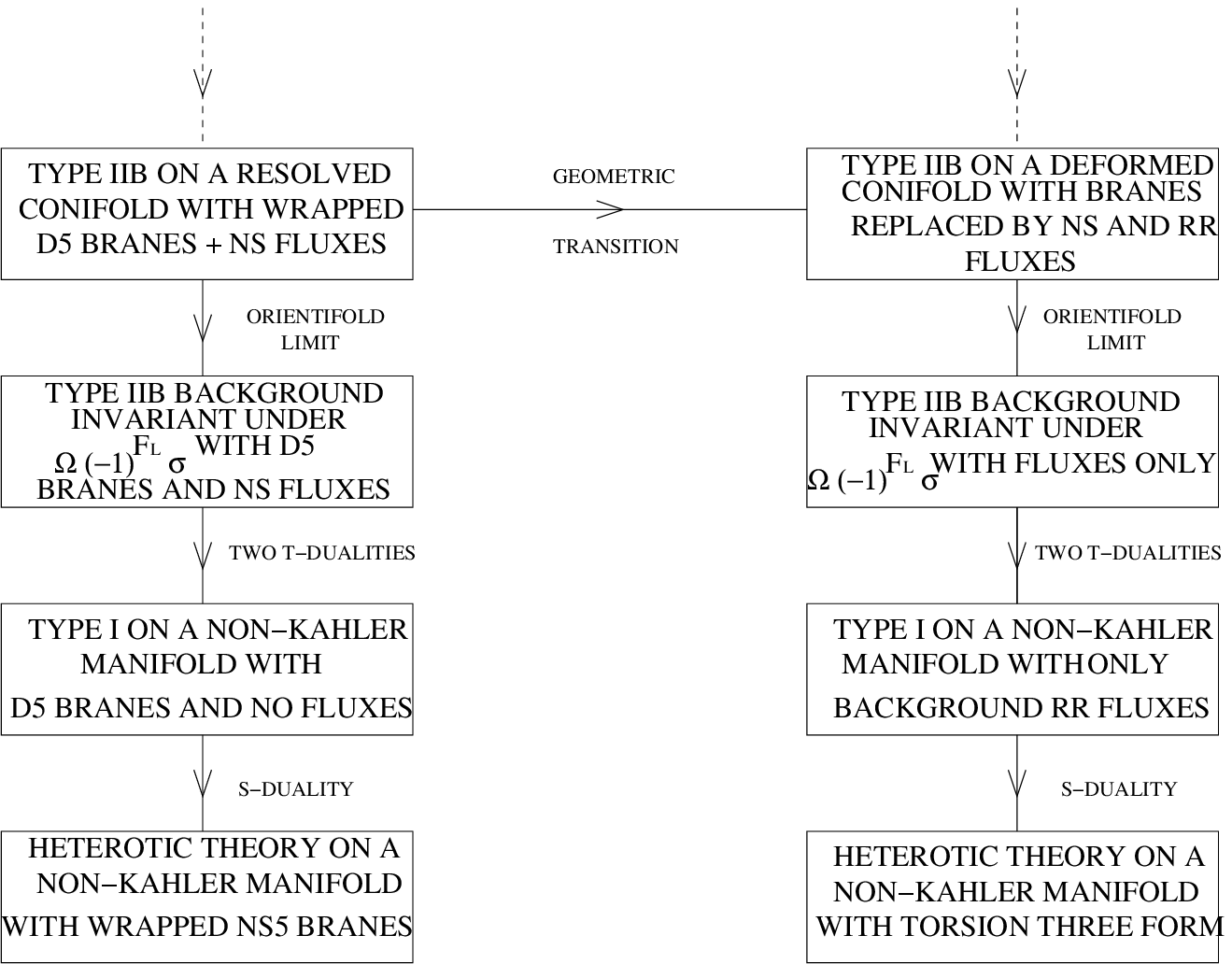}}\nobreak

\vskip.1in \noindent The boxes represent the duality chains and we
will follow the arrows. The dotted arrows are the earlier duality
chains encountered in \bdkt. Here we confirm the mechanism of the 
geometric transition for type IIB theory. The type I
heterotic duals will be obtained by going to the orientifold
corner of the type IIB compactification. It is tempting to argue
that there is a possible geometric transition between the two
sides, although we will not make such a claim herein.

Our starting point will be the type IIA background which we had
discussed in great detail in \bdkt. We will summarize the complete
background both before and after the geometric transition in
section 2. These backgrounds will be used to go to the mirror
picture. We shall see that the mirror manifolds in the type IIB
theory both before and after the geometric transition has taken
place are complex K\"ahler manifolds, even though the
corresponding type IIA manifolds were non--K\"ahler! This has been
predicted earlier in \kleb, \mn, \vafai, and will be derived here
from a SYZ type analysis  \syz. Section 3 contains the full
analysis. Section. 3.1 gives a short summary of the type IIB
background before the geometric transition. This was studied
earlier in \pandoz. New results are presented starting in section
3.2, where we derive the Klebanov--Strassler kind background.

Having obtained the type IIB background, we then go to the
orientifold limit. This is achieved by uplifting the type IIB
backgrounds to F--theory. F--theory naturally incorporates $D7$
branes and $O7$ planes. This will be explained at the beginning of
section 4. We derive the metric of the type IIB theory {\it at}
the orientifold point and use this to reach the type I theory. The
type I metric is derived in section 4.1. Section 4.2 is devoted to
the analysis after the geometric transition. The type I theory
will show two different backgrounds: one in which we have $D5$
branes wrapped on two--cycles of a non--K\"ahler manifold, and
another one in which we have RR three-forms on another
non--K\"ahler manifold. However, it is not clear to us whether
this implies that if we shrink the two cycle with wrapped $D5$
branes then we get another manifold with no branes but RR fluxes.
More work needs to be done to reach any concrete conclusion. These
details appear in sec. 4.

In section 5. we go to the heterotic background. In section 5.1 we
derive the non--K\"ahler manifold with $NS5$ branes wrapped on a
two cycle. In section 5.2 we derive another non--K\"ahler manifold
with no branes but three--form torsion. The manifolds that we get
in the heterotic theory are new examples of non--compact complex
non--K\"ahler manifolds. So far most of the examples presented in
the literature were {\it compact} complex non--K\"ahler manifolds
\drs\kat\bbdg\bbdp\bbdgs \lustu \lustd\foot{It has been brought to 
our notice that ref. \serone\ studies some non-compact examples that satisfy
torsional equation.}.  
We prove the consistency
of these manifolds by verifying that they satisfy the torsional
constraint \HULL \rstrom.

Section 6 is devoted to studying some generic properties of the
field theory, superpotential and geometric transitions for the
type I and heterotic theories.

{\bf Note added}: While we were typing up our results, a paper
appeared on the archive \micutwo\ that has some overlap with our
work.

\newsec{Supergravity Analysis in Type IIA}

In \bdkt\ an exact supergravity description of the type IIA
geometric transition of \vafai\ appeared where the backgrounds
both before and after the transition were presented. In this
section we will summarize the main results. The type IIA
background was obtained from the type IIB background involving
$D5$ branes wrapping a two cycle of the resolved conifold.

The metric of $D5$ branes wrapped on a two cycle of the resolved
conifold is \eqn\metreso{ds^2 = (dz + \Delta_1~{\rm cot}~\theta_1
~dx + \Delta_2~{\rm cot}~\theta_2 ~dy)^2 + \vert dz_1 \vert^2 +
\vert dz_2 \vert^2.} Here we have replaced the metric of the two
spheres of the resolved conifold with two tori with complex
structures $\tau_1$ and $\tau_2$. The complex one forms $dz_i,~ i
= 1, 2$, are defined as \eqn\compleone{dz_1 = dx -
\tau_1~d\theta_1, ~~~~~~ dz_2 = dy - \tau_2~d\theta_2.} For the
case that was studied in \bdkt, the complex structures $\tau_i$
can be chosen to be either integrable or non--integrable, and they
are in general given by \eqn\inteornot{\tau_1 = f_1 + {i\o 2}
\sqrt{\gamma \sqrt{h}}, ~~~~~~~~ \tau_2 = f_2 + {i\o 2}
\sqrt{(\gamma + 4a^2) \sqrt{h}}.} Here, $f_i$ are not yet
specified and $\gamma, a$ are defined in \pandoz. For the case of
an {\it integrable} complex structure, i.e. when the
transformation equations (5.49) and (5.50) of \bdkt\ can be
integrated to become finite transformations, the metric for the
mirror type IIA manifold is given by \eqn\fiiamet{\eqalign{&
ds_{IIA}^2 = ~~ g_1~\left[(dz - {b}_{z\mu}~dx^\mu) + \Delta_1~{\rm
cot}~\hat\theta_1~(dx - b_{x\theta_1}~d\theta_1) + \Delta_2~{\rm
cot}~ \hat\theta_2~(dy - b_{y\theta_2}~d\theta_2)+ ..\right]^2 \cr
& ~ + {[} g_2' d\theta_1^2 + g_2 (dx - b_{x\theta_1} d\theta_1)^2]
+ [ g_3' d\theta_2^2 + g_3 (dy - b_{y\theta_2} d\theta_2)^2{]}+
{\rm sin}~\psi~ {[}g_4 (dx - b_{x\theta_1}d\theta_1)d \theta_2 +
\cr & ~~~~~~~~~~~~~ + g_4' (dy - b_{y\theta_2} d\theta_2)
d\theta_1 {]} + ~{\rm cos}~\psi~ {[}g_4' d\theta_1 d\theta_2 - g_4
(dx - b_{x\theta_1} d\theta_1) (dy - b_{y\theta_2} d\theta_2)].}}
The variables appearing in the above metric are defined in \bdkt\
with a choice of $g_4', g_2'$ and $g_3'$ given by \eqn\gfourg{g_4'
= 2 \sqrt{\langle \alpha \rangle_1 \langle \alpha \rangle_2}~A_1
B_1,~~~~ g_3' = \langle \alpha \rangle_2 (1 + A_1^2), ~~~~ g_2' =
\langle \alpha \rangle_1 (1 + B_1^2).} Observe that with this
choice of integrable complex structure, we almost obtain the
correct type IIA metric. In fact, we see that when we identify
\eqn\repmen{ \langle \alpha \rangle_1 = \langle \alpha \rangle_2
=\alpha, ~~~~~ A_1 = A, ~~~~~ B_1 = B} then the metric \fiiamet\
becomes exactly the non--K\"ahler deformation of the deformed
conifold. This can be achieved directly by choosing a
non--integrable complex structure on the type IIB side (basically
related to the finite transformation that we discussed in \bdkt).


\subsec{{$\underline{\rm \bf {Background~ before~ geometric~
transition}}$}}


\noindent The background i.e. the metric, the $H_{NS}$ field, the
coupling $g_A$ and the gauge field $A$, before the geometric
transition can be given precisely. The metric has the following
form: \eqn\fiitaamet{\eqalign{& ds_{IIA}^2 =~~ g_1~\left[(dz -
{b}_{z\mu}~dx^\mu) + \Delta_1~{\rm cot}~\hat\theta_1~(dx -
b_{x\theta_1}~d\theta_1) + \Delta_2~{\rm cot}~ \hat\theta_2~(dy -
b_{y\theta_2}~d\theta_2)+ ..\right]^2 \cr & + g_2~ {[} d\theta_1^2
+ (dx - b_{x\theta_1}~d\theta_1)^2] + g_3~[ d\theta_2^2 + (dy -
b_{y\theta_2}~d\theta_2)^2{]} +  g_4~{\rm sin}~\psi~{[}(dx -
b_{x\theta_1}~d\theta_1)~d \theta_2 \cr & ~~~~~~~~~ + (dy -
b_{y\theta_2}~d\theta_2)~d\theta_1 {]} + ~g_4~{\rm
cos}~\psi~{[}d\theta_1 ~d\theta_2 - (dx - b_{x\theta_1}~d\theta_1)
(dy - b_{y\theta_2}~d\theta_2)].}} We see that in the absence of
the $B$ fields $b_{x\theta_1}$ and $b_{y\theta_2}$ the above
metric reduces to a K\"ahler deformed conifold, as expected. In
the presence of $B$ fields the fibration structure is well
defined. We will see later in other examples, that the $B$--field
dependent fibration structure is {\it universal}. The three--form
$H$ is given by \eqn\hfine{\eqalign{H_{NS} & = -\sqrt{\alpha^3}
A~(A~dA + B~dB) \wedge d\theta_1 \wedge dz + \sqrt{\alpha^3} (A~dA
+ B~dB) \wedge dy \wedge d\theta_2 \cr &  + \sqrt{\alpha}~dA
\wedge d\theta_1 \wedge dz + \sqrt{\alpha^3}~B~(A~dA + B~dB)
\wedge ({\rm sin}~\psi ~dy - {\rm cos}~\psi~d\theta_2) \wedge dz
\cr &  - \sqrt{\alpha^3}~(A~dA + B~dB)\wedge dx \wedge d\theta_1 ~
 -\sqrt{\alpha} ~dB \wedge ({\rm sin}~\psi ~dy - {\rm cos}~\psi~d\theta_2)
 \wedge dz,}}
where $H_{NS}$ is the finite part of the three--form ${\hat
H}_{NS}$ defined as $\epsilon^{-1/2} H_{NS}$, with $\epsilon \to
0$. All other fields are finite. The large background three--form
can be traced back to the large complex structure on the mirror
type IIB side. This is of course consistent with the requirement
of mirror symmetry according to which a large complex structure is
related to a large base manifold (which in turn is crucial to
apply the SYZ conjecture). In fact, as we will soon show, the
$B_{NS}$ field when written in terms of $\langle \alpha
\rangle_{1,2}$ instead of $\alpha$ becomes pure gauge and
therefore the corresponding $H_{NS}$ is zero! This will have an
important consequence when we go to the type IIB mirror side.

Finally, we need to specify the gauge fields and the coupling
constant. The gauge fields are associated with $D6$ branes wrapped
on a three--cycle of the non--K\"ahler manifold. In terms of the
mirror analysis that we performed in \bdkt\ the $D6$ branes are
precisely the $D5$ branes wrapped on the resolution $P^1$ of the
K\"ahler resolved conifold in type IIB theory. The background now
is given by \eqn\gaupot{A\cdot dX \equiv \Delta_1~{\rm
cot}~\hat\theta_1~(dx - b_{x\theta_1}~d\theta_1) - \Delta_2~{\rm
cot}~\hat\theta_2~(dy - b_{y\theta_2}~d\theta_2), ~~~ g_A = {g_B
\o \sqrt{1- {\epsilon\o \alpha}}}} and the three--form field is
identically zero. This completely specifies the type IIA
background before geometric transition.


\subsec{{$\underline{\rm \bf {Background~ after~ geometric~
transition}}$}}


\noindent The background after geometric transition can also be
explicitly determined. The metric now takes the following form
\eqn\iiafinal{\eqalign{ds^2 = & {1\o 4}\left(2g_2 - {g_4\o
\xi}\right)\left[d\theta_1^2 + (dx - b_{x\theta_1}~d\theta_1)^2
\right]  + {1\o 4} \left(2g_2 + {g_4\o
\xi}\right)\left[d\theta_2^2 + (dy -
b_{y\theta_2}~d\theta_2)^2\right]  \cr & + e^{2\phi}\left[dz +
\Delta_1~{\rm cot}~\hat\theta_1~(dx - b_{x\theta_1}~d\theta_1) +
\Delta_2 ~{\rm cot}~\hat\theta_2~(dy -
b_{y\theta_2}~d\theta_2)\right]^2.}} In the above equation we
observe the familiar fibration structure again. In the absence of
$b_{x\theta_1}$ and $b_{y\theta_2}$ the metric is exactly the
resolved conifold metric. In the presence of the non trivial
fibration the metric becomes a non--K\"ahler deformation of the
resolved conifold. As we saw above, before the geometric
transition the metric is a non--K\"ahler deformation of the
deformed conifold. The existence of these manifolds had been
anticipated earlier in the literature and their explicit form was
determined in \bdkt. Compactifications of the type IIA theory on
these backgrounds result in consistent theories.

Furthermore, the three-form $H_{NS}$ is given by
\eqn\hfinetwo{\eqalign{H_{NS} & = -\sqrt{\alpha^3} A~(A~dA + B~dB)
\wedge d\theta_1 \wedge dz + \sqrt{\alpha^3} (A~dA + B~dB) \wedge
dy \wedge d\theta_2 \cr &  + \sqrt{\alpha}~dA \wedge d\theta_1
\wedge dz + \sqrt{\alpha^3}~B~(A~dA + B~dB) \wedge ({\rm sin}~\psi
~dy - {\rm cos}~\psi~d\theta_2) \wedge dz \cr &  -
\sqrt{\alpha^3}~(A~dA + B~dB)\wedge dx \wedge d\theta_1 ~
 -\sqrt{\alpha} ~dB \wedge ({\rm sin}~\psi ~dy -
 {\rm cos}~\psi~d\theta_2) \wedge dz,}}
where $H_{NS}$ is the same finite part of the three-form that we
had before the geometric transition took place. Thus this field is
simply a spectator in this scenario and plays no other role. As
discussed above, this field is exactly zero when written in terms
of $\langle\alpha\rangle_i$, so we really do not have to worry
about it. On the other hand, both the metric and the gauge fluxes
have changed completely (along with the dilaton). The gauge field
and the coupling constant are given by
 \eqn\gaufinalii{{\cal A}\cdot dX =  2\Delta_1~{\rm
cot}~\hat\theta_1~(dx - b_{x\theta_1}~d\theta_1), ~~~~~~ g_A =
2^{-{3\o 2}} e^{-{\phi \o 2}} \alpha^{-{3\o 4}}.} At this point
the background has no $D6$ branes anymore because these are
replaced by fluxes, as expected in a geometric transition. To see
explicitly that the $D6$ branes have disappeared we need to
consider how the corresponding M--theory harmonic form behaves.
Some aspects of this were discussed in \bdkt. The existence of a
{\it normalizable} M--theory harmonic form is an indication of the
existence of localized gauge fluxes \imamura. Similarly, for
manifolds that are non--K\"ahler the existence of a normalizable
harmonic form indicates the existence of gauge fluxes \robbins. To
show that such localized gauge fluxes do not exist after the
geometric transition, one has to show that the harmonic forms
become non--normalizable. This analysis has still not been done
explicitly, although some details have appeared in \bdkt.

\newsec{Supergravity Analysis in Type IIB}

In the previous section we presented the supergravity analysis for
the type IIA string before and after the geometric transition. In
this section we will discuss the similar analysis for the type IIB
side. The geometry on the type IIB side before the geometric
transition is already known. It is a K\"ahler resolved conifold
with $D5$ branes wrapped on the resolution two cycle of the
resolved conifold with $B_{NS}$ and $B_{RR}$ (or the corresponding
field strength $H_{NS}$ and $H_{RR}$), turned on. The $B_{RR}$
component is related to the $D5$ brane source. Most of the details
of this background can be extracted from \pandoz\ where this
background was derived. An alternative derivation in which a
fourfold with fluxes is the starting point can be found in our
previous paper \bdkt. In the following we discuss the properties
of this background.

\subsec{{$\underline{\rm \bf {Background~ before~ geometric~
transition}}$}}


\noindent The metric and tensor fields before the geometric
transition are \eqn\metresconi{\eqalign{ds^2 = & ~h^{-1/2}
ds^2_{0123} +  \gamma'\sqrt{h}~ dr^2 + 
(dz + \Delta_1~{\rm cot}~\theta_1~ dx + \Delta_2~{\rm cot}~\theta_2~
dy)^2 + \cr & ~~~+ \left({\gamma \sqrt{h} \o 4} d\theta_1^2 + dx^2\right)
 + \left({(\gamma + 4a^2)\sqrt{h} \o 4} d\theta_2^2 + dy^2 \right)}}
\eqn\bfico{B_{NS} = {\cal J}_1 ~d\theta_1 \wedge dx + {\cal J}_2~
d\theta_2 \wedge dy, ~~F_5 = K(r)~(1 + \ast) ~dx \wedge dy \wedge
dz \wedge d\theta_1\wedge d\theta_2} \eqn\hrrbg{\eqalign{& {H} =
c_1~(dz \wedge d\theta_2 \wedge dy - dz \wedge d\theta_1 \wedge
dx)}} 
\noindent The various quantities appearing in the above equations
are defined in \pandoz,\bdkt. The type IIB axion--dilaton
vanishes. This background is the IR limit of the corresponding
${\cal N} =1$ gauge theory which is supported on the wrapped $D5$
brane(s).


\subsec{{$\underline{\rm \bf {Background~ after~ geometric~
transition}}$}}


\noindent In the previous section we discussed the background
before geometric transition and in the following we will determine
its form after geometric transition. From the duality chain
presented in \bdkt, the strategy becomes clear. One can try to
construct the type IIB background as the mirror of the type IIA
background after geometric transition, i.e. the mirror of
\iiafinal. The only subtlety which arises in this approach is that
on the type IIA side the isometry along the $\psi$ (or $dz$)
direction is broken by the $B_{NS}$ fields even though all other
fields preserve this background.

In order to restore the isometry in the $\psi$ direction we will
perform the coordinate transformation presented in equation (4.33)
of \bdkt. This transformation eliminates the $\psi$ dependence of
$B_{NS}$. The two spheres of the type IIA metric \iiafinal\ are
invariant under this transformation. On the other hand, the $dz$
fibration is not. A similar situation appeared before in \bdkt. In
order to solve this problem we assumed that in the delocalization
limit the $\psi$ dependence in the $dz$ fibration can be easily
absorbed. In this way the metric plus the complete type IIA
background will have the full isometries along the $x, y$ and $z$
directions. Similarly we will assume in the present situation that
after making the transformation in equation (4.33) of \bdkt\ we
recover the background with full isometries. As before, we would
like to remind the readers that this is the only assumption we
make.

\noindent Using this, we can write the type IIA $B_{NS}$ field as:
\eqn\bfieaf{ \epsilon~{B_{NS}\o \sqrt{\alpha}} = dx\wedge
d\theta_1 - dy\wedge
    d\theta_2 +(Ad\theta_1-Bd\theta_2)\wedge dz}
where $\alpha = (1+A^2+B^2)^{-1}$ as defined earlier. Moreover we
define \eqn\alpze{\alpha^{-1}_0 =
j_{xx}j_{yy}-j_{xy}^2=CD+(CB^2+DA^2)e^{2\phi}} with $j_{mn}$ being
the components of the type IIA metric \iiafinal. After the mirror
transformation, the type IIB metric is \eqn\iibmiror{\eqalign{
ds^2 = & {1\o G_{zz}}\big(dz+{\cal B}_{\mu z}dx^\mu+{\cal
B}_{xz}dx
    +{\cal B}_{yz}dy\big)^2 - \frac{1}{G_{zz}}\big(G_{z\mu}dx^\mu+G_{zx}dx
    +G_{zy}dy\big)^2 \cr
   & G_{\mu\nu}~dx^\mu dx^\nu +2G_{x\nu}~dx~ dx^\nu +2G_{y\nu}~dy~ dx^\nu
    +2G_{xy}~dx~ dy + G_{xx}~dx^2 + G_{yy}~dy^2.}}
The components $G_{mn}$ are given by the following matrix
elements: \eqn\twoacomp{\eqalign{G & = \pmatrix{G_{xx} & G_{xy} &
G_{xz} & G_{x\theta_1} & G_{x\theta_2} \cr \noalign{\vskip -0.20
cm}  \cr G_{xy} & G_{yy} & G_{yz} & G_{y\theta_1} & G_{y\theta_2}
\cr \noalign{\vskip -0.20 cm}  \cr G_{xz} & G_{yz} & G_{zz} &
G_{z\theta_1} & G_{z\theta_2} \cr \noalign{\vskip -0.20 cm}  \cr
G_{x\theta_1} & G_{y\theta_1} & G_{z\theta_1} &
G_{\theta_1\theta_1}& G_{\theta_1\theta_2} \cr \noalign{\vskip
-0.20 cm}  \cr G_{x\theta_2} & G_{y\theta_2} & G_{z\theta_2} &
G_{\theta_1\theta_2} & G_{\theta_2\theta_2}} \cr \noalign{\vskip
-0.25 cm}  \cr & = \pmatrix{\alpha_0 D_1 & -\alpha_0 e^{2\phi} AB
& 0 & -\alpha_0 \sqrt{\alpha} D_1 & -\alpha_0 \sqrt{\alpha}
e^{2\phi} AB \cr \noalign{\vskip -0.20 cm}  \cr
 -\alpha_0 e^{2\phi} AB & \alpha_0 C_1 & 0 & \alpha_0 \sqrt{\alpha} e^{2\phi} AB
 & \alpha_0 \sqrt{\alpha} C_1 \cr
\noalign{\vskip -0.20 cm}  \cr 0 & 0  & \alpha_0 e^{2\phi} CD  & 0
& 0 \cr \noalign{\vskip -0.20 cm}  \cr -\alpha_0 \sqrt{\alpha} D_1
&  \alpha_0 \sqrt{\alpha} e^{2\phi} AB & 0 & C + \alpha_0
\sqrt{\alpha} D_1  & \alpha_0 \alpha e^{2\phi} AB \cr
\noalign{\vskip -0.20 cm}  \cr
 -\alpha_0 \sqrt{\alpha} e^{2\phi} AB & \alpha_0 \sqrt{\alpha} C_1 & 0
 & \alpha_0 \alpha e^{2\phi} AB   &
D + \alpha_0 \alpha C_1}}} where $A$ and $B$ have been defined in
\bdkt\ and $C$ and $D$ were related to the mirror type IIA metric:
\eqn\canddinmiror{\eqalign{ & C = \frac{g_2}{2}-\frac{g_4}{4\xi}~
= ~
    \frac{\alpha}{2}\left(1+B^2-\frac{AB}{\sqrt{(1+A^2)/(1+B^2)}}\right) \cr
 &  D =  \frac{g_2}{2}+\frac{g_4}{4\xi} ~ = ~
    \frac{\alpha}{2}\left(1+B^2+\frac{AB}{\sqrt{(1+A^2)/(1+B^2)}}\right).}}
$C_1$ and $D_1$ are short hand notation for \eqn\conedone{C_1 ~=~
C + A^2 e^{2\phi}, ~~~~~~ D_1 = D + B^2 e^{2\phi}.} Using the
above form of the metric components \twoacomp, the $dz$ fibration
structure can be shown to take the following form:
\eqn\dzfib{\eqalign{\frac{1}{G_{zz}} &
    \Big[dz-A\big[\alpha_0 e^{2\phi}D~dx -
    \epsilon^{-1}\sqrt{\alpha}~(1+\alpha_0 e^{2\phi}D)~d\theta_1\big] \cr
    &- B\big[\alpha_0 e^{2\phi}C~dy
    +\epsilon^{-1}\sqrt{\alpha}~(1+\alpha_0
    e^{2\phi}C)~d\theta_2\big]\Big]^2.}}
The $dz$ fibration structure can be re--grouped so that it takes a
simpler form, in the following way
\eqn\dzgroup{\eqalign{\frac{1}{G_{zz}} & \Big[ (dz + A
  \epsilon^{-1}
  \sqrt{\alpha}~ d\theta_1 - B \epsilon^{-1}
  \sqrt{\alpha}~ d\theta_2) - \cr
  & A \alpha_0~ e^{2\phi} D (dx - \epsilon^{-1}
  \sqrt{\alpha}~ d\theta_1)
   - B \alpha_0~ e^{2\phi} C (dy + \epsilon^{-1}\sqrt{\alpha}~
  d\theta_2)\Big]^2.}}

The fibration structure has many interesting properties. First,
the $dx$ and the $dy$ directions are fibered over the base in the
expected way due to the background $B_{NS}$ fields on the type IIA
side. The $dx, dy$ fibrations are
 \eqn\dxdyfib{ dx - \epsilon^{-1}
\sqrt{\alpha}~d\theta_1, ~~~~~~~~~~ dy + \epsilon^{-1}
\sqrt{\alpha}~d\theta_2.} Does this mean that in type IIB theory
we get again a non--K\"ahler manifold in the closed string side?
Furthermore, due to small $\epsilon$, the non--K\"ahlerity would
be very large!

Fortunately, this is not the case as the fibration structure that
we see above actually {\it cancels} the transformations that we
did earlier in the open string side of type IIB to get the mirror
type IIA manifolds! Thus even though there are non--trivial
fibrations, these are actually quite harmless and therefore will
not produce any non--K\"ahlerity in the manifold. So we seem to
recover a K\"ahler manifold (at least at the fibration level) in
the type IIB picture. This is again supported by the other
components of the metric which are: \eqn\otcommet{\eqalign{ ds^2 =
& \Big[C~d\theta_1^2 + \alpha_0 D_1\big(dx- \epsilon^{-1}
\sqrt{\alpha}~d\theta_1\big)^2\Big]
    + ~\Big[D~d\theta_2^2 + \alpha_0 C_1\big(dy+\epsilon^{-1} \sqrt{\alpha}~
    d\theta_2\big)^2\Big]\cr
  &  + ~\Big[0\cdot d\theta_1 d\theta_2 -
   2\alpha_0 AB e^{2\phi}\big(dx- \epsilon^{-1}
\sqrt{\alpha}~ d\theta_1\big)
    \big(dy+\epsilon^{-1} \sqrt{\alpha}~d\theta_2\big)\Big].}}
Looking carefully at the metric we see that our old problem has
come back to haunt us again. The metric has no $d\theta_1
d\theta_2$ factor. Whatever $d\theta_1 d\theta_2$  factors were
created via mirror transformations have gone into the fibration.
This means that the type IIA manifold that we started with doesn't
really suffice, and we need to put a non--trivial complex
structure on it before we even start doing the mirror operation.

In the following we will discuss the three parts of the metric,
$z$--fibration and the two lines in \otcommet, separately. Let us
denote the three terms by \eqn\namemetric{ ds^2_{IIB} = ds_z^2 +
ds_T^2 +
  ds_{12}^2,}
where $ds_z^2$ stands for the $z$--fibration, $ds_T^2$ refers to
the first line in \otcommet, coming from the two tori defined
below, and $ds_{12}^2$ describes the $\theta_1$-$\theta_2$ cross
terms.

\noindent Let us define two tori with complex structures $\tau_1$
and $\tau_2$ in the following way: \eqn\twotor{dz_1 \equiv  dx -
\tau_1 d\theta_1, ~~~~~~~~~~ dz_2 \equiv dy - \tau_2 d\theta_2}
where $\tau_1$ and $\tau_2$ are of the form: \eqn\tautau{\tau_1 =
f_1 + b_{x\theta_1} + i, ~~~~~~~~~~~~~~~ \tau_2 = f_2 +
b_{y\theta_2} + i.} Here, $f_1$ and $f_2$ are still arbitrary and
will be fixed below. The type IIA metric \iiafinal\ can now be
written with non--trivial complex structures, as $ds^2 =
j_{mn}~dx^m dx^n$, with $j_{mn}$ being given by:
\eqn\twoacompnow{\eqalign{j & = \pmatrix{j_{xx} & j_{xy} & j_{xz}
& j_{x\theta_1} & j_{x\theta_2} \cr \noalign{\vskip -0.20 cm}  \cr
j_{xy} & j_{yy} & j_{yz} & j_{y\theta_1} & j_{y\theta_2} \cr
\noalign{\vskip -0.20 cm}  \cr j_{xz} & j_{yz} & j_{zz} &
j_{z\theta_1} & j_{z\theta_2} \cr \noalign{\vskip -0.20 cm}  \cr
j_{x\theta_1} & j_{y\theta_1} & j_{z\theta_1} &
j_{\theta_1\theta_1}& j_{\theta_1\theta_2} \cr \noalign{\vskip
-0.20 cm}  \cr j_{x\theta_2} & j_{y\theta_2} & j_{z\theta_2} &
j_{\theta_1\theta_2} & j_{\theta_2\theta_2}} \cr \noalign{\vskip
-0.25 cm}  \cr & = \pmatrix{C_1 & e^{2\phi} AB & e^{2\phi} A &
-b_{x\theta_1} C_1 - f_1 C & -e^{2\phi} b_{y\theta_2} AB \cr
\noalign{\vskip -0.20 cm}  \cr
 e^{2\phi} AB & D_1 & e^{2\phi} B & -e^{2\phi} b_{x\theta_1} AB
 & -b_{y\theta_2} D_1 - f_2 D \cr
\noalign{\vskip -0.20 cm}  \cr e^{2\phi} A & e^{2\phi} B  &
e^{2\phi} - \epsilon &  -e^{2\phi} b_{x\theta_1} A & -e^{2\phi}
b_{y\theta_2} B \cr \noalign{\vskip -0.20 cm}  \cr -b_{x\theta_1}
C_1 - f_1 C &  -e^{2\phi} b_{y\theta_2} AB & -e^{2\phi}
b_{x\theta_1} A & C F_1 + b^2_{x\theta_1}C_1 & e^{2\phi} AB
b_{x\theta_1} b_{y\theta_2} \cr \noalign{\vskip -0.20 cm}  \cr
 -e^{2\phi} b_{y\theta_2} AB & -b_{y\theta_2} D_1 - f_2 D  & -e^{2\phi}
 b_{y\theta_2} B  & e^{2\phi} AB b_{x\theta_1} b_{y\theta_2}  &
D F_2 + b^2_{y\theta_2}D_1}}} $F_1$ and $F_2$ used in the above
metric are defined as follows: \eqn\fonftw{F_1 = 1 + f_1^2 + 2 f_1
b_{x\theta_1}, ~~~~~~~~ F_2 = 1 + f_2^2 + 2 f_2 b_{y\theta_2}} and
the other quantities have been defined in \bdkt.

In the above matrix components, observe that we have again shifted
the $j_{zz}$ components by $\epsilon$, i.e. \eqn\jzzf{ j_{zz} ~=~
{}^{\rm lim}_{\epsilon~ \to ~0} ~~(e^{2\phi} - \epsilon).} The two
spheres (which we have now taken as two tori) have the following
metric \eqn\twto{ ds_{T(IIA)}^2 = C ~ \vert dz_1 \vert^2 + D ~
\vert dz_2 \vert^2} with $C,~D$ defined above. The $dz$ fibration
structure changes as \jzzf.

After the mirror transformation on the metric \twoacompnow, some
of the components of the mirror metric are not affected by the
non--trivial complex structure. They are \eqn\gxxyy{G_{xx} =
\alpha_0 j_{yy} = \alpha_0 (D + e^{2\phi} B^2), ~~~~~~~~~~ G_{yy}
= \alpha_0 j_{xx} = \alpha_0 (C + e^{2\phi} A^2)} alongwith the
other two components: \eqn\gtheta{G_{x\theta_1} = \alpha_0
(B_{y\theta_1} j_{xy} - B_{x\theta_1} j_{yy}) ~~~~~ G_{x\theta_2}
= \alpha_0 (B_{y\theta_2} j_{xy} - B_{x\theta_2} j_{yy})} and
similarly $G_{y\theta_1}$ and $G_{y\theta_2}$. Here $B_{mn}$
denote the $B_{NS}$ fields as it appeared earlier in \bfieaf. With
the choice of an integrable complex structure in type IIB {\it
before} geometric transition, the $B_{NS}$ fields take the
following form: \eqn\bfinow{{\epsilon~B_{NS}} = \sqrt{\langle
\alpha \rangle_1}~ dx\wedge d\theta_1 - \sqrt{\langle \alpha
\rangle_2} ~ dy\wedge d\theta_2 +(A \sqrt{\langle \alpha
\rangle_1}~ d\theta_1-B\sqrt{\langle \alpha \rangle_2} ~
d\theta_2)\wedge dz} where $\langle \alpha \rangle_i$ have already
been defined in sec. 2. One can now easily see that this $B$ field
is actually a pure gauge! In other words $B_{NS} = d \Lambda$. We
will derive the form of $\Lambda$ below.

Our next goal is to get a non-vanishing $d\theta_1 d\theta_2$ term
in the metric component $ds_{12}^2$. According to the arguments of
\bdkt, this can be achieved by using a non--trivial complex
structure. However, the metric on the type IIA side \iiafinal\ is
much more involved since it is non--K\"ahler. Therefore, it is
necessary to go through the whole argument once again to check if
it applies to the present case.

Using a complex structure more complicated than $\tau = i$ will
result in non--vanishing metric components $G_{z \theta_i}$ etc.
As a consequence the metric along the $\theta_i \theta_j$ ($i,j =
1, 2$) directions will have the following form (excluding the
$B$--dependent terms which go into the $z$--fibration):
\eqn\metfonu{ds^2_{\theta_i \theta_j} = \left( G_{\theta_i
\theta_j} - {G_{z\theta_i} G_{z\theta_j} \o G_{zz}}
\right)~d\theta_i d\theta_j.} Let us compute this term by term.
First we consider the $G_{zz}$ component. Choosing $j_{zz} =
e^{2\phi} - \epsilon$, we obtain: \eqn\gzzcg{G_{zz} = j_{zz} -
\alpha_0 [ j_{yy} j_{xz}^2 + j_{xx} j_{yz}^2 - 2 j_{xz} j_{yz}
j_{xy}] = \alpha_0~ CD~ e^{2\phi} - \epsilon.} The $\e$ dependence
is expected from the analysis of \bdkt. However, the finite and
non--zero part is more involved than the one of \bdkt. As a result
all the subsequent components will become more involved because of
the non--K\"ahler nature of the starting manifold. Similarly the
$G_{z\theta_i}$ component is \eqn\gzvb{G_{z\theta_i} =
j_{z\theta_i} - \alpha_0 [ j_{x\theta_i} j_{xz} j_{yy} +
j_{y\theta_i} j_{yz} j_{xx} - j_{xy} (j_{y\theta_i} j_{xz} +
j_{x\theta_i} j_{yz})].} For $i = 1,2$ one obtains \eqn\nozsdf{
G_{z\theta_1} = \alpha_0~f_1 ~e^{2\phi} ~CAD, ~~~~~~~~~~~~~~
G_{z\theta_2} = \alpha_0~f_2~ e^{2\phi}~DBC.} Observe that
$f_1=f_2=0$ implies a vanishing of $G_{z\theta_1}$ and
$G_{z\theta_2}$. This is consistent with what we already know from
our earlier analysis.

The other components, like $G_{xx}, G_{yy}$ for example, remain
unchanged from their values with trivial complex structure of
$\tau = i$. In the following we will compute $G_{\theta_1
\theta_2}$. Using mirror symmetry this component is given by
\eqn\gtt{G_{\theta_1 \theta_2} = j_{\theta_1 \theta_2} -
\alpha_0~j_{xy} ~B_{x\theta_1} B_{y \theta_2} - \alpha_0 ~[j_{yy}
j_{x\theta_1} j_{x\theta_2} + j_{xx} j_{y\theta_1} j_{y\theta_2} -
j_{xy} (j_{y\theta_1} j_{x\theta_2} + j_{x\theta_1}
j_{y\theta_2})].} The above relation is a special case of results
obtained previously in \bdkt. In deriving \gtt\ we have inherently
assumed a particular form for the $B_{NS}$ field, namely
$B_{x\theta_2} = B_{y\theta_1} = 0$. However, the cross components
are non--vanishing. This is where \gtt\ would differ from the
equivalent formula that we had in the type IIB case of \bdkt.  The
existence of non--trivial cross terms is related to
non--K\"ahlerity as well as non--trivial complex structures of the
two tori.
 For $\tau = i$, the above relation turns into
\eqn\gvft{G^{\rm old}_{\theta_1 \theta_2} =
-\alpha_0~j_{xy}~B_{x\theta_1}~B_{y\theta_2}} where we have
assumed $f_i = 0$.  Including non--vanishing values for $f_1$ and
$f_2$ results in a simple extension of our previous results in
\bdkt, i.e.: \eqn\bgh{G_{\theta_1 \theta_2} = - \alpha_0
~j_{xy}~B_{x\theta_1}~B_{y\theta_2} + \alpha_0~e^{2\phi}~ABCD~f_1
f_2.} To obtain the final answer we still have to compute the
$G_{z\theta_i}$ terms as in \metfonu\ (for $i =1$ and $j = 2$),
and we need to determine the functions $f_1$ and $f_2$. The
relation \metfonu\ implies \eqn\fonum{ds^2_{\theta_1 \theta_2} =
-2 \left( \alpha_0~j_{xy}~B_{x\theta_1}~B_{y\theta_2} +
f_1f_2~AB~\epsilon\right) d\theta_1 d\theta_2.} Now we are close
to obtaining the final expression for $ds_{12}^2$. The first term
in \fonum\ is the usual term we discussed before. The second term
is important. We want the single $\theta_1$--$\theta_2$ cross
term, that does not arise from any fibration, to have the same
pre--factor as the $x$--$y$ cross term from the metric obtained
with vanishing $f_1$ and $f_2$. This is achieved by choosing $f_1,
f_2$ to satisfy the following relation: \eqn\foftwr{ f_1 ~f_2 ~ =
~ - \epsilon^{-1} ~\alpha_0~ e^{2\phi}.} This is our first non
trivial conclusion. The above relation is reminiscent to a result
obtained previously in \bdkt. If we set $f_i \equiv \pm
\epsilon^{-1/2} \beta_i$, then $\beta_1 \beta_2 =
\alpha_0~e^{2\phi}$. With this choice, the metric for the
$\theta_1 \theta_2$ direction becomes simple, and is given by
\eqn\mettitii{ds^2_{\theta_1 \theta_2} = -2 \alpha_0
\left(j_{xy}~B_{x\theta_1}~B_{y\theta_2}
-e^{2\phi}~AB\right)~d\theta_1 d\theta_2 = -2\alpha_0
\left(j_{xy}~B_{x\theta_1}~B_{y\theta_2}-j_{xy}\right)~d\theta_1
d\theta_2.} The last identification stems from the type IIA
metric, and should remind the reader of a similar coefficient in
the type IIA case before geometric transition. Finally, the first
term in \mettitii\ can be combined with the metric components in
the $x,y$ direction to give us the complete metric for the cross
terms: \eqn\comet{ds_{12}^2 = 2\alpha_0~j_{xy}
~\left[d\theta_1~d\theta_2 - \left(dx - \epsilon^{-1}
\sqrt{\langle\alpha\rangle_1}~d\theta_1\right) \left(dy +
\epsilon^{-1}
\sqrt{\langle\alpha\rangle_2}~d\theta_2\right)\right].} Here we
have substituted the values of the $B$ fields in \bfinow\ to get
the fibration structure in the desired form. This is the full
structure of the cross terms in the metric.

In the following we will obtain the fibration structure in the
$dx$ and $dy$ directions. As discussed above, this non--trivial
fibration is rather harmless because when \comet\ is written in
terms of original type IIB coordinates {\it before} the geometric
transition, the metric is K\"ahler. To make this discussion
precise, let us rewrite the $dx$ fibration more suggestively (a
similar argument will go through for the $dy$ coordinate). First,
we need the functional form for $\langle \alpha \rangle_1$. This
can be obtained from appendix 1 of \bdkt\
\eqn\julie{\langle\alpha\rangle_1 = {1\o 1 + \Delta_1^2~{\rm
cot}^2~\theta_1 + \Delta_2^2~{\rm cot}^2~\langle\theta_2\rangle}
\equiv {1\o a + b~{\rm cot}^2~\theta_1}.} Here
$\langle\theta_2\rangle$, $a$ and $b$ are taken to be constants,
and we have already defined $\Delta_i$. These are all constants
for the case considered in \bdkt, $\delta r = 0$.

Using the above form of $\langle\alpha\rangle_1$, we will rewrite
the $dx$ fibration structure.  We will assume that $b
> a$. The $b < a$ case can be treated similarly. The
fibration structure takes the following form:
\eqn\figut{\eqalign{d \hat x ~~& ~~\equiv ~~ dx - \epsilon^{-1}
~\sqrt{\langle\alpha\rangle_1}~d\theta_1 \cr & ~~ = ~~d\left[x +
{\epsilon^{-1}\sqrt{a}\o \sqrt{b-a}}~{\rm ln}~\left(\sqrt{b-a \o
a}~{\rm cos}~ \theta_1 + \sqrt{{\rm cos}^2~\theta_1 - {b\o a} {\rm
sin}^2~\theta_1}\right)\right]}} where $\hat x$ is related to the
coordinate that we started with on the resolved conifold side
before the geometric transition. After a simple sign change the
$dy$ fibration can be treated similarly. The final result for the
metric is \eqn\metsiol{ds_{12}^2 = 2~\alpha_0~AB~e^{2\phi}
~(d\theta_1 d\theta_2 - d\hat x d \hat y).} For completeness, let
us also define \eqn\defzhat{d\hat{z} = dz+ \epsilon^{-1}
A\sqrt{\alpha}~d\theta_1 - \epsilon^{-1}
B\sqrt{\alpha}~d\theta_2.}

In order to obtain the final result for the metric we still have
to determine $f_1$ and $f_2$ (or equivalently, $\beta_1,
\beta_2$). This can be done using the reasoning laid out in \bdkt.
The functions $f_1, f_2$ can be determined from the metric of the
two spheres (or equivalently, two tori, as we did everything
assuming them to be tori). Let us first consider the torus
described by the coordinates ($\theta_2, y$). For a trivial
complex structure on the type IIA side the metric along the tori
direction can be extracted from \otcommet. For non--vanishing
$f_2$, the $G_{\theta_2\theta_2}$ component is
\eqn\gthth{G_{\theta_2\theta_2} = j_{\theta_2 \theta_2} +
\alpha_0~[ j_{xx}~B^2_{y\theta_2} - j_{yy}~j_{x\theta_2}^2 -
j_{xx}~j_{x\theta_2}^2 + 2j_{xy}~j_{x\theta_2}~ j_{y\theta_2}]}
After using a non--vanishing value for $f_2$, terms proportional
to $f_2$ and $f_2^2$ appear. As a result one obtains
\eqn\terG{G_{\theta_2\theta_2} = {\cal A}_0~f_2^0 + {\cal A}_1~f_2
+ {\cal A}_2~f_2^2} which terminates at this order because of the
mirror formula \gthth. As it is also clear that there would be no
negative powers of $f_2$. A similar analysis can be performed for
the component $G_{\theta_1 \theta_1}$, where the series will be
determined by $f_1$. The coefficients ${\cal A}_i$ are
\eqn\calai{\eqalign{&{\cal A}_0 = D + \alpha_0~(C + e^{2\phi}
A^2)~B^2_{y\theta_2}, ~~~~~~~~~~~  {\cal A}_2 = D - \alpha_0~(C +
e^{2\phi}B^2)~D^2 \cr & {\cal A}_1 = 2~D~b_{y\theta_2} -
2~D~\alpha_0~(D+e^{2\phi}B^2)(C + e^{2\phi} A^2)~b_{y\theta_2} +
2~D~\alpha_0~A^2B^2~e^{4\phi}~b_{y\theta_2}}} Let us consider
these coefficients carefully. The first coefficient ${\cal A}_0$
also appeared for a trivial complex structure. The second
coefficient ${\cal A}_1$ can be shown to vanish. The third
coefficient ${\cal A}_2$ simplifies after substituting the values
for $\alpha_0$ etc. As a result the additional terms in
$G_{\theta_2\theta_2}$ for non zero $f_2$ can now be concisely
written as: \eqn\calath{{\cal A}_2 ~=~ \alpha_0 CDB^2 e^{2\phi},
~~~~~~~~~ {\cal A}_1 ~=~ 0.} The above is of course not the
complete answer, as the metric along the $\theta_2\theta_2$
direction is also influenced by $G_{z\theta_2}$. We will evaluate
this contribution below. But first, let us consider the other
component $G_{\theta_1\theta_1}$. This is written in terms of
$f_1$ with coefficients ${\cal B}_i$, $i=0,1,2$, given by
\eqn\calbth{{\cal B}_0 ~ = ~ C + \alpha_0~(D + e^{2\phi}
B^2)~B^2_{x\theta_1}, ~~~~ {\cal B}_1 ~=~ 0, ~~~~ {\cal B}_2 ~=~
\alpha_0 CD A^2 e^{2\phi}.} Now we are getting closer to the final
result. The metric along the $\theta_1^2$ or the $\theta_2^2$
directions will get contributions from $G_{z\theta_i}$ and
$G_{zz}$. This implies (again excluding the contributions that are
absorbed into the $z$--fibration $ds_z^2$):
\eqn\mettott{\eqalign{&ds^2_{\theta_1\theta_1} = C +
\epsilon^{-2}\alpha_0\alpha (D + e^{2\phi} B^2)  - \beta_1^2 A^2
\cr & ds^2_{\theta_2 \theta_2} = D + \epsilon^{-2}\alpha_0 \alpha
(C + e^{2\phi} A^2) - \beta_2^2 B^2.}} In order to complete our
result we still need to determine $\beta_1$ and $\beta_2$. We take
the results of \bdkt\ as a guide. Requiring the terms
$ds^2_{\theta_2\theta_2}$ and the $yy$ component of the metric to
be equal, we can perform the coordinate transformation (4.33) of
\bdkt. After this transformation the metric takes the deformed
conifold form. Therefore to get the value of $\beta_2$ we have to
identify some parts of $ds^2_{\theta_2 \theta_2}$ with $G_{yy}$.
The part of $ds^2_{\theta_2 \theta_2}$ that is relevant is of
course the one that does not go into the $dy$ fibration structure: those
that are independent of $b_{y\theta_2}$ or $\alpha$ in \mettott.
In other words, we have \eqn\dforb{ D - \beta_2^2 B^2 = \alpha_0
(C + e^{2\phi} A^2).} Combining the above equation with \foftwr\
determines both $\beta_1$ and $\beta_2$. The result is
\eqn\betaiii{\eqalign{& \beta_1 = {\sqrt{\alpha_0}~ e^{2\phi} \o
\sqrt{e^{2\phi}~ CD - {(C + e^{2\phi} A^2)(1-D^2) \o B^2}}}\cr &
\beta_2 = \sqrt{\alpha_0 ~e^{2\phi}~ CD - {\alpha_0~ (C +
e^{2\phi} A^2)(1-D^2) \o B^2}}}} The above functional form for
$\beta_2$ shows that the coefficients of $ds^2_{\theta_2\theta_2}$
and $G_{yy}$ are the same. On the other hand, the functional form
for $\beta_1$ implies that the coefficients of
$ds^2_{\theta_1\theta_1}$ and $G_{xx}$ are not same. In the case
studied in \bdkt\ the coefficients for both the tori were made
equal by going to a {\it canonical} basis. The choice of the
canonical basis was possible there because we took the case where
$\delta r =0$ and therefore the coefficients for the tori metric
were basically constants and could be absorbed in the definition
of $d\theta_i$. Here, however, the metric of the tori in type IIA
have coefficients that are functions of all the coordinates. So a
simple redefinition cannot be made; implying that we can bring
only one of the tori into the desired form. This causes no problem
of course, because we are anyway making coordinate transformation
(i.e the transformation equation 4.33 of \bdkt) to one of the
tori.

The final result for the type IIB metric is
\eqn\fiibmet{\eqalign{ds^2 ~ = & ~~ h_1~[d\hat z + a_1~ d\hat x +
a_2~ d\hat y]^2 + h_2~[d{\hat y}^2 + d\theta_2^2] + h_4~[d{\hat
x}^2 + h_3~d\theta_1^2]~ + \cr & ~~~~~~~ + h_5~{\rm
sin}~\psi~[d\hat x ~d\theta_2 + d\hat y ~d\theta_1] + h_5~{\rm
cos}~\psi~[d\theta_1 d\theta_2 - d\hat x~ d\hat y]}} with the
$h_i$ defined below. This should be a K\"ahler manifold as it
lacks the $B$ field dependent fibration structure, which made the
type IIA manifold non--K\"ahler. This metric is similar to the
Klebanov--Strassler \kleb\ background. In fact, there is a simple
reason for this. The $B_{NS}$ field we discussed on the type IIA
side in \bfinow\ is pure gauge and can be written in terms of $d
\Lambda$. Defining $\Delta = \sqrt{1 - k^2 {\rm sin}^2 \theta_1}$
and $k = \sqrt{b-a \o b}$, we see that \eqn\totder{\eqalign{&
\sqrt{\langle \alpha \rangle_1}~dx \wedge d\theta_1 = {1\o k
\sqrt{b}}~d\left[{\rm ln}~(k {\rm cos}~\theta_1 + \Delta)\right]
\wedge dx \cr & A~\sqrt{\langle\alpha\rangle_1}~d\theta_1 \wedge
dz = {1\o k \sqrt{b}}~ d\left[\sqrt{b}~{\rm arctan}~{k~{\rm
sin}~\theta_1 \o \Delta} \right] \wedge dz}} with similar
relations for the $\theta_2$ coordinates. From the above relation
we see that $\Lambda$ can be defined as follows:
\eqn\deflamb{\Lambda = \left[{\rm ln}~(k {\rm cos}~\theta_1 +
\Delta)\right] \wedge dx + \left[\sqrt{b}~{\rm arctan}~{k~{\rm
sin}~\theta_1 \o \Delta} \right] \wedge dz - \big\{\theta_1
\leftrightarrow  \theta_2,~ x\leftrightarrow y\big\}} up to an
overall constant factor. Since the $B_{NS}$ field can be gauged
away, the type IIB metric remains K\"ahler. On the other hand, the
original type IIB $B_{NS}$ field given in \pandoz, \bdkt\ is {\it
not} a gauge artifact and therefore {\it does} make the type IIA
manifolds non--K\"ahler --- both before and after geometric
transition.

The two tori appearing in the type IIB metric have complex
structures $\tau_2 = i$ and $\tau_1 = i \sqrt{h_3}$. Using this we
can define the corresponding metric as \eqn\metrope{ds^2 =
h_2~\vert dz_2\vert^2 + h_4~\vert dz_1 \vert^2, ~~~ dz_2 = d\hat y
+ i ~d\theta_2, ~~~ dz_1 = d\hat x + i\sqrt{h_3}~d\theta_1.} Here
we have used the notation \eqn\defjam{\eqalign{& h_1 = {e^{-2\phi}
\o \alpha_0~CD}, ~~ h_2 = \alpha_0~(C + e^{2\phi} A^2), ~~ h_4 =
\alpha_0~(D + e^{2\phi} B^2), ~~ h_5 = 2~\alpha_0~e^{2\phi} AB \cr
& ~~~~~~~~~~~ h_3 = {C - \beta_1^2 A^2 \o \alpha_0~(D + e^{2\phi}
B^2)}, ~~~ a_1 = - \alpha_0~e^{2\phi}~AD, ~~~ a_2 = -
\alpha_0~e^{2\phi}~BC}.}

So far we have been mostly concentrated on deriving the background
metric. In the following we will determine other background
fields, like $B_{NS}, B_{RR}$, the axion and the dilaton. We will
first consider the $B_{NS}$ field. After applying mirror symmetry,
$B_{NS}$ is found to be: \eqn\bbfielc{\eqalign{ {B^b} = & ~~
\left( {\tilde B}_{\mu\nu} + {2 {\tilde B}_{z[\mu} G_{\nu]z} \over
G_{zz}} \right) dx^\mu \wedge dx^\nu + \left( {\tilde B}_{\mu x} +
{2 {\tilde B}_{z[\mu} G_{x]z} \over G_{zz}}\right)
 dx^\mu \wedge dx  \cr
& ~~ \left( {\tilde B}_{\mu y} + {2 {\tilde B}_{z[\mu} G_{y]z}
\over G_{zz}}
 \right) dx^\mu \wedge dy
+ \left( {\tilde B}_{xy} + {2 {\tilde B}_{z[x} G_{y]z} \over
G_{zz}} \right) dx \wedge dy \cr & ~~ + {G_{z \mu} \over G_{zz}}
dx^\mu \wedge dz + {G_{z x} \over G_{zz}} dx \wedge dz + {G_{z y}
\over G_{zz}} dy \wedge dz.}} The above relation has already been
derived in \bdkt. In the present case we have to take into account
that before T--duality the $B_{NS}$ field \bfinow\ changes due to
the introduction of a non--trivial complex structure in \tautau:
\eqn\bfichange{\epsilon{B_{NS}\o \sqrt{\alpha}} =
  dx\wedge d\theta_1 -  dy\wedge
  d\theta_2 +(A d\theta_1-B d\theta_2)\wedge dz + AB
  (f_1 + f_2)~d\theta_1 \wedge d\theta_2.}
Here we have taken the limit depicted earlier in \repmen\ i.e.
$\langle \alpha \rangle_1 = \langle \alpha \rangle_2 =\alpha$. If
we do not consider this limit then this will again be a gauge
artifact. The various components of ${\tilde B}$ in the above
relation \bbfielc\ become \eqn\varvak{\eqalign{{\tilde B}_{NS} & =
\pmatrix{{\tilde B}_{xx} & {\tilde B}_{xy} & {\tilde B}_{xz} &
{\tilde B}_{x\theta_1} & {\tilde B}_{x\theta_2} \cr
\noalign{\vskip -0.20 cm}  \cr {\tilde B}_{yx} & {\tilde B}_{yy} &
{\tilde B}_{yz} & {\tilde B}_{y\theta_1} & {\tilde B}_{y\theta_2}
\cr \noalign{\vskip -0.20 cm}  \cr {\tilde B}_{zx} & {\tilde
B}_{zy} & {\tilde B}_{zz} & {\tilde B}_{z\theta_1} & {\tilde
B}_{z\theta_2} \cr \noalign{\vskip -0.20 cm}  \cr {\tilde
B}_{\theta_1 x} & {\tilde B}_{\theta_1 y} & {\tilde B}_{\theta_1
z} & {\tilde B}_{\theta_1\theta_1}& {\tilde B}_{\theta_1\theta_2}
\cr \noalign{\vskip -0.20 cm}  \cr {\tilde B}_{\theta_2 x} &
{\tilde B}_{\theta_2 y} & {\tilde B}_{\theta_2 z} & {\tilde
B}_{\theta_2\theta_1} & {\tilde B}_{\theta_2\theta_2}}}}
\eqn\vabfr{\eqalign{& = \pmatrix{0 & 0 & -\alpha_0 e^{2\phi} AD &
b_{x\theta_1}~{E}_3 & -\alpha_0 e^{2\phi} ABD~f_2 \cr
\noalign{\vskip -0.20 cm}  \cr
 0 & 0 & -\alpha_0 e^{2\phi} BC & -\alpha_0 e^{2\phi} ABC~f_1 &
 b_{y\theta_2}~{E}_4 \cr \noalign{\vskip -0.20 cm}  \cr
\alpha_0 e^{2\phi} AD & \alpha_0 e^{2\phi} BC  & 0  &
-\epsilon^{-1} A \sqrt{\alpha}~{E}_2 & \epsilon^{-1} B
\sqrt{\alpha}~{E}_1 \cr \noalign{\vskip -0.20 cm}  \cr
-b_{x\theta_1}~{E}_3  & \alpha_0 e^{2\phi} ABC~f_1 & \epsilon^{-1}
A \sqrt{\alpha}~{E}_2  & 0 & \epsilon^{-1}\sqrt{\alpha} AB ~F_3
 \cr \noalign{\vskip -0.20 cm}  \cr
 \alpha_0 e^{2\phi} ABD~f_2  & -b_{y\theta_2}~{E}_4  & -\epsilon^{-1} B
 \sqrt{\alpha}~{E}_1  & -\epsilon^{-1}\sqrt{\alpha} AB ~F_3 & 0}.}}
Here we have defined the quantities $E_i$ and $F_3$ in terms of
$C_1$ and $D_1$ as follows: \eqn\defei{\eqalign{& E_1 = 1 +
\alpha_0 e^{2\phi} C, ~~~~~~~~~~ E_3 = 1 + {\alpha_0 f_1 C ~D_1 \o
b_{x\theta_1}} \cr
 & E_2 = 1 + \alpha_0 e^{2\phi} D, ~~~~~~~~~~ E_4 =
 1 + {\alpha_0 f_2 D~C_1 \o b_{y\theta_2}} \cr
 & F_3 = f_1E_1+f_2E_2.}}
Substituting \varvak\ in \bbfielc\ we find that the
$\theta_1\theta_2$ component of the $B$ field,
$B^b_{\theta_1\theta_2}$, vanishes. Note, that this could not have
been achieved without taking the change in the $B$--field
\bfichange\ into account. The type IIB $B_{NS}$ becomes:
\eqn\finbo{B^b = (b_{x\theta_1} + f_1)~dx \wedge d\theta_1 +
(b_{y\theta_2} + f_2)~dy \wedge d\theta_2 + (Af_1~d\theta_1 + B
f_2~d\theta_2) \wedge dz.} Observe that the above structure of the
B--field might look different from the original type IIB field we
considered in \bdkt\ before the geometric transition. This is due
to the non--zero values for the $z$ components of the original
type IIA B--field and also the underlying non--K\"ahler nature of
the type IIA background. The non--K\"ahler nature which arises due
to the non--trivial fibrations in the $x, y$ and $z$ directions
causes cross terms in the type IIA metric. These cross terms
eventually become the $B$ fields on the type IIB side. However, as
we discussed before, if the $B$ field components with coefficients
$f_1, f_2$ are pure gauge, the final result will be simply given
by \eqn\fontb{B^b = b_{x\theta_1}~d\hat x \wedge d\theta_1 +
b_{y\theta_2}~d\hat y \wedge d\theta_2} which, along with the
K\"ahler metric \fiibmet\ will determine a background similar to
the one studied by Klebanov and Strassler in \kleb.

Similarly, we can determine the RR background fields. The RR
fields are the axion $\tilde\phi$, the RR three form and the five
form field strength. It is easy to see that \eqn\axdil{ \phi ~ = ~
0, ~~~~~~~ \tilde\phi ~ = ~ 0.} The vanishing of $\tilde\phi$ is a
consequence of mirror symmetry. To see that the dilaton also
vanishes we have to consider the limit \repmen, where, without
loss of generality we impose the following value for the
expectation values of $\alpha$ and $\phi_b$ ($\phi_b$ is the type
IIA dilaton {\it before} the geometric transition):
\eqn\expva{\langle \alpha \rangle = \left( 64
e^{2\langle\phi_b\rangle}\right)^{-{1\o 3}}.} In fact,
$\langle\phi_b\rangle$ is close to zero \bdkt\ and therefore
\expva\ fixes the expectation value of $\alpha$ to a positive
integer.

$B_{RR}$ can now easily be determined using mirror symmetry.
{}From the exact supergravity background on the type IIA side we
see that the gauge field has two components: along the $x$ and
$\theta_1$ directions \gaufinalii. In the M--theory analysis of
\bdkt\ this choice of gauge field was the simplest one. Of course,
we could also consider other components, like $y$ and $\theta_2$.
With this choice \gaufinalii, the background $B_{RR}$ field (in
new coordinates) becomes \eqn\bgbrr{ B_{RR} = -2A~d\hat{y}\wedge
d\hat{z} ~-~
\frac{2e^{2\phi}A^2\sqrt{\alpha}}{\epsilon~C_1}~\left(d\hat{y}-
\frac{e^{2\phi}AB}{C_1} ~d\hat{x}\right)\wedge d\theta_1.} This
can be compared with \kleb. It is also easy to see that the five
form background contains a factor of $b_{x\theta_1}$ or
$b_{y\theta_2}$. This is crucial for the type IIB Bianchi identity
to work out correctly.

Before we end this section, let us summarize the full type IIB
background that we got by making a mirror transformation on the
type IIA background: \eqn\finalmente{\eqalign{& ds^2 ~ =  ~~
h_1~[dz + a_1~ dx + a_2~ dy]^2 + h_2~[dy^2 + d\theta_2^2] +
h_4~[d{x}^2 + h_3~d\theta_1^2]~ + \cr & ~~~~~~~~~~~~~~~ + h_5~{\rm
sin}~\psi~[dx ~d\theta_2 + dy ~d\theta_1] + h_5~{\rm
cos}~\psi~[d\theta_1 d\theta_2 - dx~ dy] \cr & B_{NS} =
b_{x\theta_1}~dx \wedge d\theta_1 + b_{y\theta_2}~dy \wedge
d\theta_2, ~ B_{RR} = -2A~dy \wedge dz, ~~ \phi = \tilde\phi = 0}}
where we have omitted the $~\hat{}~$ on the coordinates to avoid
cluttering of formulae.


\newsec{Supergravity Analysis in Type I}

To determine the supergravity backgrounds in type I theory we need
to reconsider the whole scenario again. First, the background that
we gave in the type IIB theory (or even in type IIA) will never
allow us to go to type I and correspondingly to the heterotic
theory. This is because type II theories have larger
supersymmetries and to obtain the chiral nature of type I we have
to cut down half of the world sheet supersymmetries.

One way to cut down supersymmetries is by introducing an
orientifold operation $\Omega$. However, as is well known, the
operation that we need to generate the desired action so that we
get to the type I/heterotic theory, is not simply $\Omega$, but a
more involved one given by \eqn\orient{ \Omega \cdot (-1)^{F_L}
\cdot \sigma,} where $(-1)^{F_L}$ reverses the space-time fermion
numbers and $\sigma$ is a pure orbifold action which reverses an
even number of directions.

The type IIB background that we studied in the previous section
does not have an orientifold action. But we know how to introduce
such an action: by lifting the type IIB background to F--theory.
As studied in detail by Sen, the F--theory lift of a type IIB
background has an orientifold limit in the moduli space which is
determined in terms of some number of $D7$ branes and $O7$ planes
located on the six dimensional base \senFone. The F--theory lift
requires a four--fold which is a non--trivial $T^2$ fibration over
a six--dimensional base. To study the type I theory both before
and after geometric transition, we need the six--dimensional base
to be {\it close} to a resolved conifold or a deformed
conifold\foot{Recall that, in general, the four--fold base is {\it
not} a Calabi--Yau manifold. Therefore, when we say the base is a
resolved (or deformed) conifold we mean some metric that is close
to the resolved (or deformed) conifold metric. It will soon be
clear from the analysis below, what this means exactly. For the
time being we will continue calling the base a resolved (or
deformed) conifold.}. Let us first consider the resolved conifold
base.

The non--trivial $T^2$ fibration over a six--dimensional base
needs to degenerate over some points on the base. Since the $T^2$
can degenerate over an even dimensional space we have the
following four possibilities

\noindent (a) The $T^2$ degenerates over the full six--dimensional
base.

\noindent (b) The $T^2$ degenerates over a four--dimensional
subspace of the six--dimensional base.

\noindent (c) The $T^2$ degenerates over a two--dimensional
subspace of the six--dimensional base.

\noindent (d) The $T^2$ does not degenerate over the base at all.

Looking at the four choices, we can easily  eliminate (a), (b) and
(d). For the first case, if the $T^2$ degenerates over the full
six--dimensional base, then the $\sigma$ action of \orient\ will
act on the full resolved conifold. This would imply that we need
{\it six} T--dualities to convert \orient\ to simply $\Omega$
\eqn\sixt{\Omega \cdot (-1)^{F_L} \cdot \sigma ~~~~
{}^{~T_6}_{\longrightarrow} ~~~~ \Omega,} i.e. six T--dualities to
go to the type I theory. Looking at the resolved conifold metric,
we see that we can only have three isometries (along the $X,Y$ and
$Z$ directions)\foot{We will use the upper case letters to denote
the coordinates, and lower case letters to write the metric
components unless mentioned otherwise.}. Therefore, the first
possibility cannot take us to the type I theory and therefore also
not to the heterotic theory. Similarly, the next choice of four
T--dualities fails because we do not have a sufficient number of
isometries.

The fourth choice allows fluxes and branes because of
non--compactness. However, any number of T--dualities will only
keep us in the type II theory and will never take us out of this
cycle. In other words, we can never go to the type I or heterotic
theories using the choice (d).

The final choice (c) where the fiber degenerates over a
two--dimensional subspace of the base seems to suffice for our
purpose. There is one subtlety, as there are three possibilities
for the two--dimensional subspace with coordinates: $XY, YZ$ and
$XZ$. The question is then which one to choose, as we have equal
isometries along the $X, Y$ and $Z$ directions. The answer to this
question is found by looking at the base of the four--fold {\it
after} the geometric transition, which changes from a resolved
conifold to a deformed conifold, as described in \dotd. The
deformed conifold base does not have three isometries because it
only has isometries along the $X$ and $Y$ directions. This would
mean that if we choose the two--dimensional base to be along the
$X$ and $Y$ directions then (both before and after the geometric
transition) two T--dualities will take us to the type I theory
(and then to heterotic). In this way we can uniquely fix the
four--fold in F--theory to be a $T^2$ fibration over a
two--dimensional base with coordinates $X$ and $Y$. In fact, the
fiber torus will degenerate 24 times over the base $X,Y$ making
this a $K3$ manifold. This chain of arguments will allow us to
determine the metric for the four--fold. Our first guess would be
that the four--fold is nothing but a non--trivial product of a
$K3$ manifold and a four--cycle which we can specify precisely.
The four--cycle, which we shall henceforth call $S_4$, will be
different before and after the geometric transition. The metric
for $S_4$ can be easily identified. Before the geometric
transition the metric can be written as
\eqn\sfourmet{ds^2_{S_{4r}} = ~ \gamma'\sqrt{h} dr^2 + (dz +
\Delta_1~{\rm cot}~\theta_1~ dx + \Delta_2~{\rm cot}~\theta_2
~dy)^2 + {\gamma \sqrt{h} \o 4}~d\theta_1^2 + {(\gamma +
4a^2)\sqrt{h} \o 4}~d\theta_2^2,} where $\gamma, \gamma'$ are
defined in \pandoz, \bdkt. The function $h$ is the overall warp
factor coming from the back--reaction of the branes and fluxes on
the geometry. Similarly, the metric of $S_4$ after the geometric
transition is given by \eqn\metgtsfour{\eqalign{ds^2_{S_{4d}} ~ =
& ~~ h_1~[dz + \hat\Delta_1~{\rm cot}~\theta_1~ dx + \hat\Delta_2~
{\rm cot}~\theta_2~dy]^2 + h_2~d\theta_2^2 + h_4~h_3~d\theta_1^2~
+ \cr & +h^{1/2} ~\gamma'~dr^2 +  h_5~{\rm sin}~\psi~[dx
~d\theta_2 + dy ~d\theta_1] + h_5~{\rm cos}~\psi~[d\theta_1
d\theta_2 - dx~ dy],}} where $h_i$ was already defined in \defjam.
The other new coefficients $\hat\Delta_i$ are defined as
$\hat\Delta_1 = -\alpha_0 ~e^{2\phi}~D~\Delta_1$ and $\hat\Delta_2
= -\alpha_0~e^{2\phi}~C~\Delta_2$.

Looking at the metric of $S_{4r}$ we see that it is of the form
$S_3 \times S^1$, where $S_3$ has coordinates ($\theta_1,
\theta_2, r$) and $S^1$ has coordinate $z$. Naively, one is led to
conclude that the F--theory four--fold should be of the form $S_3
\times S^1 \otimes K3$ with a metric $ds^2 = ds^2_{S^1}~+ ~
ds^2_{K3}~ + ~ds^2_{S_3}$, where $\otimes$ denotes the non-trivial
$S^1$ fibration over the $P^1$ base of $K3$ with coordinates $X,
Y$. The full metric of this system has also been worked out
earlier in \bb,\drs, \dotd. Unfortunately the situation is much
more involved here and this naive choice of metric does not quite
suffice for our case, as we shall explain below.


\subsec{{$\underline{\rm \bf {Background~ before ~ geometric~
transition}}$}}


As we discussed above, for the type I picture, we have to go to
the orientifold corner of the F--theory moduli space. This has
been discussed by \senFone, so we will be brief. Another relevant
reference is \gkp, that discusses the F--theory lift of the
Klebanov--Strassler solution. However, in none of the above
references the {\it metric} for the orientifold corner of the
moduli space has been computed. One might naively think that we
can write the metric as mentioned above, but due to the $Z_2$
action this is a little subtle. The $Z_2$ action arises from the
Weierstrass equation \eqn\weierst{y^2 = x^3 + x f(u) + g(u),}
where $u$ is the coordinate of the base ${\cal M}$ and $f, g$ are
some specific polynomials given in \senFone. It gives rise to a
type IIB compactification on ${\cal M}$ modded out by $Z_2 \equiv
\Omega \cdot (-1)^{F_L} \cdot \sigma$, where $\sigma$ is a pure
orbifold operation \eqn\sigorb{\sigma: ~ X~\to ~ - X, ~~~~ Y~\to ~
- Y.} Now looking at \sfourmet\ we see that the metric does not
possess this symmetry. In fact, it is broken by the components
$j_{xz}$ and $j_{yz}$. Furthermore, the existence of these
components is already problematic if we want to go to the type I
theory. After two T--dualities these components will transform
into $B_{NS}$ fields and we know that the type I theory should
{\it not} have $B_{NS}$ fields!

A way out of this problem is to untwist the $dz$ fibration. This
can be achieved by going to the double cover, where we write the
metric along the $z$ direction simply as $dz^2$. Recall that the
twist of the $dz$ direction originates from having {\it two}
Taub--NUT spaces. A T--dual version gives us two NS5 branes
related to this \uranga, \dmconi. Alternatively the two Taub--NUT
spaces become two intersecting $D6$ branes in M--theory \tatar.

To obtain a metric with the required isometries that is also
invariant under \sigorb, we first remove the wrapped $D5$ branes.
This can be achieved by setting the harmonic function $h = 1$. The
metric of the resolved conifold after removing the wrapped $D5$
branes becomes \eqn\metresco{ds^2 = ds^2_{xy} ~+~ ds^2_Z ~+~
ds^2_{S_3},} with the notation used in earlier sections and in
\bdkt. We will keep the metric along the $X,Y$ and $Z$ directions
as \eqn\metxy{\eqalign{& ds^2_{xy} = (1 + A^2) ~dx^2 ~ + ~ (1 +
B^2)~ dy^2 ~+ ~ 2 AB~dx~dy, \cr & ds^2_Z = dz^2 ~ +~ 2A~dx~dz ~+~
2B~dy~dz,}} where $A, B$ are defined as before with the
restriction $h =1$. From the above metric we see that the
problematic components are $dx ~dz$ and $dy ~dz$ which are not
invariant under \sigorb. The metric along the $S_3$ direction can
be written as \eqn\metsthree{ds^2_{S^3} = ~ \gamma' dr^2 + {\gamma
\o 4}~d\theta_1^2  + {\gamma + 4a^2 \o 4}~d\theta_2^2} which is
the same as before except that we have set $h =1$.

As discussed, the above metric (which is simply the resolved
conifold metric without wrapped five branes) cannot suffice. We
are looking for a metric in type IIB that has the following
properties:

\noindent (a) It should be invariant under \sigorb.

\noindent (b) It should preserve some number of supersymmetries
(i.e. ${\cal N} = 1$).

\noindent (c) It should be close to the original type IIB metric
in terms of its form.

\noindent (d) It should allow wrapped $D5$ branes along with some
number of $D7$ branes and $O7$ planes, and

\noindent (e) After two T--dualities, it should look close to the
one obtained from T--dualizing the resolved conifold.

\noindent Let us start with (a) and (c). A generic metric that is
invariant under \sigorb\ can be written in terms of two complex
tori with complex coordinates $\chi_1$ and $\chi_2$ in the
following way: \eqn\comtor{ ds^2 = {d}_1~\vert d\chi_1 \vert^2 +
{d}_2 ~\vert d\chi_2 \vert^2 + {d}_3 ~dz^2 + {d}_4~ dr^2} where
${d}_i$ are some numerical coefficients that could in general be
functions of the radial and angular coordinates, i.e.
$$d_i \equiv d_i(\theta_1, \theta_2, r)$$
\noindent which  is motivated from the fact that the metric should
have the required isometries along $X, Y$ and $Z$ directions. The
complex coordinates $\chi_i$ above are defined in the following
way: \eqn\chicom{d\chi_1 = dx + \tau_1~ dy, ~~~~~~~~~~ d\chi_2 =
d\theta_1 + \tau_2 ~ d\theta_2} where $\tau_i$ are the complex
structures of the given tori.

Expanding the above metric and keeping ${\rm Re}~\tau_2 = 0$, we
see that the components can be written as \eqn\metexp{ds^2 =
d_1(dx^2 + \vert \tau_1 \vert^2 ~dy^2 + 2 {\rm Re}~\tau_1~dx~dy) +
d_2 (d\theta_1^2 + \vert \tau_2\vert^2~d\theta_2^2) + d_3~dz^2 +
d_4~dr^2.} We make the following observations regarding various
issues here:

\noindent $\bullet$ The non--trivial coefficients $d_i$ cannot be
arbitrary as the fourfold (which is a $T^2$ fibration) over the
base \comtor\ has to be a Ricci--flat K\"ahler manifold i.e. a CY
fourfold. In the orientifold limit, the fourfold is $T^4/{\cal
I}_4 \times S_4$ where $T^4/{\cal I}_4$ is the orbifold limit of
$K3$\foot{In fact it is a $T^2$ fibration over $T^2/{\cal I}_2$
base. As the reader can easily verify, this $T^2/{\cal I}_2$ is
precisely a base with four fixed points \senFone, \dmconstant.}
and $S_4$ is a four--cycle parameterized by $\theta_{1},\theta_2,
r, z$.

\noindent $\bullet$ There are three regions of interest here: (1)~
When we are {\it at} the orientifold point. The predicted metric
at that point is the above metric \comtor. We expect the
background axion--dilaton ($\tilde\phi, \phi$) to be exactly
cancelled in this background. (2)~ When we perturb one of the $D7$
branes away from the $O7$ plane. This creates a non--trivial
axion--dilaton in the background. The metric is then calculated
from the Seiberg--Witten curve. (3)~When we take the
axion--dilaton far away so that there would be a local region
where the expectation values of the axion--dilaton would be zero.
In this paper we will only discuss (1) and (3), and leave (2) for
future work.

\noindent $\bullet$ The full non--perturbative correction to this
model can be worked out using the analysis of \senFone, \banks.
This non--perturbative correction arises when we move the $D7$
branes away from the orientifold seven planes. This is very well
known, so we will not repeat the discussion. In the end, this
would convert the $T^2/{\cal I}_2$ parameterized by $X, Y$ to a
$P^1$ with 24 singularities. These 24 singularities are the 24
zeroes of the discriminant of the Weierstrass equation \weierst.

Let us concentrate on the region when we have moved the seven
branes far away. The generic metric along the $X,Y$ direction will
be a $P^1$ with fixed points. If we call the $X,Y$ plane the
$u$-plane\foot{This is the same $u$-plane as that of
Seiberg--Witten theory \sw.}, then the metric of the $P^1$ will be
\eqn\metpo{ds^2 = C~ \sum_i {du d\bar u \o ~~\vert u - u_i
\vert^{1\o 2}}} where $u \equiv X + iY$ and $u_i$ are the points
where the seven branes are located and $C$ is an appropriate
constant. Consider now the case when we keep all the seven branes
far away. In terms of the metric \metpo\ this will imply:
\eqn\bgimpl{ds^2 = C~ {}^{\rm lim}_{u_i \to \infty}~\left(\sum_i
{du d\bar u \o ~~\vert u - u_i \vert^{1\o 2}}\right) ~ \to ~
\sum_i {du d\bar u \o ~~\vert u_i \vert^{1\o 2}}, ~~~~~ \langle
\tilde\phi \rangle \approx 0, ~~~ \langle \phi \rangle \approx 0.}
This means there is a region in our space where we could see the
two cycle of the resolved conifold on which we wrap $D5$ branes.
Since in this limit the metric along $X, Y$ directions is
approximately flat, the two cycle is given by ($y, \theta_2$)
which is precisely the two cycle that we encountered in the
resolved conifold setting. The metric of this region cannot be a
Calabi--Yau due to back reactions from the seven branes. These
back reactions can be calculated using an analysis similar to the
one presented in \grana, \afm.

So far we discussed the situation when we put the seven branes far
away. Let us come back to the case when we are at the orientifold
point. The generic analysis using non trivial values for the $d_i$
in \comtor\ is very involved. So, as a first approximation we will
take all the $d_i$ to be constants. These constants are not
arbitrary, and their choice will be motivated by the approximation
that we used in \bgimpl, namely deforming the orientifold picture
by moving the seven branes, so that the metric comes close to
resembling a resolved conifold.

As we discussed in \bgimpl, we can identify the two cycles of the
original resolved conifold. This is what we meant earlier by being
{\it close} to the resolved conifold.

To identify various values here, we will use the approximation
that we took in \bdkt, namely, we set the radial coordinate to a
fixed value. Thus, $\gamma$ etc. will be defined w.r.t. that
value. We would like to remind the reader that this is just to
simplify the ensuing analysis. A more detailed analysis can indeed
be performed, but we will not do so here. Let us now define the
generic complex structure of the two tori as: \eqn\gencosto{\tau_1
~ = ~ {1\o 1 + A_1^2}\left[ c A_1 B_1 +
i~\sqrt{{\langle\alpha\rangle}^{-1} + (1-c^2)A_1^2 B_1^2}\right],
~~~~~ \tau_2 = i \sqrt{1 + {4a^2 \o \gamma}}} where $A_1, B_1$ are
measured w.r.t. some average values of $\theta_1$ and $\theta_2$,
respectively, as we had mentioned earlier in section 2 and also in
\bdkt. We have also introduced an integer $c$ that can take values
1 or 0. The precise value of $c$ will be determined by the choice
of fluxes. We will soon evaluate this using the type IIB
superpotential \potgvw. For the choice of $c = 1, 0$, the complex
structure of the two tori become \eqn\comtwto{\eqalign{& c =
1:~~~~~ \tau_1 ~ = ~ {1 \o 1 + A_1^2} \left[A_1 B_1 + {i \o
\sqrt{\langle \alpha \rangle}}\right], ~~~~~~~ \tau_2 = i \sqrt{1
+ {4a^2 \o \gamma}}\cr & c = 0:~~~~~ \tau_1 ~ = ~i
{\sqrt{{\langle\alpha\rangle}^{-1} + A_1^2 B_1^2}\o 1 + A_1^2},
~~~~~~~~~~~~~~~~~ \tau_2 = i \sqrt{1 + {4a^2 \o \gamma}}.}}
Furthermore, if we now allow the following values for the
coefficients $d_i$ in \comtor\ \eqn\valai{d_1 ~=~ 1 + A_1^2,
~~~~~~ d_2 ~=~ {\gamma \o 4}, ~~~~~~ d_3 ~=~ 1, ~~~~~~ d_4 =
\gamma'} we see that the metric \comtor\ comes very close to
\metresco. In the limit \repmen\ they will resemble exactly, but
now without the problematic cross terms! In fact with the generic
complex structure \gencosto\ the metric along the $xy$ directions
take the form: \eqn\genmetn{d_1~\vert d\chi_1\vert^2 ~=~
(1+A_1^2)~dx^2 + (1+B_1^2)~dy^2 + 2 c A_1 B_1~dx~dy.} By
construction, the metric preserves supersymmetry because it is
written in terms of tori with non--trivial complex structure. To
show that this metric preserves the precise amount of
supersymmetry as that of the original resolved conifold, we have
to project our space with $Z_2^m$ orbifold action, $m$ being some
specific integer. Unfortunately, these $Z_2$ actions are {\it not}
visible in the metric, so one cannot infer them from the above
analysis.

We have already taken care of the points (a), (b) and (c). Now let
us deal with point (d). We require our metric \comtor\ to allow
wrapped $D5$ branes and also some number of $D7$ branes and $O7$
planes. A way to achieve this would be to lift the background to
F--theory or M--theory. In M--theory the back--reaction of branes
and fluxes can be easily evaluated along the lines proposed in
\dotd. The fluxes are of course the $B_{NS}$ fields and the
$B_{RR}$ fields that form the sources of the wrapped $D5$ branes.
The final answer is gained by simply replacing the initial
harmonic function $h$ with a different harmonic function $H$ that,
from type IIB point of view, satisfies \eqn\warpyh{\quabla~H =
H_{NS} \wedge H_{RR} + {\pi^2 \alpha'^2 \o 4} \sum_{i = 1}^{24}
~b_i ~{\rm tr}(R_i \wedge R_i) ~\delta^2 (z - z_i)} where $i$ sums
the number of branes and planes, each contributing a factor of
$b_i$ amount of ${\rm tr}(R_i \wedge R_i)$. These $b_i$'s have
been worked out in \djm, \moral, \stef. 
The positions of these branes and planes
on the $X, Y$ spaces are denoted by complex coordinates $z_i$.
Wrapped $D5$ branes, and $D3$ branes will have additional
contributions to the warp factor. These details appeared in \drs.

The direction $dr$ is non--compact and therefore the metric allows
many wrapped $D5$ branes with no restrictions from the anomaly
condition \svw. The $2d$ surface with metric $ds^2_{xy}$ will have
points where the F--theory fiber degenerates, creating $D7$ branes
and ($p,q$) seven branes. In special limits, these seven branes
behave like {\it non--dynamical} $O7$ planes. Since $O$--planes
are perturbative, these limits are at small type IIB coupling
\senFone, \dmconstant. The fact that this configuration of $D5$
branes and seven branes preserves supersymmetry can be easily seen
from the number of intersection points. In the figure below:
\vskip.1in

\centerline{\epsfbox{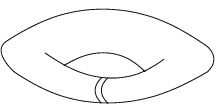}}\nobreak

\vskip.1in \noindent the $D5$ branes wrap a torus along
($\theta_2, y$) and extend along $x^{0, 1, 2, 3}$. The seven
branes are stretched along $\theta_2$ (given by the strip in the
above figure) plus ($\theta_1, r, z$) directions inside the
manifold along with the usual $x^{0, 1, 2, 3}$ directions. Such a
configuration preserves supersymmetry. This way we take care of
point (d).

To further clarify the issue, we compare our results with the ones
of \grana,~where D7 branes together with fractional D5 branes were
considered. They dealt with an ${\cal N} = 2$ theory as a $Z_2$
orbifold of the D3/D7 solution. The D3 branes were orthogonal to
$R^4/Z_2~\times~R^2$ and the D7 branes were wrapped on the
$R^4/Z_2$ and a point in $R^2$. At the orbifold point the
singularity can be resolved by blowing up a $P^1$ cycle. If the
$P^1$ cycle is of vanishing size and with a finite NS flux through
it, then the D5 branes which are wrapped on $P^1$ are fractional
D3 branes. A T--dual on the angular direction of the vanishing
$P^1$ cycle gives a type IIA brane configuration with NS branes
and D4 branes \karchlust.~The  ${\cal N} = 2$ configuration of
\grana\ can be rotated to one with  ${\cal N} = 1$ supersymmetry,
where the $R^4/Z_2$ is replaced by a conifold singularity. In this
case the fractional D3 branes are D5 branes wrapped on a vanishing
$P^1$ cycle at the conifold point. The T--dual is taken with
respect to an $S^1$ in the base of the conifold and the T--dual
picture is an elliptic model with circular D4 brane
\uranga\dmconi.~ The worldvolume of the D7 branes contains the
vanishing cycle.

It would then appear that the D7 branes and D5 branes have two
common directions if the vanishing cycle is made finite. This
would raise serious doubts as the configuration would be
non--supersymmetric. But this is not an allowed process, the
reason being discussed at length in \dotu.~ We used the fact that
in the conifold case, the distance between the NS branes in the
T--dual picture is given by the flux of NS field through the
vanishing 2--cycle. For the resolved conifold the distance between
the NS branes in the T--dual picture is given by the size of the
$P^1$. As these are two independent quantities, we cannot identify
them. There is no continuous way to take the D7 on the conifold
with fluxes to the D7 branes on the resolved conifold, as they
have at least one different radial direction. Therefore we can
have a D5--D7 intersection on one internal direction since the
resolved conifold metric will allow this
configuration\foot{We would like to
thank Radu Roiban and Bogdan Florea for discussions on this and
related issues.}. It would be very interesting to discuss in
detail the supergravity solution by considering the back--reaction
of the D7 branes, in the spirit of \grana~(see also
\pandovaman~for recent developments in this direction).

Before we discuss case (e), we should raise the important question
whether the metric that we derived at the orientifold point
\comtor\ or \metexp\ shows any geometric transition or not. Can we
shrink the two cycle on which we have wrapped $D5$ branes and go
to a background that has only fluxes with possible $D7$ branes and
$O7$ planes? In other words, away from the orientifold point the
type IIB metric with wrapped $D5$ branes on a resolved conifold
shows geometric transition. What happens {\it at} the orientifold
point? Since, at the orientifold point we have removed the cross
terms in the metric, it is a--priori not clear whether one can see
the geometric transition or not. This might look a little
disappointing, as we seem to lose the main reason of going to the
type I scenario. One way out of this would be not to go to the
orientifold point at all. But away from orientifold point we do
not have any ways to go to type I!

This dilemma is resolved nicely by examining the metric in type I
after two T--dualities on the metric \comtor. We find an
interesting surprise here. Imagine we perform two T--dualities on
the metric away from the orientifold point. Of course the final
metric will remain in type IIB, but the two metrics\foot{One by
T--dualizing \metresco\ with $D5, D7$ and ($p,q$) seven branes
inserted in, and the other by T--dualizing \metexp\ with $D5, D7$
branes and $O7$ planes inserted in.} that we get by doing this
look exactly the same up to a possible warp factor along the $dz$
direction! Thus, in this sense, the type I background that we will
find after two T-dualities, will be {\it dual} to the type IIB
background before geometric transition.

To confirm this claim, let us explicitly T--dualize the metric
\comtor. Before we can do this we first require the following two
steps:

\noindent $\bullet$ Insertion of the wrapped $D5$ branes along with
the $D7$ and $O7$ planes, and

\noindent $\bullet$ Specify the background $H_{NS}$ field at the
orientifold point.

\noindent The first step is easy. We write the metric \comtor\ but
now with the harmonic function $H$ put in. This amounts to simply
replacing $h$ in the original definition of $A_1, B_1$ by $H$,
without any other changes.

\noindent For the second step, i.e. determining the value of the
background $H_{NS}$, we recall that the only $H_{NS}$ field that
is invariant under the orientifold projections should have one of
its components\foot{Recall that away from the orientifold point we
can always excite other components of the $B$ field that do not
have any legs along the duality directions. At the orientifold
point these components are projected out.} along the duality
directions $X$ or $Y$. Although the original choice of $B_{NS}$
field \bfico\ satisfies this condition, we will adopt a more
generic ansatz. This is because the original choice \bfico\ is
suited for the specific metric that we had earlier. Since we are
changing the metric by choosing all $d_i$ in \metexp\ constant, we
require a more general ansatz for the $B$ field.
 Let us therefore take
\eqn\bnsf{B_{NS}=b_{xi}~dx~ \wedge ~ d\zeta^i ~+~ b_{yj}~ dy ~
\wedge ~ d\zeta^j} where $\zeta^i = (\theta_1, \theta_2, z, r)$,
generically. Since the components of $B$ field have one leg along
the duality directions, we see that this would survive the
orientifold action. We will also take both $b_{xi}$ and $b_{yj}$
to be functions of ($\theta_1, \theta_2, r$)\foot{The reason why
the components are independent of functions along the $Z$
direction is because of mirror symmetry.}.

The analysis including all the components along the ($\theta_1,
\theta_2, z, r$) directions is pretty involved, so we will
simplify the analysis by first taking \eqn\onechoice{b_{xi} \equiv
b_{x\theta_1}, ~~~~ b_{yj} \equiv b_{y\theta_2}, ~~~~ c = 1.}
These simplifying choices do not change any of the dualities that
we will perform, as one can easily put back the other components
to get the complete result. On the other hand these choices will
be useful to see the fibration structure in the type I scenario in
a simple way. The final metric after two T--dualities is similar
to the one worked out in \drs, \kat, and is given by:
\eqn\metinhet{\eqalign{ds^2 = & {1\over g_{yy}} (dy + {B}_{\mu y}
dx^\mu + {B}_{xy} dx + {B}_{zy} dz)^2 -{1\over g_{yy}} (g_{y\mu}
dx^\mu + g_{yx} dx + g_{yz} dz )^2 \cr & + g_{\mu\nu} dx^\mu
dx^\nu + 2 g_{x \nu} dx dx^\nu + 2 g_{z \nu} dz dx^\nu + 2 g_{xz}
dx dz + g_{xx} dx^2 + g_{zz} dz^2,}} where $g_{mn}$ is the metric
after one T--duality along $X$ direction: \eqn\gmn{g_{mn}~ =~
\pmatrix{{1\o 1+A_1^2} &0 & 0 & -{b_{x\theta_1}\o 1+A_1^2} & 0\cr
\noalign{\vskip -0.20 cm} \cr
 0 & {1 \o \langle\alpha\rangle(1+A_1^2)} & 0 & 0
 & 0 \cr
\noalign{\vskip -0.20 cm}  \cr 0 & 0  &1 & 0  & 0 \cr
\noalign{\vskip -0.20 cm}  \cr -{b_{x\theta_1}\o 1+A_1^2} &  0& 0
& {\gamma \o 4} + {b^2_{x\theta_1}\o 1+A_1^2} & 0 \cr
\noalign{\vskip -0.20 cm} \cr
 0& 0 & 0 & 0 & {\gamma+4a^2}\o{4}}.}
Similarly, the $B_{NS}$ field after one T--duality appearing in
\metinhet\ is given by: \eqn\bmn{B_{mn} ~ =~ \pmatrix{0 & -{A_1
B_1\o 1+A_1^2} & 0 & 0 & 0\cr \noalign{\vskip -0.20 cm}  \cr
 {A_1 B_1 \o 1+A_1^2} & 0 & 0 & -{b_{x\theta_1} A_1 B_1 \o 1+A_1^2}
 & b_{y\theta_2} \cr
\noalign{\vskip -0.20 cm}  \cr 0 & 0  & 0 & 0  & 0 \cr
\noalign{\vskip -0.20 cm}  \cr 0 &  {b_{x\theta_1} A_1 B_1 \o
1+A_1^2}& 0 & 0  & 0 \cr \noalign{\vskip -0.20 cm}  \cr
 0& -b_{y\theta_2}  & 0
 & 0 & 0}}
where the components of the matrices for the metric and the $B$
field follow the same order as used so far. The final metric in
type I is now found to be: \eqn\metinI{\eqalign{ds^2 ~ = ~&
\langle\alpha\rangle(1 +A_1^2)~ (dy - b_{y\theta_2}~d\theta_2)^2 +
\langle\alpha\rangle (1 + B_1^2)~ (dx - b_{x\theta_1}~d\theta_1)^2
+ \gamma'\sqrt{H} dr^2\cr ~&~~~ - 2 \langle\alpha\rangle A_1
B_1~(dx - b_{x\theta_1}~d\theta_1)(dy - b_{y\theta_2}~d\theta_2) +
dz^2 + d_2 \vert d\chi_2\vert^2,}} with vanishing $B_{NS}$. The
above metric follows exactly the criteria laid down in \bdkt,
namely, whenever we see $dx$ and $dy$ we have to replace them with
the $b$ dependent fibration $(dx - b_{x\theta_1}~d\theta_1)$ and
$(dy - b_{y\theta_2}~d\theta_2)$, respectively. The only other
change is the warp factors in front of the metric components. By
construction this metric can be easily shown to be non--K\"ahler.
A similar fibration structure, leading to a  non--K\"ahler metric,
was proposed earlier in a series of papers \drs, \kat, \bbdg,
\bbdgs (see also \kachru).

Let us now show that this metric is indeed dual to some type IIB
metric. We will explicitly T--dualize the type IIB metric {\it
away} from the orientifold point, i.e. our starting metric will be
\metresco. Here we use the approximation \bgimpl, i.e. the seven
branes are far away and therefore the base metric can be
approximated by a resolved conifold. The $B_{NS}$ field remains
the same with components $b_{x\theta_1}$ and $b_{y\theta_2}$. As
we have seen, these components not only survive the orientifold
projections, but are also solution to the equation of motion.
After two T--dualities the metric will be similar to \metinhet,
but we will remain in type IIB theory. The values of $g_{mn}$
appearing in \metinhet\ are: \eqn\gmnnow{g_{mn}~ =~ \pmatrix{{1\o
1+A^2} &0 & 0 & -{b_{x\theta_1}\o 1+A^2} & 0\cr \noalign{\vskip
-0.20 cm} \cr
 0 & {1 \o \alpha (1+A^2)} & {B \o 1 + A^2} & 0
 & 0 \cr
\noalign{\vskip -0.20 cm}  \cr 0 & {B \o 1+A^2}  & {1 \o 1+A^2} &
0  & 0 \cr \noalign{\vskip -0.20 cm}  \cr -{b_{x\theta_1}\o 1+A^2}
&  0& 0 & {\gamma \o 4} + {b^2_{x\theta_1}\o 1+A^2} & 0 \cr
\noalign{\vskip -0.20 cm}  \cr
 0& 0 & 0
 & 0 & {\gamma+4a^2 \o 4}}}
Comparing with \gmn\ we see that there are some specific changes.
First, the $dz^2$ component is now no longer 1 but is more
complicated. We have also developed a new component along the $dy
dz$ direction, that was absent in \gmn. Furthermore, all the other
components have remained the same, but are not measured w.r.t. the
expectation values any more, they will coincide with the
approximation \repmen. On the other hand, the $B_{NS}$ field turns
out to be: \eqn\bmnnow{B_{mn} ~ =~ \pmatrix{0 & -{A B\o 1+A^2} &
-{A \o 1 +A^2} & 0 & 0\cr \noalign{\vskip -0.20 cm}  \cr
 {A B \o 1+A^2} & 0 & 0 & -{b_{x\theta_1} A B \o 1+A^2}
 & b_{y\theta_2} \cr
\noalign{\vskip -0.20 cm}  \cr {A\o 1 + A^2} & 0  & 0 &
-{b_{x\theta_1} A \o 1+A^2}  & 0 \cr \noalign{\vskip -0.20 cm}
\cr 0 &  {b_{x\theta_1} A B \o 1+A^2}& {b_{x\theta_1}A \o 1 + A^2}
& 0  & 0 \cr \noalign{\vskip -0.20 cm}  \cr
 0& -b_{y\theta_2}  & 0
 & 0 & 0}}
Again, comparing with \bmn, we see that we have developed new
components along $dz \wedge d\theta_1$ and $dx \wedge dz$
directions. These new components are related to the cross terms in
the type IIB metric away from the orientifold point. The final
metric in type IIB after two T--dualities therefore will take the
following form: \eqn\metinIIB{\eqalign{ds^2 ~ = ~& \alpha(1 +
A^2)~ (dy - b_{y\theta_2}~d\theta_2)^2 + \alpha (1 + B_1^2)~ (dx -
b_{x\theta_1}~d\theta_1)^2 + \gamma'\sqrt{H} dr^2 \cr ~& - 2~
\alpha A B ~(dx - b_{x\theta_1}~d\theta_1)(dy -
b_{y\theta_2}~d\theta_2) + \alpha dz^2 + d_2 \vert
d\chi_2\vert^2}} with the corresponding $B_{NS}$ given by
\eqn\bnsIIB{{B}^{(2)}_{NS} = -\alpha A(dx-b_{x\theta_1}
d\theta_1)\wedge dz -\alpha B(dy-b_{y\theta_2} d\theta_2)\wedge
dz.} Compare this metric with the one that we got for type I
theory \metinI. We see that if we impose the limit \repmen, these
two metrics are exactly the same up to one possible warp factor
for the $dz^2$ directions. For the metric \metinI, the $dz^2$ term
comes with coefficient 1, whereas in \metinIIB\ the $dz^2$ term
comes with coefficient $\langle\alpha\rangle$ when we use \repmen.
This if we redefine the $z$ direction in \metinI\ as \eqn\defz{ z
~=~ \sqrt{\langle\alpha\rangle}~z'} then the two metrics will
coincide exactly. The metric in type IIB \metinIIB\ shows
geometric transition because it is T--dual to the original type
IIB metric that showed geometric transition. Using this analogy,
we see that the type I metric will form the precise dual to the
type IIB metric and therefore this will provide the geometric
transition dual to type IIB in the type I theory.

In terms of our previous discussion, after the two T--dualities
the three regions with D7 and O7 overlap such that it is natural
to obtain an identical metric after taking 2 T--dualities for each
of them. In type I there is no distinction between regions which
are close or far away from the orientifold plane as the
orientifold plane is an O9 plane and there is no extra direction
to separate from it.

At this point we have to study what this background represent in
the type I theory. Since the background originates directly from
type IIB via T--dualities, it should be related to wrapped $D5$
branes on this non--K\"ahler background. In fact, as we will
discuss soon, under an S--duality this will be transported to
heterotic theory on a non--K\"ahler background with NS5 branes.
The NS5 branes are the sources of torsion in this scenario. This
fits exactly with the picture developed in \gauntlett. We will
discuss more on this connection in section 5. But first we need
the other fields, i.e. the RR three form and the dilaton for the
type I background.

To get the $H_{RR}$ for the type I we cannot simply T--dualize
$H_{RR}$ of type IIB \hrrbg, because $H_{RR}$ of type IIB is in
fact not invariant under the orientifold action. The problematic
components are: \eqn\hrrprob{ c_2~{\rm cot}~\hat\theta_1~dx \wedge
d\theta_2 \wedge dy - c_3~{\rm cot}~\hat\theta_2 ~dy \wedge
d\theta_1 \wedge dx} that pick up minus signs under the
orientifold action. The coefficient $c_2, c_3$ are defined in
\bdkt, \pandoz. The only components that could remain invariant
under the orientifold action will be the ones that have one leg
along $X$ or $Y$ directions. Let us therefore use the following
ansatz for the RR field in type IIB theory {\it at} the
orientifold point: \eqn\ansbrr{\eqalign{ H^b_{RR}& ~ = ~ {\cal
H}^b_{ykl}~dy~\wedge~d\zeta^k~\wedge~d\zeta^l~ +~ {\cal
H}^b_{xij}~dx~\wedge~d\zeta^i~\wedge~d\zeta^j \cr & ~=~ {\cal
H}^b_{yz\theta_2}~ dy ~\wedge ~dz ~\wedge ~d\theta_2 + {\cal
H}^b_{xz\theta_1}~ dx ~\wedge ~dz ~\wedge ~d\theta_1 ~ + \cr &
~~~~~~~~~~ + ~ {\cal H}^b_{xzr}~ dx ~\wedge ~ dz ~\wedge~ dr~ +~
{\cal H}^b_{yzr}~dy ~\wedge ~dz ~\wedge ~dr}} along with the
$B_{NS}$ field given by \bnsf\foot{Observe that we have chosen
only few components of the three-form field. This is for
simplicity. We can always choose all the twelve components. None
of the analysis will be affected by this choice.}. The
axion--dilaton is zero because at the orientifold point the seven
brane charges are cancelled exactly. Along with the metric
\comtor, \comtwto\ and \valai, the above set of background will
provide the complete solution of the type IIB system at the
orientifold point.

Before we move ahead, we need to see whether the choices of
$H^b_{RR}$ and $H_{NS}$ are compatible with the metric that we
have at the orientifold point \comtor. A careful look at the
metric tells us that the metric is actually a toroidal orbifold
with non--trivial complex structures. This situation is now
similar to the one that was seen earlier in \kat, namely, a
toroidal orbifold compactification of type IIB theory with fluxes.
As in \kat, the fluxes will fix the complex structures of the type
IIB manifold \comtor\ at the orientifold point. One choice of
$H_{NS}$ and $H^b_{RR}$ fluxes that would do the job are constant
fluxes (this doesn't mean that the $B_{NS}$ and $B_{RR}$ are
constants). For such fluxes, there is a constraint coming from the
choice of the components: \eqn\constrain{
B^{NS}_{[xm}~B^{RR}_{ny]} = 0} which was shown in \kat, \bbdgs.
This constraint has implications when we determine the type I RR
field.

Performing two T--dualities on this background will determine the
type I theory for us. We have already used this to get the metric.
For the generic choice of $B_{NS}$ field, the fibration structure
of the type I metric will simply change to \eqn\fibtye{ (dx -
b_{x\theta_1}~d\theta_1)~ \to ~(dx - b_{xi}~d\zeta^i), ~~~~~ (dy -
b_{y\theta_2}~d\theta_2) ~ \to ~ (dy - b_{yj}~d\zeta^j).} The
question now is whether there are other changes in the metric when
we choose constant background fluxes. It turns out when the
background fluxes are arbitrary and not constants, the value of
$c$ is typically 1. For constant background fluxes, as the one
chosen in \kat, \drs\ the value of $c$ turns out to be
\eqn\cvalue{ c ~ = ~ 0.}
This can be easily shown by minimising the type IIB superpotential 
\potgvw. It turns out that 
\fibtye\ and \cvalue\ are only changes that we have to do 
to the metric. 

Let us now evaluate the RR fields of type I.   
Something interesting happens now. When
we perform two T-dualities on the type IIB background, the
components of type I RR field that are {\it orthogonal} to the
duality directions $X, Y$ take the following form\foot{See \kat\
for a derivation of this.}: \eqn\brrnowis{B^I_{mn} ~ =
~a_0~D^+_{xymn} ~+~ a_1 ~B^{NS}_{[xm}~B^{RR}_{ny]} ~+~
a_2~B^{NS}_{x[m}~B^{RR}_{n]x} ~+~ a_3~B^{NS}_{y[m}~B^{NS}_{n]y}}
with $D^+$ being the four--form that has both legs along the
duality directions. This vanishes trivially because in the
presence of NS and RR three--forms we do not require the five form
to be self--dual. The $a_i$ now take the following values:
\eqn\bgai{a_0 ~=~ 1, ~~~~ a_1 ~ = ~ 6, ~~~~~ a_2 ~ = ~ {c A_1 B_1 \o 1 +
A_1^2}, ~~~~~ a_3 ~ = ~ - 2 \tilde\phi} where $\tilde\phi$ is the
type IIB axion background. Imposing the constraints \constrain\
and \cvalue\ we see that the type I three--form fields that do not
have any legs\foot{The type I $B$ field that has both legs along
the duality directions also vanish because for this $a_0 = a_1 =
a_2 = 0, ~~ a_3 = -\tilde\phi$.}
 along the
duality directions vanish! Therefore
performing two T--dualities on
\ansbrr\ (and the coupling) gives the following type I RR field (and
coupling):
\eqn\brri{\eqalign{& H^I_{RR}~ =
~ {\cal H}^b_{xz\theta_1}~ dy ~\wedge ~dz ~\wedge ~d\theta_2 -
{\cal H}^b_{yz\theta_2}~ dx ~\wedge ~dz ~\wedge ~d\theta_1 ~ + \cr
& ~~~~~~~~~~ + ~{\cal H}^b_{xzr}~ dy ~\wedge ~ dz ~\wedge~ dr~ -~
{\cal H}^b_{yzr}~dx ~\wedge ~dz ~\wedge ~dr \cr & ~~~ g^I ~ = ~
\sqrt{\langle \alpha \rangle},}} which along with the metric
\metinI\ will determine the type I dual background before
geometric transition in type IIB theory.

There are several issues that need to be mentioned at this point.
First, observe that the dilaton in type I theory no longer
vanishes, even though the corresponding type IIB theory has a
vanishing dilaton--axion. Second, we still have to determine the
background values of $H^b_{RR}$. We shall determine them below and
compare the result using torsional constraints \HULL, \rstrom\
later when we go to the heterotic theory. Third, in determining
the $H^I_{RR}$ we see the $dx, dy$ terms {\it do not} have the
expected fibration structures that the type I metric possesses.
This phenomenon was also observed in \kat, \drs\ where the metric
showed the fibration structure but the $B$ field did not. The
reason for this is simple. Any $z\theta_i$--cross terms in the $B$
field would in fact come from a term that is proportional to
\constrain\ as we saw in \brrnowis\ (see also \kat\ for a
derivation of this). Since this term is identically zero, the
$B_{RR}$ lacks any fibration structure. This reasoning would
remain true for completely generic $B_{NS}$ and
$B_{RR}$\foot{Note, that $B_{NS}$ and $B_{RR}$ are still
constrained by the orientifold action, so the most generic case
would only allow $xz$, $x\theta_1$, $x\theta_2$ and $xr$
components (plus $x \leftrightarrow y$)}: \constrain\ would serve
to eliminate all components that do not have one leg along $X$ or
$Y$, ergo to forbid any fibration structure. Fourth, the
background that we determined for type I is in fact incomplete,
because we have not specified the gauge bundle yet. The gauge
bundle has to satisfy the usual Donaldson--Uhlenbeck--Yau (DUY)
equation \eqn\duy{g^{a\bar b} F^i_{a\bar b} ~ = ~ 0} where $g$ is
the type I metric \metinI\ written in terms of complex
coordinates, and $i$ counts the number of D9 branes along with the
Wilson lines on it; along with another constraint on ${\rm tr}~F^i
\wedge F^i$: \eqn\trff{{\rm tr}~F^i \wedge F^i ~ = ~ {\rm tr}~R^i
\wedge R^i - i \del\bar\del J} which was first discussed in some
details in \bbdgs. The solutions to these two equations are rather
non--trivial, so we will not address them here.
 Furthermore, the fact that we can write the
DUY equations in terms of $a, \bar b$ means that the complex
structure in type I theory is {\it integrable} even though the
manifold is non--K\"ahler. We already encountered such kind of
manifolds in \drs, \kat, \bbdg, \bbdgs. The difference is that
those manifolds were compact. Here we find new complex
non--K\"ahler manifolds that are non--compact. On the other hand,
the non--K\"ahler manifolds encountered in \bdkt\ for the type IIA
theory were non--complex and non--K\"ahler. It is interesting to
see that as we move down the duality chain we go from
non--complex, non--K\"ahler in type IIA to complex K\"ahler in
type IIB and then to complex non--K\"ahler in type I (and also in
heterotic)!

To determine the background $B^b_{RR}$ we can use the type IIB
supergravity equation of motion. We already know the background
$B^b_{NS}$ and the metric along with the axion--dilaton.
Therefore, the background supergravity equations are much simpler
here. In fact, they become even simpler if we use the
supersymmetry variation equations. The second order differential
equations for the $B^b_{RR}$ become first order equations:
\eqn\first{H^b_{RR} ~\equiv ~ {\cal H}^b ~ = ~ \ast ~H^b_{NS}} as
conjectured first in \kleb, and later shown to occur from the
primitivity equation\foot{The primitivity equation basically tells
us what choices of fluxes on a given fourfold preserve
supersymmetry \bb.} in M--theory \dotd. Now using the $B^b_{NS}$
field and \first, we can determine the non--zero components of
$H_{RR}$ as\foot{Again, the generic choice will involve twelve
equations. We will however continue using these components only.}:
\eqn\hrrfrom{\eqalign{& {\cal H}^b_{ryz} =
\alpha_1~\del_{[\theta_2}b_{x\theta_1]}, ~~~~~~~~~~~ {\cal
H}^b_{\theta_2 yz} = \alpha_2 ~\del_{[r} b_{x\theta_1]} \cr &
{\cal H}^b_{\theta_1 xz} = \alpha_3~\del_{[r} b_{y\theta_2]},
~~~~~~~~~~~ {\cal H}^b_{rxz} = \alpha_4~\del_{[\theta_1}
b_{y\theta_2]}}} where the coefficients $\alpha_i$ are determined
from the Hodge star and the epsilon tensor that appears in \first.



\subsec{{$\underline{\rm \bf {Background~ after ~ geometric~
transition}}$}}


Having given the background before geometric transition, we now
determine the corresponding metric after geometric transition in
type I theory. To do so we have to go through the details that we
discussed for the case before geometric transition again. Some
preliminary analysis has been done earlier in \gkp\ and \dotd\
where fourfolds with bases having a conifold singularity were
discussed (see also \giveon). In \gkp\ the base of the fourfold
was shown to be a hypersurface given by a quartic equation in
$P^4$: \eqn\gkpepn{z_5^2\left(\sum_{i =1}^4 z_i^2\right) - t^2
z_5^4 + \sum_{i =1}^4 z_i^4 = 0} where $z_i$ are the homogeneous
coordinates of $P^4$ and $t$ is a real parameter. In \dotd, a
similar but compact model was constructed with Euler number $\chi
= 19728$. The \gkp\ model, that is non--compact has Euler number
$\chi = 1728$.

The fourfold can be constructed by a $T^2$ fibration over this
hypersurface. At a generic point this allows ($p,q$) seven branes.
The above hypersurface can be analyzed at the locus where $z_5 =
1$. We see that \gkpepn\ reduces over this locus to
\eqn\gkpred{\sum_{i = 1}^4 (z_i^4 + z_i^2) = t^2} which, when
taking the real part of $z_i$, becomes the equation for an $S^3$
of a deformed conifold. This is again the limit with the seven
branes far away. The metric {\it away} from the orientifold plane
 can therefore be approximated by the one given in \fiibmet, with
the warp factors $h_i$ as defined in \defjam. At the orientifold
point, this metric clearly cannot suffice, because of the
non--zero $dx~ dz, dy~ dz, dx ~d\theta_2$ and $dy ~d\theta_1$
components. The components $dx ~d\theta_2$ and $dy ~d\theta_1$ can
be eliminated by the usual trick that we applied earlier, namely
using the transformation equation (4.33) of \bdkt. In this way the
starting type IIB background away from the orientifold point (and
with $h$ replaced by $H$ in the warp factors $h_i$) can be
approximated as: \eqn\straiib{\eqalign{ ds^2 = &~ h_1[dz + a_1 dx
+ a_2 dy]^2 + h_2 [dy^2 + d\theta_2^2] + h_4 [ dx^2 + h_3
d\theta_1^2] \cr & + h_5[d\theta_1d\theta_2 - dx dy] +
\gamma'\sqrt{H}~dr^2\cr  B_{NS} = &~ b_{x\theta_1}~dx \wedge
d\theta_1 + b_{y\theta_2} ~dy \wedge d\theta_2}} which is a
K\"ahler background and is in fact precisely the
Klebanov--Strassler background under the transformation (4.33) of
\bdkt\ (and with the $dz$ direction delocalized). We have also
used $x,y$ to denote the coordinates instead of $\hat x, \hat y$
to avoid clutter.

At the orientifold point we still cannot use the above metric
because of the cross terms $dx~dz$ and $dy~dz$. The metric that we
require would follow the criteria laid down earlier (before
geometric transition). The metric will again look like \comtor:
 \eqn\comtornow{ ds^2 = {b}_1~\vert d\chi_1
\vert^2 + {b}_2 ~\vert d\chi_2 \vert^2 + {b}_3 ~dz^2 + {b}_4~
dr^2} but now the complex structures will be different. In fact,
due to the relation (which is easy to see from \defjam):
\eqn\rela{h_5 = 2 a_1 a_2 h_1 ~~ => ~~ {\rm Re}~\tau_1 = 0} but
${\rm Re}~\tau_2 \ne 0$. This is opposite to what we had earlier:
${\rm Re}~\tau_1 \ne 0$ and ${\rm Re}~\tau_2 = 0$ for generic $c$.
This also means that \comtornow\ naturally has $c = 0$.

At this point we see that the $b_i$ defined above are again
functions of the radial and the angular coordinates i.e. $b_i
\equiv b_i(\theta_1, \theta_2, r)$. The generic analysis with
non--trivial functions of $b_i$ is complicated and therefore we
will approximate these coefficients by constants. These constant
values will be motivated by the fact that in some region, when we
move the seven branes away, the space is given by the hypersurface
\gkpred\ and therefore the metric will resemble \straiib. Now
imposing the condition \repmen, and taking the same symbol $h_i$
to represent the expectation values of $h_i$'s, we can write the
complex structure of the two tori in \comtor\ as:
\eqn\cotwton{\tau_1 ~ = ~ i \sqrt{h_2+a_2^2h_1 \o h_4+a_1^2h_1},
~~~~ \tau_2 = {1 \o h_3 h_4} \left[a_1 a_2 h_1 + i \sqrt{h_2 h_3
h_4 - a_1^2 a_2^2 h_1^2} \right].} The coefficients $b_i$ in
\comtornow\ will be given by: \eqn\coefb{b_1 ~=~ h_4+a_1^2h_1,
~~~~~ b_2 = h_3 h_4, ~~~~~ b_3 = h_1, ~~~~~ b_4 = \gamma'
\sqrt{H}} where $H$ has already been derived earlier. The metric
after two T--dualities in type I will again have the same form as
\metinhet, but now $g_{mn}$ will be given by the following matrix:
\eqn\gmnafter{g_{mn}~ =~ \pmatrix{{1\o h_4 + a_1^2 h_1} &0 & 0 &
-{b_{x\theta_1}\o h_4 + a_1^2 h_1} & 0\cr \noalign{\vskip -0.20
cm} \cr
 0 & {h_2 + a_2^2 h_1} & 0 & 0
 & 0 \cr
\noalign{\vskip -0.20 cm}  \cr 0 & 0  & h_1 & 0  & 0 \cr
\noalign{\vskip -0.20 cm}  \cr {-b_{x\theta_1}\o h_4 + a_1^2 h_1}
& 0& 0 & h_3 h_4 + {b^2_{x\theta_1}\o h_4 + a_1^2 h_1} & {h_5/2}
\cr \noalign{\vskip -0.20 cm}  \cr
 0& 0 & 0
 & {h_5/2} & h_2}}
where $h_i$ are the expectation values of $h_i$, unless mentioned
otherwise. Observe that we have taken the $B_{NS}$ field to be
\bnsf\ again with the same simplifying choices that we made earlier.
Observe also that the terms in the above matrix are
different from the ones that we had in \gmn\ or \gmnnow. This is
expected from the choice of type IIB background \straiib\ itself .
On the other hand, the $B$ field appearing in \metinhet\ is much
simpler than its cousin \bmn, and \bmnnow. In matrix form it is
given by: \eqn\bmnlater{B_{mn} ~ =~ \pmatrix{0 & 0 & 0 & 0 & 0\cr
\noalign{\vskip -0.20 cm}  \cr
 0 & 0 & 0 & 0
 & b_{y\theta_2} \cr
\noalign{\vskip -0.20 cm}  \cr 0 & 0  & 0 & 0 & 0 \cr
\noalign{\vskip -0.20 cm}  \cr 0 &  0 & 0 & 0  & 0 \cr
\noalign{\vskip -0.20 cm}  \cr
 0& -b_{y\theta_2}  & 0
 & 0 & 0}}
where as we see, most of the components are zero. The fact that
only $b_{y\theta_2}$ appears is clear, as the other component of
the $B$ field $b_{x\theta_1}$ dissolves in the metric. Using these
matrices, the final type I metric can be written in a concise way
as: \eqn\tyime{\eqalign{ds^2  = & {1\o h_2 + a_2^2 h_1} (dy -
b_{y\theta_2} d\theta_2)^2 + {1\o h_4 + a_1^2 h_1}(dx -
b_{x\theta_1} d\theta_1)^2 \cr & ~~~~~~~~~~~ + h_1 dz^2 +  h_3 h_4
\vert d\chi_2 \vert^2 + \gamma' \sqrt{H}~ dr^2}} with, of course,
vanishing $B_{NS}$.

To show that this metric is indeed the geometric transition dual
of the type I metric in \metinI\ we have to show that it is also
in some sense dual to the IIB metric after geometric transition.
So, we need to find the metric away from the orientifold point,
after two T--dualities. The final metric is of course \metinhet,
but now $g_{mn}$ is given by: \eqn\gmnnowafter{g_{mn}~ =~
\pmatrix{{1\o h_4 + a_1^2 h_1} &0 & 0 & {-b_{x\theta_1}\o h_4 +
a_1^2 h_1} & 0\cr \noalign{\vskip -0.20 cm}  \cr
 0 & {h_2 + a_2^2 h_1} & a_2 h_1 & 0
 & 0 \cr
\noalign{\vskip -0.20 cm}  \cr 0 & a_2 h_1  & {h_1 h_4 \o h_4 +
a_1^2 h_1} & 0  & 0 \cr \noalign{\vskip -0.20 cm}  \cr
-{b_{x\theta_1}\o h_4 + a_1^2 h_1} &  0& 0 & h_3 h_4 +
{b^2_{x\theta_1}\o h_4 + a_1^2 h_1} & h_5/2 \cr \noalign{\vskip
-0.20 cm}  \cr
 0& 0 & 0
 & h_5/2 & h_2}}
where $h_i$ no longer represent the expectation values any more.
Observe that we have, as before, developed non--zero cross terms
$dy ~dz$. And similarly the $dz^2$ term has become more
complicated. The $B$ field also develops more components from the
ones that we had in \bmnlater, and is now given by
 \eqn\bmnafter{B_{mn} ~ =~ \pmatrix{0 & 0 & -{a_1 h_1 \o h_4 + a_1^2 h_1} & 0 & 0\cr
\noalign{\vskip -0.20 cm}  \cr  0 & 0 & 0 & 0
 & b_{y\theta_2} \cr \noalign{\vskip -0.20 cm}
 \cr {a_1 h_1 \o h_4 + a_1^2 h_1} & 0  & 0 &  {b_{x\theta_1} a_1 h_1 \o h_4 + a_1^2 h_1} & 0
\cr \noalign{\vskip -0.20 cm}  \cr 0 &  0 & -{b_{x\theta_1} a_1
h_1 \o h_4 + a_1^2 h_1} & 0  & 0 \cr \noalign{\vskip -0.20 cm} \cr
 0& -b_{y\theta_2}  & 0
 & 0 & 0}}
where again $h_i$ are not the expectation values any more. With
these, the final type IIB metric after two T-dualities will now be
given by: \eqn\tymeup{\eqalign{ds^2  = & {1\o h_2 + a_2^2 h_1} (dy
- b_{y\theta_2} d\theta_2)^2 + {1\o h_4 + a_1^2 h_1}(dx -
b_{x\theta_1} d\theta_1)^2  + \gamma' \sqrt{H}~ dr^2\cr &
~~~~~~~~~~~ + h_1 \left[ {h_2 h_4 - a_1^2 a_2^2 h_1^2 \o (h_4 +
a_1^2 h_1) (h_2 + a_2^2 h_1)}\right] dz^2 + h_3 h_4 \vert d\chi_2
\vert^2.}} Comparing \tymeup\ and \tyime\ we see that imposing
\repmen\ the two metrics are exactly identical up to a possible
warp factor for the $dz$ direction. Thus \tyime\ will form the
dual to type IIB theory after geometric transition.

We now want to consider the corresponding RR background and
coupling in type I theory after geometric transition. For this
again we have to look at the RR background in type IIB theory at
the orientifold point, that is invariant under the orientifold
action. Let us therefore make the following ansatz for the RR
background in type IIB theory: \eqn\brrnow{\eqalign{ H^b_{RR} = &
\tilde{\cal H}^b_{yz\theta_2}~ dy ~\wedge ~dz ~\wedge ~d\theta_2 +
\tilde{\cal H}^b_{xz\theta_1}~ dx ~\wedge ~dz ~\wedge ~d\theta_1 ~
+ \cr & \tilde{\cal H}^b_{xzr}~ dx ~\wedge ~ dz ~\wedge~ dr~ +~
\tilde{\cal H}^b_{yzr}~dy ~\wedge ~dz ~\wedge ~dr}} which is
similar to the $H^b_{RR}$ choice that we made earlier. The values
for $\tilde H^b$ will be given by an equation similar to \hrrfrom\
with ${\cal H}^b$ replaced by $\tilde{\cal H}^b$ with no other
changes. We will be able to compare this with another calculation
later when we go to the heterotic side. The type I coupling and
the RR form can now be determined as \eqn\Ibrr{\eqalign{&
H^I_{RR}~ = ~ \tilde{\cal H}^b_{xz\theta_1}~ dy ~\wedge ~dz
~\wedge ~d\theta_2 - \tilde{\cal H}^b_{yz\theta_2}~ dx ~\wedge ~dz
~\wedge ~d\theta_1 ~ + \cr & ~~~~~~~~~~ + ~\tilde{\cal H}^b_{xzr}~
dy ~\wedge ~ dz ~\wedge~ dr~ -~ \tilde{\cal H}^b_{yzr}~dx ~\wedge
~dz ~\wedge ~dr \cr & ~~~ g^I  = (h_4+a_1^2h_1)^{-1/2}
(h_2+a_2^2h_1)^{-1/2} = \sqrt{C'D'}.}} The coupling turns again
out to be constant as $C'$ and $D'$ are to be understood as
expectation values, i.e. $C$ and $D$ at fixed radius and angle.
Thus, they will be independent of functions of the internal
coordinates of the manifold. The complete background can be
determined once we specify the solution to DUY equation \duy\ as
before.


\newsec{Supergravity Analysis in Heterotic Theory}

Now that we have the type I background, it is time to find the
corresponding heterotic background. This is of course straight
forward, as the heterotic background is simply the S--dual of type
I above. But there is more to it. Since the heterotic background
comes from S--dualizing the type I background, it inherits the $B$
field of type I. This is now interesting because the $B$ field or
equivalently the $H$ field will in fact make this a torsional
background. Therefore we are in the realm of \HULL, \rstrom,
\bbdg, \bbdgs\ where a detailed study of such backgrounds was
performed, albeit, for a compact space. Here we have new torsional
backgrounds that are non--compact, non--K\"ahler but complex.

Having a complex non--K\"ahler background means that we can use
the mathematical construction developed in \bbdg, \GP, \lust,
\louis, \bbdgs. Since all these details have been extensively
described there, we will avoid discussing them here. Our
description will therefore be the bare minimum necessary for
exposing the subject.

{}From the detailed studies done in \bbdg,\bbdgs\ (see also \HULL)
it is clear that these torsional backgrounds allow two different
choices of spin connections $\omega_+$ and $\omega_-$, which are
determined by $\pm{1\o 2} H$. The question now is to see whether
an equivalent of standard embedding \chsw\ is also allowed here.
As discussed in \bbdg\ standard embedding here means the following
equation: \eqn\semd{ A ~ = ~ \omega_+} where $A$ is the gauge
bundle satisfying DUY equations \duy, \trff. The fact that the other spin
connection does not appear in this equation has been explained in
\bbdg. Existence of this equation implies that the ($0,2$)
supersymmetric non--linear sigma model (see second reference of
\HULL) defined on this background satisfying the necessary
conditions is finite to {\it all} orders\foot{To two loop orders
it has been shown in \hullt.}. However, for the torsional
background that was studied in \bbdg, \bbdgs\ and also the one
that we are going to study here, the equation \semd\ is not
allowed. For the compact case, this was argued in \bbdg\ (see also
\hari). Allowing \semd\ would make the warp factor trivial and
therefore the manifold would become K\"ahler. We believe a similar
argument will go through for the non--compact case, too. This does
not imply that the ($0,2$) sigma model is not finite here. The
fact that now $dH \ne 0$ and the manifold is non--compact modify
the argumentation. A full analysis of this has not been done yet
and we hope to address it in the near future.

\subsec{{$\underline{\rm \bf {Background~ before ~ geometric~
transition}}$}}

After this general discussion, let us go back to the precise model
in question. The supergravity background for the heterotic case
can be determined by S--dualizing the type I background. The
metric for our space will be given as \eqn\metinHH{\eqalign{ds^2
~ = ~& \sqrt{\langle\alpha\rangle}(1 + A_1^2)~ (dy -
b_{yj}~d\zeta^j)^2 + \sqrt{\langle\alpha\rangle} (1 +
B_1^2)~ (dx - b_{xi}~d\zeta^i)^2  + {\gamma'\sqrt{H}\o
\sqrt{\langle\alpha\rangle}} dr^2\cr ~&~~~ - 2
\sqrt{\langle\alpha\rangle} A_1 B_1~(dx -
b_{xi}~d\zeta^i)(dy - b_{yi}~d\zeta^j) + {dz^2 +
d_2 \vert d\chi_2\vert^2 \o \sqrt{\langle\alpha\rangle}}.}}
\noindent which looks similar to the type I metric that we had in
\metinI, as it should. The metric has the usual fibration along
the $dx$ and the $dy$ directions and the base is given by
($\theta_1, \theta_2, r, z$) coordinates. The background $B$ field
(whose components depend on $r, \theta_1$ and $\theta_2$ only)
that will serve as the torsion and the coupling constant are given
by: \eqn\bandcoup{\eqalign{& H^{\rm het}~ \equiv ~ H~=~ {\cal
H}^b_{xz\theta_1}~ dy ~\wedge ~dz ~\wedge ~d\theta_2 - {\cal
H}^b_{yz\theta_2}~ dx ~\wedge ~dz ~\wedge ~d\theta_1 ~ + \cr &
~~~~~~~~~~~~~ + ~{\cal H}^b_{xzr}~ dy ~\wedge ~ dz ~\wedge~ dr~ -~
{\cal H}^b_{yzr}~dx ~\wedge ~dz ~\wedge ~dr \cr & ~~~~ g^{\rm het}
~ =~ {1\o \sqrt{\langle\alpha\rangle}}~.}} Along with the solution
to the DUY equation this will specify the complete background.

Let us now use the heterotic superpotential \poth\ to determine
the $B^{\rm het}$. As was discussed in \bbdp, \lustu, one can
minimize the superpotential \poth\ to determine the background
torsional equation as: \eqn\toreq{ H ~ = ~ \ast~dJ} where $J$ is
the fundamental two form. This equation also appeared in
\gauntlett\foot{In fact in \gauntlett\ the torsional equation is
$H = e^{2\phi}\ast d(e^{-2\phi}J)$ for manifolds with $SU(3)$
holonomy. This reduces to the torsional equation that we proposed
when the dilaton is a constant which, for our case, is true.}.
Having a non--K\"ahler manifold means that it will support torsion
$H$. For a complex manifold, this equation can be re--written as
\eqn\streqn{H ~ = ~ i (\del - \bar\del) J} which is the familiar
way the torsional equation first appeared in \rstrom, \HULL,
\hari. In the analysis to follow we will, however, use \toreq\
because it is written in terms of real variables. To proceed
further, we first need the vielbeins for our space. They are given
by: \eqn\vielbeins{\eqalign{& e^1 =  {1\o
\sqrt{2}}{\langle\alpha\rangle}^{1\o 4}(1 + A_1^2)^{1\o 2}
\Big[(dy - b_{yj}~d\zeta^j) + \gamma_2 (dx -
b_{xi}~d\zeta^i) \Big] \cr & e^2 = {1\o
\sqrt{2}}{\langle\alpha\rangle}^{1\o 4} (1 + A_1^2)^{1\o 2}
\Big[(dy - b_{yj}~d\zeta^j) + \gamma_3 (dx -
b_{xi}~d\zeta^i) \Big]\cr & e^3 = {d_2^{1\o 2} \o
\langle\alpha\rangle^{1\o 4}}~d\theta_1, ~~ e^4 =
{\vert\tau_2\vert d_2^{1\o 2} \o {\langle\alpha\rangle}^{1\o
4}}~d\theta_2, ~~ e^5 = {\gamma^{'{1\o 2}} H^{1\o 4} \o
{\langle\alpha\rangle}^{1\o 4}} dr, ~~ e^6 = {dz \o
{\langle\alpha\rangle}^{1\o 4}}}}
where we have already defined
$d_2$ and $\tau_2$. The other variables $\gamma_2$ and $\gamma_3$
are given as: \eqn\gamdef{\gamma_2 = -{1\o 1 + A_1^2}\left[ A_1
B_1 \pm {1\o \sqrt{\langle\alpha\rangle}}\right], ~~~~ \gamma_3 =
-{1\o 1 + A_1^2}\left[ A_1 B_1 \mp {1\o
\sqrt{\langle\alpha\rangle}}\right].} Having the vielbeins, we now
need to determine the fundamental form $J$, so that we can
evaluate $dJ$, the deviation from K\"ahlerity. The fundamental
form is given by \eqn\fundform{J = e^1 \wedge e^2 + e^3 \wedge e^4
+ e^5\wedge e^6 = \left[J_1 \right]_{b_{x\theta_1} = b_{y\theta_2}
= 0} + \left[J_2\right]} where $dJ_2$ measures the actual deviance
from K\"ahlerity because $dJ_1 = 0$ by definition. Now we can
easily evaluate $J_2$ from the vielbeins \vielbeins, it is found
to be: \eqn\jtwo{J_2 = - \left[b_{xi}~dy \wedge d\zeta^i -
b_{yj}~dx \wedge d\zeta^j + b_{xi}
b_{yj}~d\zeta^i \wedge d\zeta^j \right].}
Using \toreq\
we get the following results for the three form $H$:
\eqn\bghinhet{\eqalign{&
H_{rxz} = \del_{[r} B^{\rm het}_{xz]} = \eta_1 ~\del_{[\theta_2}
b_{x\theta_1]}, ~~~ H_{\theta_2 xz} = \del_{[\theta_2} B^{\rm
het}_{xz]} = \eta_2 ~\del_{[r} b_{x\theta_1]} \cr & H_{\theta_1 yz} =
\del_{[\theta_1} B^{\rm het}_{yz]} = \eta_3 ~\del_{[r} b_{y\theta_2]},
~~~ H_{ryz} = \del_{[r} B^{\rm het}_{yz]} = \eta_4 ~\del_{[\theta_1}
b_{y\theta_2}]}} along with another relation: \eqn\anrela{ H_{xyz}
~ = ~ \eta_5~\del_{[x}B^{\rm het}_{yz]} ~ = ~ 0} which vanishes
because we have taken the $B^{\rm het}$ to be functions of
($\theta_1, \theta_2, r$) and independent of ($x, y, z$). The
variables $\eta_i$ can be evaluated from the epsilon tensor and
Hodge star.

To compare \bghinhet\ with \hrrfrom, observe that
\eqn\obsbfie{H^{\rm het}_{xz(\theta_1,r)} ~ = ~ - {\cal H}^b_{yz(\theta_2, r)}, ~~~~~~ H^{\rm
het}_{yz(\theta_2, r)} ~=~ {\cal H}^b_{xz(\theta_1, r)},} which is basically \bandcoup. Plugging this
in \bghinhet\ we get: \eqn\bghnoww{\eqalign{& {\cal H}^b_{ryz} = -
\eta_1~\del_{[\theta_2}b_{x\theta_1]}, ~~~~~~~~~~~ {\cal
H}^b_{\theta_2 yz} = - \eta_2 ~\del_{[r} b_{x\theta_1]} \cr & {\cal
H}^b_{\theta_1 xz} = \eta_3~\del_{[r} b_{y\theta_2]}, ~~~~~~~~~~~
{\cal H}^b_{rxz} = \eta_4~\del_{[\theta_1} b_{y\theta_2]}.}}
Comparing \bghnoww\ with \hrrfrom, we see that if we impose the
following conjectured conditions on $\alpha_i$ and $\eta_i$
\eqn\condonae{\alpha_1 = -\eta_1, ~~~~ \alpha_2 = -\eta_2, ~~~~
\alpha_3 = \eta_3, ~~~~ \alpha_4 = \eta_4}
then we obtain perfect
equality!  The reason why we called this  {\it conjectured
condition} is because $\alpha_i$ are constants as we had used the
limit \repmen\ to go to the metric at the orientifold point. Away
from this limit, $\alpha_i$ do not need to be constants. In that
case they have to be determined using the $B$ fields, especially
$b_{xi}$ and $b_{yj}$. Therefore, we expect
$\alpha_i$ to be functions of $b_{xi}$ and $b_{yj}$.
On the other hand $\eta_i$ are in fact functions of
$b_{xi}$ and $b_{yj}$, as can be easily seen by
working out the Hodge star. This is exactly what the equality
\condonae\ implies here.

\subsec{{$\underline{\rm \bf {Background~ after ~ geometric~
transition}}$}}

After the geometric transition in type IIB theory the dual metric
from the type I picture can be easily brought in the heterotic
framework via S--duality in the same manner as before. The metric
is now: \eqn\tyhet{\eqalign{ds^2  = & {1\o (h_2 + a_2^2
h_1)\sqrt{C'D'}} ~(dy - b_{yj}~ d\zeta^j)^2 + {1\o (h_4 +
a_1^2 h_1)\sqrt{C'D'}}~(dx - b_{xi} ~d\zeta^i)^2 \cr &
~~~~~~~~~~~ + {h_1 \o \sqrt{C'D'}} ~dz^2 +  {h_3 h_4 \o
\sqrt{C'D'}} ~\vert d\chi_2 \vert^2 + {\gamma' \sqrt{H} \o
\sqrt{C'D'}}~ dr^2.}} which is again similar to the type I picture
developed earlier, i.e. \tyime. The torsion and the coupling are
\eqn\torcnow{\eqalign{& H_{\rm het}~ \equiv \tilde H ~=~ \tilde{\cal H}^b_{xz\theta_1}~
dy ~\wedge ~dz ~\wedge ~d\theta_2 -
\tilde{\cal H}^b_{yz\theta_2}~ dx ~\wedge ~dz ~\wedge ~d\theta_1 ~ + \cr
& ~~~~~~~~~~ + ~\tilde{\cal H}^b_{xzr}~
dy ~\wedge ~ dz ~\wedge~ dr~ -~ \tilde{\cal H}^b_{yzr}~dx ~\wedge ~dz ~\wedge ~dr \cr
& ~~~~~ g^{\rm het}
= { 1\o \sqrt{C'D'}}}} which, as before, along with the solution to
the DUY equation will specify the complete background.

To verify the torsional equation we can follow the same steps,
with the $H$'s replaced by $\tilde H$. It is easy to see that
torsional equations similar to \bghnoww\ are satisfied here, too.
However, for completeness we give the vielbeins for this case:
\eqn\vielsnow{\eqalign{& e^1 = (h_2 + a_2^2 h_1)^{-{1\o 2}}
(C'D')^{-{1\o 4}} (dy - b_{yj}~ d\zeta^j), ~~~~ e^2 = (h_4
+ a_1^2 h_1)^{-{1\o 2}} (C'D')^{-{1\o 4}} (dx - b_{xi}
~d\zeta^i) \cr & e^3 = {1\o \sqrt 2}(h_3 h_4)^{1\o 2}
(C'D')^{-{1\o 4}}(d\theta_1 + \gamma_4 d\theta_2), ~~~~ e^4 = {1\o
\sqrt 2}(h_3 h_4)^{1\o 2} (C'D')^{-{1\o 4}}(d\theta_1 + \gamma_5
d\theta_2)\cr & e^5 = \gamma^{'{1\o 2}} H^{1\o 4} (C'D')^{-{1\o
4}} dr, ~~~~~ e^6 = h_1^{1\o 2} (C'D')^{-{1\o 4}} dz}} where we
now see that the vielbeins $e^1$ and $e^2$ are simple because
${\rm Re} ~\tau_1 = 0$, whereas the vielbeins $e^3$ and $e^4$
contain mixed components because ${\rm Re}~\tau_2 \ne 0$. For
\vielbeins\ this was exactly the opposite. It is intriguing to see
that, for generic $c$,
before geometric transition the heterotic dual manifold has
${\rm Re}~\tau_1 \ne 0, {\rm Re}~\tau_2 = 0$ and after geometric
transition the heterotic dual manifold has ${\rm Re}~\tau_1 = 0,
{\rm Re}~\tau_2 \ne 0$. The variables $\gamma_4$ and $\gamma_5$
appearing in \vielsnow\ are defined as: \eqn\gamviel{\gamma_4 =
{\rm Re}~\tau_2 \pm {\rm Im}~\tau_2, ~~~~ \gamma_5 = {\rm
Re}~\tau_2 \mp {\rm Im}~\tau_2} where $\tau_2$ was defined in
\cotwton.

At this point let us pause a little to ask what the two
backgrounds derived above mean. It is clear that the dual
background before geometric transition is related to $NS5$ branes
wrapped on two cycles of the non--K\"ahler manifold \metinHH.
Following the duality chain, this comes from the $D5$ wrapped on a
two cycle of a resolved conifold (which survives the orientifold
operation). After the geometric transition in type IIB theory 
we get a dual background in heterotic
that has only fluxes but no branes. The question here is whether
we can shrink the two cycle on which we have wrapped $NS5$ branes
and get another background with fluxes. At a more fundamental
level, can the closed string background with fluxes compute
anything of the world volume dynamics on the wrapped $NS5$ branes?
We do not have any answer to this question yet. We will discuss a
bit more on this in the next section from the superpotential point
of view.

A similar question can also be raised for the type I setting. The
dynamics on the wrapped $D5$ branes are known. It is the same
theory as for the type IIB case. Whether the closed string
background that we gave earlier is a dual background for the $D5$
brane theory can only be clear after some concrete analysis is
performed.


\newsec{Field Theory and Geometric Transitions in Type I and Heterotic String}

We now consider the primary region of interest in the above
discussion, i.e. the orientifold point. The field theory on the D5
brane will change from the usual one to a theory describing D5
branes wrapped on a $P^1$ cycle of a resolved conifold. In the
language of type I, the paper \wisd\ discussed the field theory on
$N$ D5 branes which appear as $N$ instantons. When all the $N$ D5
branes coincide, the field theory gains an unbroken $Sp(N)$
symmetry. The T--dual type IIB picture will replace the  $N$ D5
branes with another set of $N$ D5 branes and there will exist an
unbroken $Sp(N)$ symmetry on the D5 branes.

The field theory on the $N$ wrapped D5 branes is similar to the
one on the probe D3 branes considered in type IIB in \dolo. The
difference is that the wrapped D5 branes are fixed on the $P^1$
cycle so that the antisymmetric hypermultiplet (which comes from
projecting an ${\cal N} = 2$ adjoint field) is not part of the
theory. We only have an ${\cal N} = 2, Sp(N)$ theory with four
flavors of matter field in the fundamental representation (given
by the four D7 branes which touch the D5 brane).

What is the superpotential of the theory? This can be obtained
from the field configuration present after the transition. It
involves $H_{RR}$ instead of D5 branes and $H_{NS}$. Besides, the
orientifold O7 plane and the D7 branes survive the transition.
They will be grouped identically as in the configuration before
the transition. Therefore one is still at the orientifold point
where the axion and dilaton do not vary. The well known
superpotential \eqn\super{ W_{IIB} = \int \left(H_{RR} + \tau
H_{NS}\right) \wedge \Omega,} still appears. Here $\Omega$ is the
usual holomorphic (3,0) form of the base discussed earlier. But
there is another contribution to the superpotential which comes
from the D7 branes and this can be calculated using arguments
similar to the ones in \rrj. The contribution of the D7 branes
appears as a Chern--Simons term analog to the holomorphic
Chern--Simons computation for D5 branes wrapped on holomorphic
cycles. Even though it is now more difficult to identify the
four--cycles around which the D7 branes are wrapped, the
computation of the Chern--Simons action as a function of the
complex structure of the manifold would proceed similarly to \rrj.

We now study the transformation of the type IIB superpotential
under two T--dualities. In the type I theory we get several
contributions to the superpotential. The first one appears as
$\int dB_{RR} \wedge \Omega,$ which originates from the similar
term appearing in the type IIB theory. The second one comes as $d
J \wedge \Omega,$ \bbdp, \lustu\ and is due to the fact that the
type I theory is compactified on a non--K\"ahler manifold which
has a non--zero $dJ$.

The third and the most intriguing  one appears as a holomorphic
Chern--Simons functional \eqn\cs{W^A_{CS}=\int {\rm tr} \left(A
\wedge d A + \frac{2}{3} A \wedge A \wedge A \right) \wedge
\Omega,} where $A$ is the SO(32) gauge field. One would naively
think that this term appears as an extra term compared to the type
IIB theory but, in view of the above discussion, it is not. This
contribution actually arises rather naturally because of the extra
Chern--Simons contribution coming from the field theory on the D7
branes, which survive the transition\foot{For a derivation of this
see \bbdg.}. The two Chern--Simons terms can, and should, be
mapped into each other.

The most generic superpotential for the type I theory is then
\bbdp,\lustu\ \eqn\total{W_I = \int \left( dB_{RR} + i d J\right)
\wedge \Omega + W^A_{CS} + W^\omega_{CS},} where $W^A_{CS}$ and
$W^\omega_{CS}$ are the gauge and the gravitational Chern--Simons
terms, respectively. This will describe the low--energy effective
theory on the small instanton D5 branes of the type I superstring.

Performing an S--duality to go from type I string to the heterotic
string, the D5 branes become NS5 branes, the field $H_{RR}$
becomes $H_{NS} \equiv H$ and the compactification manifold
remains non--K\"ahler. The superpotential therefore is \bbdp,
\lustu\ \eqn\hetsuper{W_{het} = \int \left(H + i dJ \right) \wedge
\Omega,} where $H$ is the usual three--form of the heterotic
theory satisfying $dH = {\rm tr} ~R \wedge R - {1 \o 30} {\rm
tr}~F \wedge F$. Furthermore, the D9 branes and $O9$ planes have
disappeared and the flavor degrees of freedom now live on the
$SO(32)$ vector bundle\foot{In the examples that we studied in
this paper the gauge group is broken to a smaller subgroup by
Wilson lines. These Wilson lines map to the {\it distances}
between the stack of $D7$ branes in the original type IIB
picture.}.

What about the possibility of having a gluino condensate in the
heterotic string as described in \rohm ? The combination of fluxes
and a gluino condensate has been considered in \lustd\gukkac\
where it was argued that the superpotential \hetsuper\ becomes
\eqn\hetsupergc{W_{het} = \int \left(H + i dJ + \Sigma\right)
\wedge \Omega,} where $\Sigma$ is the expectation value for
$tr({\bar\chi \Gamma_{mnp} \chi})$, $\chi$ being the gluino. For
compact Calabi--Yau, considered in \gukkac, this typically fixes
both the radius and the dilaton non--perturbatively\foot{There is
an interesting simplification when we go to the compact
non--K\"ahler case. Due to the torsional equation \toreq\ {\it
and} the Bianchi identity of $dH^{\rm het}$ the scaling freedom of
the fundamental two form $J$ is broken. This means that the radius
of the manifold is fixed at tree level! The fact that there exists
a potential for the radial modulus which has a minimum was shown
in \bbdp. Thus, this leaves us to fix only the dilaton
non--perturbatively.}.

The gluino condensate in equation \hetsupergc\ is different from
the gluino condensate for the field theory living on the NS5
branes. The NS5 brane is the S--dual of the type I D5 brane and
the field theory on the NS5 brane appears in the decoupling limit
as discussed in \mn. The gauge coupling constant for the theory on
the NS5 brane is inverse proportional to the volume of the $P^1$
on which the NS5 is wrapped on.

The bulk gauge coupling constant is given by $e^{\phi}$, where
$\phi$ is the four dimensional dilaton given by \gukkac
\eqn\dilf{\phi~=~\frac{\Phi}{2} - 6 \sigma,} where $\Phi$ is the
ten dimensional dilaton and $\sigma$ is the volume modulus. The
volume scalar $\rho$ is \gukkac \eqn\rhoo{\rho~=~\frac{\Phi}{2} +
2 \sigma.} Now, because the internal manifold is non--compact,
$\sigma$ is infinite so $\phi$ goes to minus infinity and $\rho$
goes to infinity. Because the gluino condensate is the exponential
of a combination of $- e^{\rho}$ and $- e^{- \phi}$, in this limit
it reduces to zero. This is indeed the expected result for a
non--compact manifold. This is because the initial type IIB
picture contains non--compact D7 branes and the gluino condensate
on the D7 branes is then zero.

\subsec{{$\underline{\rm \bf {Geometric~Transitions}}$}}

We now would like to discuss the geometric transitions which type
I strings and heterotic strings should inherit from the type IIB
theory after T--dualities and S--duality.

In type IIB theory geometric transitions, the superpotential is a
holomorphic function which is computed using topological strings
on both the open string and the closed string side. For the case
of type IIA theory on the conifold, the results have been matched
in \vafai, where the non--K\"ahlerity and non--complexity on the
closed string side was first observed. The non--zero $d \Omega$ on
the closed string side controls the field theory bare coupling
constant.

The question now is whether the same topological string
computations can be performed for our case. In view of the
original introduction of topological strings \witt, these objects
are obtained after twisting two dimensional (2,2) theories.
Progress has also been made in defining topological strings for
heterotic string theory \antoniad. Even though the left--right
symmetry of the worldsheet is absent in the heterotic case, the
F--term of the ${\cal N} = 1$ supersymmetric theory is related to
topological quantities. The topological partition function is
related in this case to F--terms like $W^{2g}$, where $W$ is now
the ${\cal N} = 1$ gauge multiplet.

It would be very interesting to go further in this direction. This
could provide us with a recipe for computing the superpotential by
using the twisted topological (0,2) theory. One should remember
that there is an extra ingredient due to the departure of the
metric from being complex and K\"ahler. In terms of topological
strings, there should be an extra change similar to the one
appearing in the recent proposal of \kapustin\ for the (2,2)
theories. A similar change due to the presence of torsion would be
naturally expressed in terms of some generalized complex geometry.
In what concerns the actual computation of the superpotential in
terms of the topological string, it should be shown that the new
open topological strings contribute in a similar way to the ones
of the A--model and B--model. Once such a picture becomes clear,
then it will be possible to see whether we can really observe a
geometric transition in heterotic or type I theory. The wrapped
$D5$ branes on a two--cycle of the non--K\"ahler manifold in type
I theory should be replaced by the other background that we gave
which had fluxes supported on another non--K\"ahler manifold. The
dynamics on the wrapped $D5$ should also be obtained from this
background\foot{We are assuming here that the theory on the
wrapped $D5$ branes is decoupled theory from the bulk dynamics.
Otherwise we have to deal with a gauge theory coupled with
gravity.}.

Another interesting future direction is to use our
results as a tool for understanding
 mirror symmetry for (0,2) theories \blu\as\sk.
As we have the metrics and we know how to transform between mirror
symmetric solutions in terms of T--dualities, one could try to
describe the mirror of the heterotic theory.

\vskip.2in

\centerline{\bf Acknowledgments}

Its our pleasure to thank Gianguido Dall' Agata, Shamit Kachru,
Amir-Kian Kashani-Poor, Dieter Luest, and Mohammed M.
Sheikh-Jabbari for many interesting discussions and useful
correspondences. The work of S.A. is supported by US DOE under
grant DE-AC03-76SF00515. The work of K.B. is supported by NSF
grant PHY-0244722, an Alfred Sloan Fellowship and the University
of Utah. The work of M.B. is supported by NSF grant PHY-01-5-23911
and an Alfred Sloan Fellowship. The work of K.D. is supported in
part by a Lucile and David Packard Foundation Fellowship
2000-13856. A.K. would like to acknowledge support from the
University of Maryland.~R.T. is supported by DOE Contract
DE-AC03-76SF0098 and NSF grant PHY-0098840.


\listrefs

\bye